\documentclass[11pt,letterpaper]{article}
\pdfoutput=1
\usepackage{jheppub}
\usepackage{mathrsfs}
\usepackage{slashed}
\usepackage{caption}
\usepackage{epstopdf}
\usepackage[bottom]{footmisc}
\usepackage{subcaption}
\usepackage{titlesec}
\usepackage{threeparttable}
\usepackage{booktabs}
\usepackage{changepage}
\usepackage[utf8]{inputenc}
\usepackage{grffile}
 
\usepackage{graphicx}  
\usepackage{dcolumn}   
\usepackage{bm}        
\usepackage{amssymb}   
\usepackage{setspace}
\usepackage{amsmath, amssymb, setspace}
\usepackage{array}
\usepackage{booktabs}
\usepackage{caption}
\usepackage{indentfirst}
\usepackage{float}
\usepackage{lmodern}
\usepackage{multirow}
\usepackage{soul}
\usepackage{braket}
\usepackage{comment}

\usepackage[draft]{pgf}

\usepackage{adjustbox} 
\renewcommand\afterProceedingsSpace{\vskip 6pt}
\renewcommand\afterTitleSpace{\vskip 8pt}
\renewcommand\afterRuleSpace{\vskip 10pt}
\renewcommand\afterEmailSpace{\vskip 6pt}

\newlength{\myimageoversize}
\newsavebox{\myimage}

\def\={~=~}
\def\+{~+~}
\def\-{~-~}  

\usepackage[most]{tcolorbox}

\usepackage{empheq}

\tcbset{colback=white!10!white, colframe=black!50!black, 
        highlight math style= {enhanced, 
            colframe=black,colback=white!10!white,boxsep=0pt}
        } 

\titleformat{\subsubsection}
 {\normalfont\fontsize{12}{17}\itshape}{\thesubsubsection}{1em}{}

 \makeatletter

\renewcommand{\l@section}[2]{%
\addvspace{5pt}
  \begingroup
  \bfseries 
  \@dottedtocline{1}{0em}{1.5em}{#1}{#2}%
  \endgroup
}

\renewcommand{\l@subsection}{%
  \@dottedtocline{2}{1.5em}{2.3em}%
}

\makeatother

\title{Gravitational Wave Diffractive Images Reveal the Structure of Astrophysical Objects}
\author[a,b]{Han Gil Choi,}
\author[c,d]{Sunghoon Jung,} 
\author[e,f,g,h]{Volodymyr Takhistov} 

\affiliation[a]{Max Planck Institute for Gravitational Physics (Albert Einstein Institute) \\ Am M\"uhlenberg 1, D-14476 Potsdam-Golm, Germany}

\affiliation[b]{Cosmology, Gravity, and Astroparticle Physics Group, Center for Theoretical Physics of the Universe,
Institute for Basic Science (IBS), Daejeon, 34126, Korea}
\affiliation[c]{Center for Theoretical Physics, Department of Physics and Astronomy,
Seoul National University, Seoul 08826, Korea}
\affiliation[d]{Astronomical Research Center,
Seoul National University, Seoul 08826, Korea}
\affiliation[e]{International Center for Quantum-field Measurement Systems for Studies of the Universe and Particles (QUP,WPI), High Energy Accelerator Research Organization (KEK), Oho 1-1, Tsukuba, Ibaraki 305-0801, Japan}
\affiliation[f]{Theory Center, Institute of Particle and Nuclear Studies (IPNS),
High Energy Accelerator Research Organization (KEK), Tsukuba, Ibaraki 305-0801, Japan} 
\affiliation[g]{Graduate University for Advanced Studies (SOKENDAI), \\
1-1 Oho, Tsukuba, Ibaraki 305-0801, Japan}
\affiliation[h]{Kavli Institute for the Physics and Mathematics of the Universe (WPI), Chiba 277-8583, Japan}

\emailAdd{hangil.choi@aei.mpg.de} 
\emailAdd{sunghoonj@snu.ac.kr}
\emailAdd{vtakhist@post.kek.jp}

\abstract{ 
Scattering patterns encode the structure of extended objects, as demonstrated by charged particle scattering on atomic nuclei. Gravitational wave (GW) diffraction  offers an analogous probe of otherwise invisible mass distributions.  
At linear order in the lens potential  we derive expressions relating the central diffractive image contribution to a gravitationally lensed GW waveform to the Fourier transform of the projected lens density. For finite mass lenses  this contribution takes the form of a point mass Born amplitude multiplied by a normalized mass form factor, analogous to that in nuclear scattering.
As the frequency of GWs from a merging compact binary source evolves  the signal samples this form factor across spatial scales, imprinting lens structure through interference with the primary image. Using full numerical wave optics calculations we quantify the observational reach for extended dark matter objects, primordial black holes halo-dressed by dark matter halos and superradiance clouds of ultralight bosons around black holes.  For the considered source populations, five year observations with next generation detectors can distinguish extended lenses from point masses, probing dark matter fractions down to $\mathcal{O}(10^{-5})$, and halo-dressed primordial black holes from matched bare black holes, probing primordial black hole dark matter fractions down to $\mathcal{O}(10^{-6})$. For sufficiently close source-lens configurations, scalar superradiance clouds can produce distinguishable diffraction signatures for boson masses of approximately $10^{-17}-10^{-12}~\mathrm{eV}$. These results provide a framework for characterizing extended dark matter and compact object astrophysical environments. 
}

\begin{document}
\preprint{\small\shortstack[r]{%
KEK-QUP-2026-0019\\
KEK-TH-2882}}
\maketitle


\section{Introduction}

Determining how matter is distributed within astrophysical objects is essential for understanding their nature and origin. Although dark matter (DM) constitutes $\sim 85\%$ of all the matter in the
Universe~\cite{Planck:2018vyg}, its nature remains unknown (e.g.~\cite{Bertone:2004pz,Gelmini:2015zpa}).
DM scenarios encompass structures spanning a wide
range of sizes and compactness, from primordial black holes (PBHs) to
extended halos, solitonic cores and other self-bound
configurations.
Their spatial mass distributions encode information about the
underlying physics and formation history. Related questions
arise in ordinary astrophysical systems including nuclear
star clusters and gas disks around black holes, whose structure
influences how compact objects form, grow, and merge
\cite{Kritos:2022non}.
Measuring the total mass alone is insufficient to distinguish
systems with different internal structures and physical origins.
Laboratory experiments based on scattering have transformed our understanding of
internal structure. Rutherford's interpretation of
$\alpha$-particle scattering~~\cite{Rutherford:1911zz} established the nuclear model of
the atom, while elastic electron-scattering measurements
resolved nuclear charge distributions~\cite{Hofstadter:1956qs}. However, distant astrophysical
objects cannot be studied in this way and their investigation
relies on naturally occurring sources and on interactions
sensitive to the matter of interest.

Gravitational waves (GWs) provide a powerful opportunity to
pursue this challenge, offering access to mass distributions of astrophysical objects
independently of their luminosity. 
Observations of binary black hole (BBH) and neutron star mergers now yield a large and steadily increasing sample of signals whose source properties can be inferred from  waveforms
\cite{LIGOScientific:2025hdt,LIGOScientific:2025slb}. A myriad of such sources is expected to be detected in the upcoming future. GW lensing offers a particularly intriguing venue in this context as it is sensitive to intervening mass distributions between the source and observer \cite{Schneider:1992bmb,Schneider:2006qyj}.
Dedicated searches for GW lensing are already underway across LIGO-Virgo-KAGRA observing runs
\cite{LIGOScientific:2021izm,LIGOScientific:2023bwz,LIGOScientific:2025cwb}. GW231123 was identified as an outlier in the O4a search for lensing-induced distortions, and its possible interpretation as a GW lensing event remains under active investigation~\cite{LIGOScientific:2025cwb,Goyal:2025eqo,Chakraborty:2025pxt,Shan:2025dcd,Cheung:2026pky}.

A distinctive feature of GW lensing arises from the wave nature of GWs, resulting in a prominent role for wave optics effects \cite{nakamura1999wave,takahashi2003wave}.  Lensing of GWs differs qualitatively from lensing of light. The long wavelengths of GWs enable diffraction and interference to remain important for comparatively light lenses~\cite{nakamura1999wave,takahashi2003wave}. While the resulting images are challenging to resolve with GW detectors~\cite{abbott2020prospects}, lensing manifests as frequency-dependent distortions of waveforms. Such wave optics distortions have motivated investigations and forecasts for compact and extended lenses such as intermediate mass black holes and PBHs~\cite{lai2018discovering,Jung:2017flg,Diego:2019rzc,Urrutia:2021qak,Basak:2021ten,Zhou:2022yeo,GilChoi:2023ahp}, low mass DM halos~\cite{dai2018detecting,Choi:2021bkx,Guo:2022dre,Fairbairn:2022xln,Tambalo:2022wlm,Savastano:2023spl,Caliskan:2023zqm,Cheung:2024ugg,Brando:2024inp}, extended dark objects of finite size such as boson stars~\cite{Blinov:2026bxj}, and microlenses embedded in strong lensing systems~\cite{Diego:2019lcd,Mishra:2021xzz,Shan:2022xfx,Shan:2024min,Seo:2025dto,Zumalacarregui:2026uqs}. 
The amplitude and phase fluctuations produced by lens populations have been studied theoretically~\cite{Macquart:2004sh,Takahashi:2005ug,Inamori:2021tlx,Mizuno:2022xxp,Mizuno:2023pru,Tanaka:2025nfw,Nakazono:2026vei} and as probes of compact objects, low mass halos, and small-scale matter fluctuations~\cite{oguri2020probing,Urrutia:2024pos,Zumalacarregui:2024ocb,Kim:2025njb,Choi:2026lsa}. 
Wave optics features can also sharpen lens-parameter inference, distinguish parametric lens models~\cite{lai2018discovering,Cremonese:2021ahz,Tambalo:2022wlm,Cheung:2024ugg,Brando:2024inp}, and break lensing degeneracies~\cite{Cremonese:2021puh,Chen:2024xal,Meena:2024glj}.

Much of the existing work proceeds model by model where a specific GW lens profile is assumed, and one asks how well its parameters can be measured or its signal discriminated from that of another model. On the other hand, in elastic electron-nucleus scattering  the nuclear charge form factor makes the probed spatial structure explicit. A similarly explicit form factor description of the central diffractive contribution in GW lensing has remained less developed. Such a description would clarify how the lens profile enters the diffraction pattern and provide a basis for comparisons beyond a fixed set of assumed profiles.
Perturbative and asymptotic analyses have established partial connections~\cite{Takahashi:2005sxa,Mizuno:2022xxp,Yarimoto:2024uew,Wu:2025rvy,Tanaka:2026bql}. At linear order in the lens potential, the diffraction signal is related to the mean convergence within a Fresnel-scale aperture around the projected source position~\cite{Choi:2021bkx}.  A distinct contribution from the lens center, interpreted as a diffracted image, was identified for particular cases of cuspy or singular mass profiles with a frequency dependence set by the central density slope~\cite{takahashi2004quasi}. In fact, such a contribution can also arise for regular lens profiles, but its relation to a general projected mass distribution has remained less explicit.

In this work we identify how the mass distribution of a lens is encoded in GW diffraction, establishing the gravitational counterpart of the role played by the charge distribution in nuclear scattering. 
At linear order in the lens potential, corresponding to the Born approximation, and when the Fresnel scale is small compared with the source impact parameter,
the central diffractive image of a lens of finite total mass is the point mass contribution multiplied by the mass form factor of the lens. 
The form factor is the Fourier transform of the projected mass distribution normalized by the total mass, evaluated at a wave-vector transfer set by the source-lens geometry.
This is the gravitational analogue of the charge form factor through which electron scattering revealed nuclear structure~\cite{Hofstadter:1956qs}. For a fixed lens-source geometry  higher frequency GWs probe finer spatial structures. Thus, a single lensed event spanning a range of frequencies enables sampling the lens form factor across spatial scales. Within these approximations, and provided that the stationary-phase evaluation is controlled the result applies to arbitrary projected mass distributions, with no central cusp or axial symmetry required. This complements the description of diffraction near the projected source position~\cite{Choi:2021bkx}. 
The correspondence between the diffraction integral and scattering amplitudes has been developed with a focus on point mass lenses~\cite{CarrilloGonzalez:2025gqm}. That treatment does not explicitly identify the projected mass form factor governing the central diffractive contribution of an extended lens.
We also determine when this structural information is observable.

The rest of this paper is organized as follows. In Sec.~\ref{sec:waveoptics} we develop the diffractive image and GW lensing form factor description of diffraction. Then in Sec.~\ref{sec:Distinction} we introduce the matched comparisons that isolate lens structure, and Sec.~\ref{sec:statistics} presents the statistical framework. Sections~\ref{sec:prospectextended}-\ref{sec:cloud} present the applications to extended astrophysical objects and Sec.~\ref{sec:conclusion} concludes.

\section{Wave Optics of Gravitational Lensing and Diffractive Images} \label{sec:waveoptics}
\begin{figure}[t]
\centering
\includegraphics[width=0.9\linewidth]{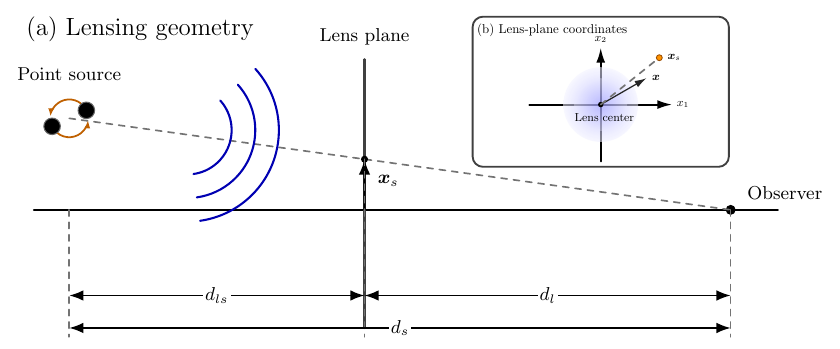}
\caption{Schematic illustration of the lensing geometry and lens-plane coordinates. (a) GWs emitted by a merging binary black hole, treated as a point source, propagate toward the observer through the lens plane. (b) The inset defines the lens-plane coordinates: $\bm{x}$ denotes a position on the lens plane, and $\bm{x}_s$ is the projected source position on the lens plane.}
\label{fig:geometry}
\end{figure}

In this section, we summarize the wave optics formalism used throughout this work and develop the connection between diffraction and the projected mass distribution of lenses. This connection provides the basis for using diffraction to distinguish lens structures in subsequent sections. Unless otherwise stated, we adopt units  $G=c=1$ throughout this work.

\subsection{Wave optics formalism and diffraction}

The central quantity in wave optics lensing is the amplification factor,
\begin{equation}
F(f) = \frac{h_L(f)}{h_0(f)},
\end{equation}
where $h_L(f)$ and $h_0(f)$ are the lensed and unlensed gravitational waveforms in the frequency domain, respectively. Assuming a point source and adopting the scalar wave, thin lens, and small angle approximations, the amplification factor can be written as~\cite{takahashi2003wave}
\begin{equation}\label{eq:F(f)}
F(f) = \frac{f(1+z_l)}{i d_\mathrm{eff}}\int d^2x e^{2\pi i \frac{f(1+z_l)}{d_\mathrm{eff}} \varphi (\bm{x})}~,
\end{equation}
where $z_l$ is the lens redshift and $d_\mathrm{eff}=d_l d_{ls}/d_s$ is the effective lens distance. Here, $d_l$, $d_s$, and $d_{ls}$ are the angular diameter distances from the observer to the lens, from the observer to the source, and from the lens to the source, respectively.

The lensing geometry and the coordinates used are illustrated in Fig.~\ref{fig:geometry}.
The wave phase is determined by the Fermat potential  
\begin{equation}
\varphi(\bm{x}) = \frac{1}{2}|\bm{x}- \bm{x}_s|^2 -\psi(\bm{x}) - \varphi_m(\bm{x}_s)~.
\end{equation}
Here, $\bm{x}$ denotes a position on the lens plane and $\bm{x}_s$ is the projected source position on the lens plane relative to the lens center. The offset $\varphi_m(\bm{x}_s)$ is chosen such that the minimum of $\varphi(\bm{x})$ is zero. 
The gravitational effect of the lens is encoded in the two dimensional lensing potential $\psi(\bm{x})$, obtained from the Newtonian gravitational potential by projection along the line of sight and geometric rescaling. It satisfies
\begin{equation}
\nabla^2_\mathrm{2d} \psi (\bm{x}) = 8\pi d_\mathrm{eff} \Sigma(\bm{x})~,
\end{equation}
where $\Sigma(\bm{x})$ is the projected surface mass density. 

For convenience we introduce the Fresnel scale $x_F$, defined by $x_F^2=d_\mathrm{eff}/[2\pi f(1+z_l)]$~\cite{Macquart:2004sh}. Its physical meaning and   role in classifying diffraction regimes are discussed below. With this, the wave optics integral of Eq.~\eqref{eq:F(f)}  is written as
\begin{equation}\label{eq:F(xF)}
F(f) = \frac{1}{2\pi i x_F^2}\int d^2x e^{i  \varphi (\bm{x})/x_F^2}~.
\end{equation}
For the axisymmetric lens profiles used in our numerical calculations, Eq.~\eqref{eq:F(xF)} reduces to \cite{takahashi2003wave}
\begin{equation}\label{eq:F(f)1d}
F(f) = e^{i [x_s^2/2- \varphi_m(x_s)]/x_F^2}\frac{1}{i x_F^2}\int_0^{\infty} dx ~ x ~ e^{i [x^2/2-\psi(x)]/x_F^2}J_0(x_s x/x_F^2)~.
\end{equation}
Here, $x=|\bm{x}|$, $x_s=|\bm{x}_s|$ is the source impact parameter, and $J_0$ is the zeroth-order Bessel function of the first kind. We employ analytic approximations to analyze lensing diffraction regimes. All lensing probabilities and observational forecasts in this work use amplification factors obtained by full numerical evaluations of the wave optics integral in Eq.~\eqref{eq:F(f)1d}.

For these numerical evaluations, we decompose $J_0$ into two Hankel functions and separate their rapidly oscillating phases. We use a one-dimensional adaptive grid, refining intervals where indicators based on phase variation or expansion errors exceed prescribed tolerances. Within each interval, we expand the phase and remaining amplitude to quadratic order and integrate analytically using complex error functions. For the tail beyond the outer grid boundary, we deform the integration contour into the complex plane, initially along the imaginary direction and then toward Gaussian decay along the steepest descent contour of the quadratic phase. This decay concentrates the tail contribution near the boundary, allowing us to evaluate it with the same analytic expressions using the expansions in the last grid interval.

\subsection{Lensing regimes and Fresnel scale}

The Fresnel scale sets the characteristic spatial scale of diffraction on the lens plane. In physical units this corresponds to
\begin{equation}\label{eq:xF}
x_F \simeq 1.24
\sqrt{
\frac{d_\mathrm{eff}}{1~\mathrm{Gpc}}
\frac{1~\mathrm{Hz}}{f(1+z_l)}
}
~\mathrm{pc}~.
\end{equation}
Thus, spatial scales probed by diffraction decrease  as $x_F\propto f^{-1/2}$. Across the frequency range of ground-based  and space-based GW observations  $x_F$ can range from subparsec scales to tens of parsecs, allowing diffraction to probe small scale structures of the lens.

For the lensing regimes we consider weak lensing when $F(f)$ can be expanded perturbatively in the lens potential, and strong lensing for the regime in which such an expansion is no longer valid\footnote{Strong lensing in this classification does not necessarily require multiple geometric images. For example, an axisymmetric Gaussian lens with $\kappa(x)=\kappa_0 e^{-x^2/(2a^2)}$ and $0<\kappa_0<1$ produces a single geometric image, yet its central geometric optics magnification $\mu(0)=(1-\kappa_0)^{-2}$ becomes arbitrarily large as $\kappa_0\rightarrow1^-$.}. Separately, we distinguish the geometric optics regime from the diffraction regime. In the geometric optics regime  $F(f)$ can be described in terms of geometric paths. In the diffraction regime a finite region of the lens plane contributes coherently and cannot be reduced to individual geometric paths.

In geometric optics GW propagation paths correspond to stationary points of $\varphi(\bm{x})$. Their positions $\bm{x}_i$ on the lens plane satisfy
\begin{equation}
\nabla\varphi(\bm{x}_i) = \bm{x}_i-\bm{x}_s-\nabla\psi(\bm{x}_i) = 0~.
\end{equation}
In weak lensing, the relevant solution remains close to the unlensed path  $\bm{x}_i\simeq\bm{x}_s$. In strong lensing, the stationary point structure can instead contain multiple solutions.

For weak lensing, the amplification factor can be expanded to linear order in the lens potential. Defining $\delta F= F-1$, the result can be written as~\cite{Choi:2021bkx}
\begin{equation}\label{eq:Fdiff}
\delta F(f) \simeq \frac{1}{i x_F^2} \int_0^{\infty} dx~x~ e^{i x^2/(2x_F^2)} \overline{\kappa}(x;x_s)~,
\end{equation}
where $\kappa(\bm{x})=4\pi d_\mathrm{eff}\Sigma(\bm{x})$ is the convergence and
\begin{equation}
\overline{\kappa}(x;x_s) = \frac{1}{\pi x^2} \int_{|\bm{x}'-\bm{x}_s|<x} d^2x'~\kappa(\bm{x}')
\end{equation}
is the aperture-mean convergence centered at $\bm{x}_s$. When $x_F\gtrsim x_s$, the projected source position is not resolved from the lens center on the Fresnel scale. The aperture-mean convergence centered at $\bm{x}_s$ can then be approximated by that centered on the lens, so the diffraction signal probes the overall projected mass distribution on scales of order $x_F$. When the aperture-mean convergence varies slowly over the relevant scale, Eq.~\eqref{eq:Fdiff} further reduces to~\cite{Choi:2021bkx}
\begin{equation}\label{eq:Fdiffdiff}
\delta F(f) \simeq \overline{\kappa}(i^{1/2}x_F)~.
\end{equation}
This provides a direct relation between the weak diffraction signal and the projected mass distribution averaged on the Fresnel scale.

In the opposite limit, $x_F\ll x_s$, the relevant stationary point remains close to the unlensed path $\bm{x}_i\simeq\bm{x}_s$. The amplification factor then approaches the geometric optics result 
\begin{equation}
F_\mathrm{geo} = \frac{1}{\sqrt{(1-\kappa)^2-\gamma^2}} \simeq 1+\kappa+\kappa^2+\frac{\gamma^2}{2}+\cdots~,
\end{equation}
where $\gamma$ is the shear at the image position. The linear geometric optics contribution is therefore $\kappa(x_s)$. Additional diffraction corrections remain at finite $x_F$. These corrections can retain information about the lens profile away from the geometric image and therefore contain nonlocal information beyond $\kappa(x_s)$. We discuss their structure below.
In strong lensing, the full stationary point structure can become relevant. 

For isolated geometric images $i$ and $j$, the  condition $f|\Delta t_{ij}|\gg1$ expresses the intuitive requirement that their relative phase be large enough for the contributions to be considered separately~\cite{takahashi2003wave}. However, equal arrival times do not by themselves imply diffraction, because spatially isolated stationary points can still admit a geometric optics description. For the purpose of systematically organizing the lensing regimes we  use $\Delta\varphi_{ij}$ as a heuristic path-based separation measure\footnote{We consider $\Delta\varphi_{ij}=\min_{\gamma_{ij}}\int_{\gamma_{ij}}|\nabla\varphi|~d\ell$, where $\gamma_{ij}$ runs over lens-plane paths from $\bm{x}_i$ to $\bm{x}_j$. The use of $|\nabla\varphi|$ makes this a positive path-based quantity rather than an arrival time difference. It is not intended to define a general transition criterion.}.

Provided that the stationary points are isolated and nondegenerate, the strong geometric optics limit is characterized by Fermat potential differences satisfying $|\varphi(\bm{x}_j)-\varphi(\bm{x}_i)|\gg x_F^2$ for every relevant image pair. The amplification factor is then
\begin{equation}
F_\mathrm{geo}(f) = \sum_i |\mu_i|^{1/2} \exp\left[ 2\pi i f t_i-i\frac{\pi n_i}{2} \right]~,
\end{equation}
where $\mu_i$ is the image magnification. The Morse index $n_i$ counts the negative eigenvalues of the Hessian of $\varphi$ at the image position and determines the image's geometric phase shift $n_i=0$, $1$, and $2$ for a minimum, saddle point, and maximum, respectively. The arrival time of the $i$th image is $t_i=(1+z_l)\varphi(\bm{x}_i)/d_\mathrm{eff}$. The relative time delay is
\begin{equation}
\Delta t_{ij} = \frac{1+z_l}{d_\mathrm{eff}} \left[ \varphi(\bm{x}_j)-\varphi(\bm{x}_i) \right]~.
\end{equation}
The geometric optics signal is therefore characterized by the image magnifications, relative time delays and Morse phase shifts.

Within the same heuristic classification, the strong diffraction limit is reached when the separation is no larger than $x_F^2$ for all relevant image pairs. In such case, the contributions cannot be reduced to separate geometric images. Between these limits  some image pairs can contribute separately while others contribute together through diffraction, and geometric image and diffraction features can coexist.

\begin{table}[t]
\centering 
\renewcommand{\arraystretch}{1.35}
\setlength{\tabcolsep}{3pt}
\footnotesize
\begin{tabular}{>{\raggedright\arraybackslash}p{0.15\textwidth}| >{\centering\arraybackslash}p{0.18\textwidth} >{\raggedright\arraybackslash}p{0.30\textwidth} >{\raggedright\arraybackslash}p{0.29\textwidth}}
\hline\hline
Lensing regime & Condition & Contributing lens-plane region & Characteristic $F(f)$ behavior \\
\hline
Weak geometric optics
& $x_F\ll x_s$
& Fresnel region around the primary image, central region of the lens
& Geometric optics contribution, beyond geometric optics (bGO) correction, central diffractive image \\
Weak diffraction
& $x_F\gtrsim x_s$
& Fresnel region around the projected source position
& Aperture-mean convergence \\
\hline
Strong geometric optics
& $x_F^2\ll\min\limits_{i\neq j}\Delta\varphi_{ij}$
& Isolated stationary points
& Image magnifications, relative time delays \\
Strong diffraction
& $x_F^2\gtrsim\max\limits_{i\neq j}\Delta\varphi_{ij}$
& Finite width neighborhood of a critical curve, part or all of the critical curve
& Contribution not reducible to separate geometric images~\cite{Ezquiaga:2025gkd}, $F(f)\propto f^{1/2}$ for an aligned axisymmetric critical ring \\
\hline\hline
\end{tabular}
\caption{Classification of the limiting GW lensing regimes by their conditions, contributing lens-plane regions and characteristic behavior of the amplification factor $F(f)$. Here $x_F$ is the Fresnel scale, $x_s$ is the source impact parameter giving the magnitude of the projected source position relative to the lens center, and $\Delta\varphi_{ij}$ is the heuristic minimum path integral of the magnitude of the Fermat potential gradient between geometric images $i$ and $j$. The strong lensing conditions apply to configurations with at least two relevant geometric images and are parametric limiting estimates.}
\label{tab:lensing_regimes}
\end{table}

For such multiple-image configurations, the strong diffraction contribution arises from a finite width neighborhood extending along part or all of the associated critical curve, over which the Fermat potential variation is small compared with $x_F^2$. Because this contribution cannot be reduced to separate geometric images  it must be evaluated through the diffraction integral~\cite{Ezquiaga:2025gkd}.

For an axisymmetric lens near the Einstein ring  the strong diffraction amplification is approximately~\cite{nakamura1999wave}
\begin{equation}
F(f) \simeq i^{-1/2} \frac{x_E}{x_F} \sqrt{ \frac{2\pi} {1-\kappa(x_E)+\gamma(x_E)} }~,
\end{equation}
where $x_E$ is the Einstein radius. Thus, $F(f)\propto x_F^{-1}\propto f^{1/2}$. For an axisymmetric lens, this picture reduces to a critical ring contribution. Near source-lens alignment $\Delta\varphi_{ij}\sim x_s x_E$ so the limiting conditions become $x_F^2\ll x_s x_E$ for geometric optics and $x_F^2\gtrsim x_s x_E$ for strong diffraction~\cite{Ezquiaga:2025gkd}. At exact alignment, the coherent contribution from the
Einstein ring yields the $|F|\propto f^{1/2}$ scaling
discussed above~\cite{Choi:2021bkx,Ezquiaga:2025gkd}.

The four limiting GW lensing regimes are summarized in Table~\ref{tab:lensing_regimes}. Eq.~\eqref{eq:Fdiff} provides a general lens profile description of the diffraction associated with the Fresnel region around the projected source position. However, in the weak lensing geometric optics regime the same linear order integral also contains a distinct contribution from the central region of the lens, far from the primary geometric image. We derive its general form below and interpret it as a \textit{diffractive image}.

\subsection{Diffractive images}

We now consider the diffraction that remains in the weak lensing geometric optics regime, $x_F\ll x_s$, where $x_s=|\bm{x}_s|$. The leading geometric optics contribution is associated with the primary image, while finite frequency effects can also retain information associated with other parts of the lens plane.

An early example of such a contribution was found in Ref.~\cite{takahashi2004quasi}, where the lens potential was treated nonperturbatively. That analysis considered expansion of the diffraction integral locally around a point that is not a stationary point of the lensing phase and found that the contribution from a regular region vanishes order by order in the high frequency expansion. Since this argument requires the lens potential and its derivatives to be well defined at the expansion point, cuspy and singular lens centers were treated separately. They were found to produce an additional contribution with a frequency dependence determined by the central density slope, which was interpreted as a diffracted image formed at the lens center.

Here we extend the description of central diffraction and relate the diffractive contribution directly to the projected mass distribution. We work at linear order in the lens potential and analyze the diffraction integral in Fourier space, following a representation used in perturbative studies of wave optics lensing \cite{Takahashi:2005sxa,Yarimoto:2024uew}. The low-$k$ expansion gives the geometric optics contribution and its finite-frequency correction associated with the primary image. At higher $k$, the oscillatory kernel has a stationary phase at a characteristic wave number set by $x_s$ and $x_F$. We show that this contribution is associated with the central region of the lens and can be written directly in terms of the Fourier space convergence $\tilde{\kappa}$. This extends the diffractive image description to a general projected mass distribution without requiring a cuspy or singular lens center.

Starting from Eq.~\eqref{eq:Fdiff}, we write the aperture-mean convergence in Fourier space as
\begin{equation}
    \overline{\kappa}(x;x_s) = \frac{1}{\pi x^2} \int \frac{d^2 k}{2\pi} \tilde{\kappa}(\bm{k}) \frac{x}{k}J_1(kx) e^{i\bm{k}\cdot\bm{x}_s}~,
\end{equation}
where
\begin{equation}\label{eq:kappatilde}
    \tilde{\kappa}(\bm{k}) = \int d^2x~\kappa(\bm{x})e^{-i\bm{k}\cdot\bm{x}}
\end{equation}
is the Fourier transform of the convergence and $J_1$ is the first-order Bessel function of the first kind. Substituting this expression into Eq.~\eqref{eq:Fdiff} and integrating over $x$ gives
\begin{equation}\label{eq:Fdiffk}
    F(f) \simeq 1+ \frac{1}{i x_F^2} \int \frac{d^2 k}{(2\pi)^2} \tilde{\kappa}(\bm{k}) e^{i\bm{k}\cdot\bm{x}_s} \frac{2}{k^2} \left( 1-e^{-ik^2x_F^2/2} \right)~.
\end{equation}

To evaluate Eq.~\eqref{eq:Fdiffk}, we first consider $k \ll 1/x_F$. Expanding the wave optics kernel gives
\begin{equation}
    \frac{1}{i x_F^2} \frac{2}{k^2} \left( 1-e^{-ik^2x_F^2/2} \right)
    = 1-\frac{i}{4}k^2x_F^2 +\mathcal{O}(k^4x_F^4)~.
\end{equation}
Since $1/x_s\ll1/x_F$, the integration domain can be extended over the full Fourier plane for these terms. Using
\begin{equation}
    \int\frac{d^2k}{(2\pi)^2} k^2\tilde{\kappa}(\bm{k}) e^{i\bm{k}\cdot\bm{x}_s}
    = -\nabla^2\kappa(\bm{x}_s)~,
\end{equation}
we obtain
\begin{equation}\label{eq:FgeobGO}
    F_{\rm GO+bGO}(f) \simeq
    1+\kappa(\bm{x}_s) +\frac{i x_F^2}{4} \nabla^2\kappa(\bm{x}_s) +\mathcal{O}(x_F^4\nabla^4\kappa)~.
\end{equation}
The first correction is the linear geometric optics (GO) contribution. The next term is a finite frequency correction associated with the primary geometric image, also known as the beyond geometric optics (bGO) correction. Ref.~\cite{takahashi2004quasi} derived the corresponding bGO correction without expanding in the lens potential. Expanding that result to linear order in the lens potential reproduces Eq.~\eqref{eq:FgeobGO}.

The leading finite frequency correction is purely imaginary. Since $\kappa$ and $\nabla^2\kappa$ are real,
\begin{equation}
    \left| 1+\kappa +\frac{i x_F^2}{4}\nabla^2\kappa \right|
    = 1+\kappa+\mathcal{O}(\kappa^2)~,
\end{equation}
where the finite frequency term contributes to the magnitude only at quadratic order in the lens potential. It therefore changes the phase of the primary image at linear order, but not its magnification.

We next consider the $k \gg 1/x_F$ part of Eq.~\eqref{eq:Fdiffk}. The term proportional to $e^{i\bm{k}\cdot\bm{x}_s}$ has no stationary point and is suppressed by rapid oscillations. The remaining phase,
\begin{equation}
    \bm{k}\cdot\bm{x}_s-\frac{k^2x_F^2}{2}~,
\end{equation}
has a stationary point at
\begin{equation}\label{eq:kdiff}
    \bm{k}_* = \frac{\bm{x}_s}{x_F^2}~.
\end{equation}
Since $x_s\gg x_F$, this point lies in the high-$k$ region. Expanding around $\bm{k}_*$ gives
\begin{equation}  \label{eq:station}
 -\frac{1}{i x_F^2}\int\limits_{k>1/x_F}\frac{d^2 k}{(2\pi)^2} \tilde{\kappa}(\bm{k})\frac{2}{k^2} e^{i(\bm{k}\cdot\bm{x}_s-k^2x_F^2/2)}
 \simeq \frac{1}{\pi x_s^2} e^{ix_s^2/(2x_F^2)} \tilde{\kappa} \left( \frac{\bm{x}_s}{x_F^2} \right)~. 
\end{equation}
The stationary phase evaluation in Eq. \eqref{eq:station} assumes that the prefactor $ (\widetilde{\kappa}(\mathbf{k})/k^2)$  varies sufficiently slowly over the contributing region around $ \mathbf{k}_*$ that has characteristic width  $  \Delta k\sim x_F^{-1} $ . Thus Eq.~\eqref{eq:station} is therefore a leading stationary phase
approximation valid under this condition. On the other hand, Eq.~\eqref{eq:Fdiffk} captures the full linear order dependence on the projected density.

Combining the contributions, we obtain
\begin{equation}\label{eq:Fdiffgeo}
\begin{split}
F(f) \simeq~ 1+
\underbrace{ \kappa(\bm{x}_s) }_{\text{geometric optics}}
+
\underbrace{ \frac{i x_F^2}{4} \nabla^2\kappa(\bm{x}_s)}_{\text{bGO}}
+
\underbrace{ \frac{1}{\pi x_s^2} e^{ix_s^2/(2x_F^2)} \tilde{\kappa} \left( \frac{\bm{x}_s}{x_F^2} \right) }_{\text{Diffractive image}}~.
\end{split}
\end{equation}
The first two terms give the linear geometric optics result, while the third term is the bGO correction around the primary image. The last term is qualitatively different as it depends on the Fourier transform $\tilde{\kappa}(\bm{k}_*)$, which integrates the projected mass distribution over the lens plane  rather than only on the density and its derivatives at the primary image. Therefore, it probes nonlocal information about the lens structure. We refer to this nonlocal contribution as the central diffractive image.

The factor $e^{ix_s^2/(2x_F^2)}$ gives the propagation phase associated with the lens center relative to the primary path at linear order in the lens potential. The detailed mass distribution modifies the amplitude and can add profile-dependent phase through $\tilde{\kappa}(\bm{k}_*)$. This includes, for example, phase structure associated with a finite boundary, while the common central propagation phase remains explicit in Eq.~\eqref{eq:Fdiffgeo}.

Eq.~\eqref{eq:Fdiff} and Eq.~\eqref{eq:Fdiffgeo} therefore provide two complementary connections between diffraction and a general projected mass distribution. Eq.~\eqref{eq:Fdiff} relates the signal to the aperture-mean convergence over a Fresnel-scale region centered at the projected source position, while Eq.~\eqref{eq:Fdiffgeo} identifies an additional contribution associated with the central region of the lens through the Fourier space convergence. These relations show how spatially separated parts of the projected mass distribution can enter the frequency-dependent lensing signal.

Eq.~\eqref{eq:kdiff} gives the spatial scale probed by the central diffractive image contribution. In the $x_F\ll x_s$ regime, the probed Fourier mode is $k_*=x_s/x_F^2$, rather than the source-independent estimate $k\sim1/x_F$ based only on the Fresnel scale, and corresponds to a real space scale of order $x_F^2/x_s$. Across a lens structure of size $R$ around the lens center  the phase therefore changes by an amount of order $k_*R=x_sR/x_F^2$. When $k_*R\lesssim1$ the phase varies sufficiently slowly for the structure to contribute coherently even though the lens center is not a stationary point. For $k_*R\gtrsim1$  phase variations across the structure become important and its internal profile can be resolved. The relevant quantity is therefore the phase variation across a finite region, rather than the phase gradient at a single point.

This finite region picture provides a complementary perspective on the central diffractive contribution identified for cuspy lenses in Ref.~\cite{takahashi2004quasi}. The local high frequency expansion used there gives no contribution around a regular non-stationary point, while cuspy or singular centers require a separate treatment and can produce a central diffractive image. Eq.~\eqref{eq:Fdiffgeo} shows that a central diffractive contribution does not in general require such a singularity. For an unresolved finite structure  $\tilde{\kappa}(k_*)\simeq\tilde{\kappa}(0)$  and its leading contribution is determined primarily by the total mass. Once the structure is resolved, the scale dependence of $\tilde{\kappa}(k_*)$ encodes its internal mass distribution.

\begin{figure}[t]
    \centering
    \includegraphics[width=0.5\linewidth]{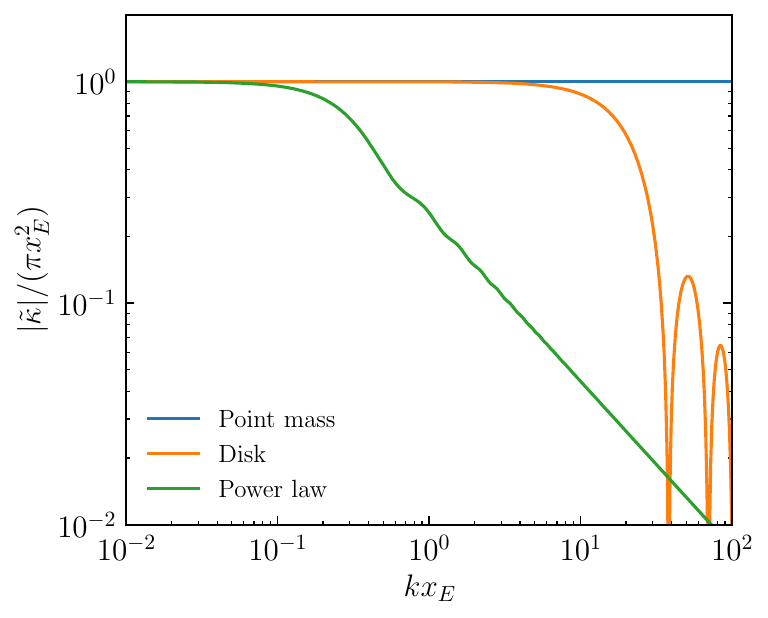}
    \caption{
Normalized Fourier space convergence $|\tilde{\kappa}(k)|/(\pi x_E^2)$ for a point mass, a uniform disk with $R=0.1x_E$, and a truncated spherical power law lens with $\rho(r)\propto r^{\gamma-1}$ for $0<r<R$ and zero density outside, where $\gamma=-1.25$ and $R=10x_E$. Here $x_E=\sqrt{4Md_{\rm eff}}$ is defined by the common total mass $M$ and serves as a reference length scale. At $k_*=x_s/x_F^2$ in the $x_F\ll x_s$ regime, the plotted ratio gives the central diffractive image magnitude relative to a point mass with the same $M$ and $x_s$. All profiles satisfy $\tilde{\kappa}(0)=\pi x_E^2$ because the same total mass contributes coherently at small $k$. The extended profiles depart from this limit near $kR\sim1$, at different $k$ because their radii differ. The disk's sharp boundary produces oscillations with a $k^{-3/2}$ envelope, while the central cusp of the power law profile gives a leading $k^{-2-\gamma}=k^{-3/4}$ tail. The truncated profile is included only as an illustrative finite size example. The halo-dressed lens model considered below uses an untruncated power law halo.
}
    \label{fig:kappa_tilde}
\end{figure}

Fig.~\ref{fig:kappa_tilde} compares the Fourier space convergences of a point mass, a uniform disk, and a truncated power law lens profile with the same total mass. The truncated profile has a spherically symmetric density $\rho(r)\propto r^{\gamma-1}$ for $0<r<R$ and zero density outside $R$, with its normalization fixed by the total mass $M$. We take $\gamma=-1.25$ and $R=10x_E$, giving an inner projected convergence $\kappa(x)\propto x^\gamma$ for $x\ll R$. The extended profiles initially follow the point mass curve, but the larger radius of the power law profile makes its departure occur at smaller $k$. Once each lens is resolved, cancellations across the lens reflect its internal mass distribution. The sharp disk boundary produces oscillations, whereas the central cusp of the power law profile results in a gradual decay. In the $x_F\ll x_s$ regime, Eq.~\eqref{eq:Fdiffgeo} samples these curves at $k_*=x_s/x_F^2$. At fixed $M$ and $x_s$, the plotted ratio therefore gives the magnitude of the central diffractive image relative to that of a point mass.

Eq.~\eqref{eq:Fdiffgeo} reproduces the known central diffractive image result for a singular isothermal sphere. For
\begin{equation}
    \tilde{\kappa}(k) = \frac{\pi x_E}{k}~,
\end{equation}
the diffractive image contribution becomes
\begin{equation}
    F_{\rm diff}(f) \simeq e^{ix_s^2/(2x_F^2)} \frac{x_F^2x_E}{x_s^3}~.
\end{equation}
This agrees with the leading large-$x_s/x_E$ limit of the central diffractive image contribution obtained in Ref.~\cite{takahashi2004quasi}.

The point mass limit follows directly from the unresolved regime discussed above. For a finite lens with total mass $M$, $k_*R\ll1$ gives $\tilde{\kappa}(k_*)\simeq\tilde{\kappa}(0)=\pi x_E^2$, so that its internal structure is not resolved by the diffractive image. As $R\rightarrow0$, this condition is satisfied at any fixed $k_*$, and the lens continuously approaches a point mass description. Substituting $\tilde{\kappa}(k)=\pi x_E^2$ into Eq.~\eqref{eq:Fdiffgeo} gives
\begin{equation} \label{eq:eqdiff}
    F_{\rm diff}^{\rm p}(f) \simeq \frac{x_E^2}{x_s^2} e^{ix_s^2/(2x_F^2)}~.
\end{equation}
This agrees in magnitude with the geometric image of a point mass lens at large $x_s/x_E$~\cite{takahashi2003wave}. The central diffractive image of an extended lens therefore connects continuously to the secondary geometric image as the lens becomes unresolved and approaches the point mass limit.

An approximate weak lensing prescription for central wave optics contributions from symmetric lens profiles was developed in Ref.~\cite{Wu:2025rvy}. Eq.~\eqref{eq:Fdiffgeo} provides a complementary description by expressing the central diffractive image directly through $\tilde{\kappa}(\bm{k}_*)$. The derivation does not require axial symmetry or regularity of the lensing potential at the center, and therefore describes regular extended profiles and singular lenses, including a point mass, within the same expression.

Eq.~\eqref{eq:Fdiffgeo} is derived to linear order in the lens potential and in the limit $x_F\ll x_s$. The linear truncation does not necessarily capture all relevant effects even when lensing remains perturbative. For example, the geometric optics amplification contains
$F_{\rm geo}\simeq 1+\kappa+\kappa^2+(\gamma^2/2)+\cdots~,$
so the effect of shear first appears at second order. This is particularly relevant for extended lenses, where the local convergence can be small or vanish outside the projected mass distribution while the shear remains nonzero. The observational calculations in this work therefore evaluate the full wave optics integral numerically, while Eq.~\eqref{eq:Fdiffgeo} provides the physical interpretation of the diffraction features.

\subsection{Images of extended and halo-dressed lenses}
\label{sec:diffprofile}

We now apply the diffractive image interpretation to two representative lens profiles consisting of a uniform disk and a point mass surrounded by a halo with a power law profile. These examples illustrate two different ways in which the lens structure appears in the geometric optics regime. For the disk, a finite lens radius introduces a scale at which the internal structure becomes resolved and the central diffractive image separates into contributions associated with the lens boundary. For the halo-dressed point mass lens, the central point mass produces a secondary geometric image near the lens center, while the extended halo adds a central diffractive contribution. We describe $\Sigma$, $\kappa$ and $\psi$ for point mass, uniform disk, uniform sphere, soliton and halo-dressed point mass lens profiles in Appendix~\ref{app:potentials}.

We first consider the uniform disk.  For a central point mass $M$ the Fourier space convergence is
\begin{equation}
    \tilde{\kappa}_{\rm p}(k) = \pi x_E^2~,
    \label{eq:kappatilde_point}
\end{equation}
where $x_E=\sqrt{4Md_{\rm eff}}$. For the uniform disk of total mass $M$ and  radius $R$ the convergence is
\begin{equation}
    \tilde{\kappa}_{\rm d}(k) = \pi x_E^2 \Big( \frac{2J_1(kR)}{kR} \Big) =  \tilde{\kappa}_{\rm p}(k) \Big( \frac{2J_1(kR)}{kR} \Big) 
    \label{eq:kappatilde_disk}
\end{equation} 
and $x_E$ is given as for point mass.

Substituting disk's convergence Eq.~\eqref{eq:kappatilde_disk} into Eq.~\eqref{eq:Fdiffgeo} gives
\begin{equation}
    F_{\rm diff}^{\rm d}(f)
    \simeq \frac{x_E^2}{x_s^2} e^{ix_s^2/(2x_F^2)} \Big(\frac{2J_1(k_*R)}{k_*R}\Big) = F_{\rm diff}^{\rm p} \Big(\frac{2J_1(k_*R)}{k_*R}\Big)~, 
    \label{eq:Fdiff_disk}
\end{equation}
where  $k_*= x_s/x_F^2$ and $F_{\rm diff}^{\rm p}$ is point mass contribution of Eq.~\eqref{eq:eqdiff}. The quantity $k_*R=x_sR/x_F^2$ determines whether the disk radius is resolved by the diffractive image. For $k_*R\lesssim1$, or equivalently $x_F\gtrsim\sqrt{x_sR}$, the disk is unresolved and
\begin{equation}
    \frac{2J_1(k_*R)}{k_*R}
    \simeq 1~.
\end{equation}
The diffractive image therefore cannot resolve the disk structure and is indistinguishable from that of a point mass with the same total mass. Combining $k_*R\lesssim1$ with the geometric optics condition $x_F<x_s$ gives
\begin{equation}
    \sqrt{x_sR}\lesssim x_F<x_s~.
\end{equation}

\begin{figure*}[t]
    \centering
    \includegraphics[width=0.48\textwidth]{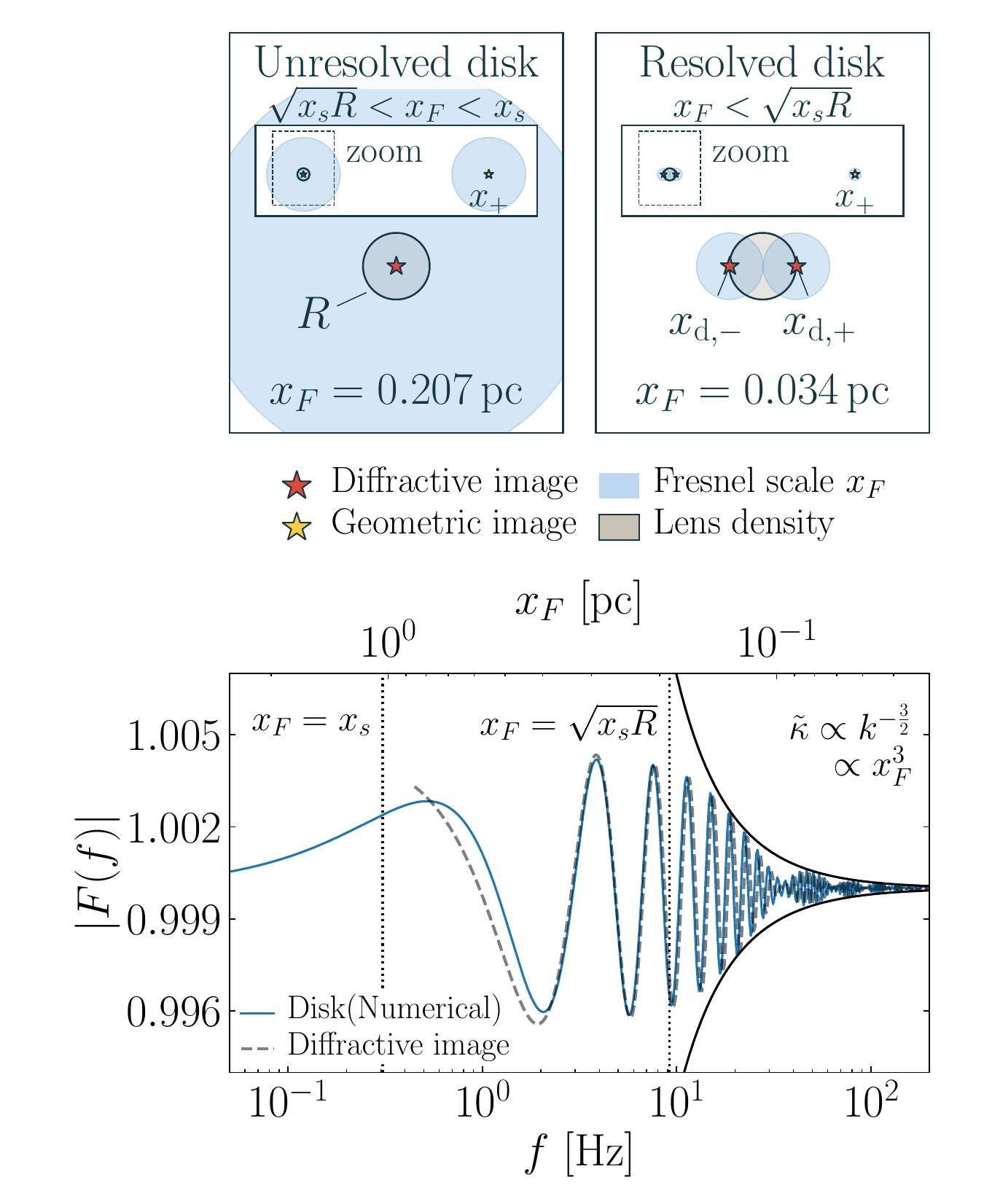}
    \includegraphics[width=0.48\textwidth]{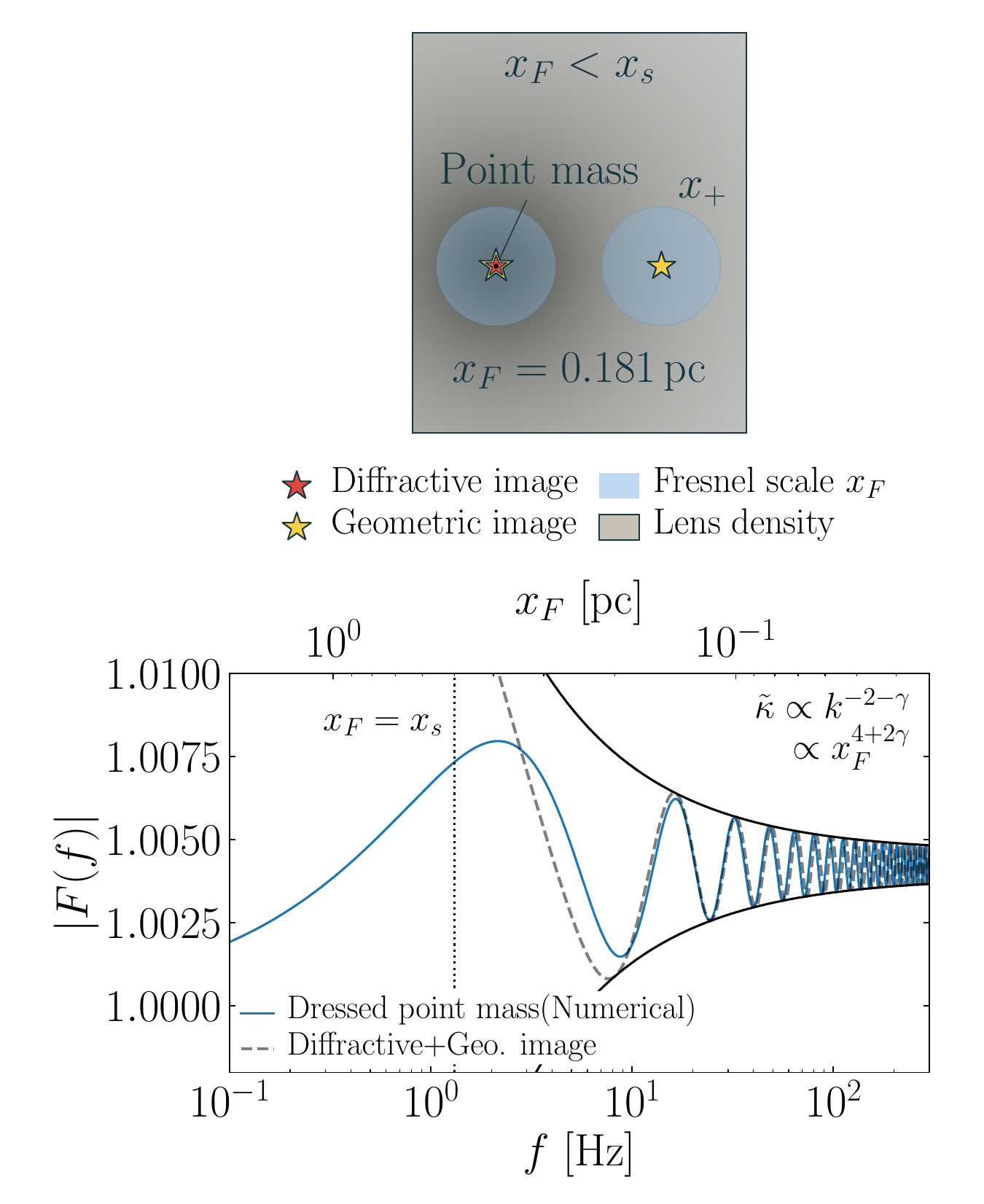}
\caption{
Diffraction image interpretation for GW lensing of a uniform disk (left) and a power law halo-dressed point mass lens (right). The upper schematics identify the lens-plane regions associated with diffraction terms, and the lower panels compare the numerical wave optics results with the analytic decompositions. For the disk, the upper schematics show the unresolved and resolved regimes separated by $x_F=\sqrt{x_sR}$. The small inset in each disk schematic shows a wider view of the lens plane, including the primary geometric image. The dashed box marks the central region enlarged below. The red stars in the resolved disk schematic mark the effective locations of rim diffractive images associated with edge-diffraction phases from opposite sides of the rim. These diffractive images are not additional stationary points or geometric images. For the disk, the gray dashed curve is $|1+F_{\rm diff}^{\rm d}|$. For the halo-dressed lens, it includes the primary geometric image, the central point mass and halo contributions, and the bGO correction. The vertical dotted lines mark $x_F=x_s$ and, for the disk, $x_F=\sqrt{x_sR}$. Resolving the disk boundary produces high frequency oscillations, whereas the halo diffractive contribution decreases toward high frequencies and the halo-dressed lens approaches its geometric optics limit. The black curves show the oscillation envelopes around the primary geometric image contribution. They include the rim contribution for the disk and the combined central point mass and halo contributions for the halo-dressed lens. For the disk, we use $M=100~M_\odot$, $z_l=0.17$, $z_s=0.3$, $x_E=0.069~{\rm pc}$, $x_s=15x_E$, and $R=0.5x_E$ as characteristic parameters. For the halo-dressed point mass lens, we use $M=2~M_\odot$, $z_l=0.17$, $z_s=0.3$, $x_s=0.5~{\rm pc}$, and $\gamma=-1.25$.
}
    \label{fig:diffimage_models}
\end{figure*}

The upper left schematic of Fig.~\ref{fig:diffimage_models} represents this coherent disk contribution as one central diffractive image. Its interference with the primary geometric image gives the point mass-like oscillations in the lower left panel. The behavior changes when $k_*R\gtrsim1$, or equivalently $x_F\lesssim\sqrt{x_sR}$. Using the large-$k_*R$ expansion of the Bessel function, Eq.~\eqref{eq:Fdiff_disk} becomes
\begin{equation} 
    F_{\rm diff}^{\rm d}(f)
    \simeq~\sqrt{\frac{2}{\pi}}\frac{x_E^2}{x_s^2}(k_*R)^{-3/2}
    \bigg[ e^{i\left(x_s^2/(2x_F^2)+k_*R-3\pi/4\right)} + e^{i\left(x_s^2/(2x_F^2)-k_*R+3\pi/4\right)}\bigg]~.
    \label{eq:Fdiff_disk_rim} 
\end{equation}
Substituting $k_{\ast}$, the frequency-dependent parts of the two phases are
\begin{equation}
    \frac{1}{x_F^2}\left(\frac{x_s^2}{2}\pm x_sR\right)~.
\end{equation}
The two phase terms can be interpreted as rim diffractive images with effective locations
\begin{equation}
    x_{{\rm d},\pm}=x_s\left[1-\sqrt{1\mp\frac{2R}{x_s}}\right]\simeq\pm R~,
    \label{eq:xd_disk}
\end{equation}
where the last expression applies for $R\ll x_s$. The effective locations of the two diffractive images lie on opposite sides of the disk rim. They are not additional stationary points of the lensing phase, but represent edge diffraction contributions generated by the sharp disk boundary. In this resolved regime, the primary geometric image interferes with the two rim diffractive images and produces the more complicated oscillatory pattern shown in the lower-left panel of Fig.~\ref{fig:diffimage_models}.

The transition to the geometric optics regime and the resolution of the disk radius are marked by vertical dotted lines at $x_F=x_s$ and $x_F=\sqrt{x_sR}$ in the lower left panel of Fig.~\ref{fig:diffimage_models}, respectively. In the resolved disk regime Eq.~\eqref{eq:Fdiff_disk_rim} gives the following scaling for the envelope of the Fourier space convergence 
\begin{equation}
    |\tilde{\kappa}_{\rm d}(k_*)|\propto(k_*R)^{-3/2}\propto x_F^3~,
\end{equation}
which determines the decreasing envelope of the rim diffractive-image contribution shown by the black curves in Fig.~\ref{fig:diffimage_models}. The gray dashed curve shows $|1+F_{\rm diff}^{\rm d}|$.

For the case of halo-dressed point mass lens described by density profile $\rho_h(r) \propto r^{\gamma - 1}$ with slope index $\gamma$, the central point mass gives $\tilde{\kappa}_p(k)$ as in Eq.~\eqref{eq:kappatilde_point}, while power law halo gives
\begin{equation}
    \tilde{\kappa}_{\rm h}(k)=\mathcal{C}_{\gamma}\mathcal{R}x_E^2(kR_h)^{-2-\gamma}~,
    \qquad
    \mathcal{C}_{\gamma}=2^{\gamma+2}\pi^{5/2}\frac{\Gamma[(\gamma+2)/2]}{\Gamma[(1-\gamma)/2]}~,
    \label{eq:kappatilde_halo}
\end{equation}
where $R_h$ is the halo scale and
$\mathcal{R}=\rho_0R_h^3/M$ parametrizes the halo amplitude with normalization density $\rho_0$.
For concreteness,  we adopt the fiducial     parameters motivated by the halo-dressed PBH GW lensing analysis of Ref.~\cite{GilChoi:2023ahp} 
\begin{equation}
    \rho_0 = 0.26~M_\odot~{\rm pc}^{-3},
    \qquad
    R_h = 0.61~{\rm pc} \left(\frac{97M}{M_\odot}\right)^{1/3},
\end{equation}
which correspond to $\mathcal{R}\simeq5.72$. We take $\gamma=-1.25$ as the characteristic power law slope.

For the halo-dressed point mass lens, the central contribution follows from the sum of Eqs.~\eqref{eq:kappatilde_point} and \eqref{eq:kappatilde_halo} 
\begin{equation}
\begin{split}
    F_{\rm c}(f)
    \simeq~
    \frac{x_E^2}{x_s^2} e^{ix_s^2/(2x_F^2)}
    \bigg[ 1 + \frac{\mathcal{C}_{\gamma}}{\pi} \mathcal{R} \left( \frac{x_s R_h}{x_F^2} \right)^{-2-\gamma} \bigg] = F_{\rm diff}^{\rm p} 
    \bigg[ 1 + \frac{\mathcal{C}_{\gamma}}{\pi} \mathcal{R}
    \left( \frac{x_s R_h}{x_F^2} \right)^{-2-\gamma} \bigg]~.
    \label{eq:Fcentral_dress}
\end{split}
\end{equation}
The first term in the brackets is the secondary geometric image produced by the central point mass in the large-$x_s/x_E$ limit given in Eq.~\eqref{eq:eqdiff}. The second is the central diffractive image produced by the power law halo. At linear order in the lens potential, the two components contribute additively to the central image. Their contributions have the same frequency-dependent propagation phase and therefore appear as a single central contribution, as illustrated in the upper right panel of Fig.~\ref{fig:diffimage_models}.

The power law halo also produces a finite frequency correction around the primary image. From Eq.~\eqref{eq:FgeobGO},
\begin{equation}
    \delta F_{\rm bGO,h} = \frac{i x_F^2}{4} \nabla^2\kappa_{\rm h}(x_s)
    = i\frac{\gamma^2}{4} \frac{x_F^2}{x_s^2} \kappa_{\rm h}(x_s)~.
    \label{eq:bgo_dress}
\end{equation}
This contribution originates from the Fresnel region around the primary image rather than from the lens center. At linear order in the lens potential it is purely imaginary and therefore changes mainly the phase, with no linear order correction to $|F|$. The visible modulation of $|F|$ is consequently dominated by the central diffractive-image contribution. The gray dashed curve in the lower right panel of Fig.~\ref{fig:diffimage_models} combines the primary geometric image, the central contribution in Eq.~\eqref{eq:Fcentral_dress}, and the bGO correction.

The pure power law halo has no finite structural scale that introduces an additional transition in the diffractive image. The only vertical line in the lower-right panel therefore marks $x_F=x_s$. The halo contribution follows
\begin{equation}
    \tilde{\kappa}_{\rm h}(k_*) \propto k_*^{-2-\gamma} \propto x_F^{4+2\gamma}~,
    \label{eq:powerlaw_envelope}
\end{equation}
which determines the halo contribution to the oscillation envelope in the lower right panel of Fig.~\ref{fig:diffimage_models}. The black curves include both this decreasing term and the point mass contribution in Eq.~\eqref{eq:Fcentral_dress}. The signal approaches the geometric optics limit of the combined point mass and halo potential at high frequencies.

Below, we consider several distinct lens models and assess whether their diffraction signatures enable distinguishing them. We first match the models in a regime where their signals are nearly degenerate, then consider lensing features to discriminate between them.

\subsection{Interpretation in terms of particle scattering}
\label{sec:scattering}

GW lensing admits a natural analogous interpretation in the language of scattering theory of particle physics (see e.g. Ref.~\cite{Peters:1974gj} for early related work). The diffraction integral in Eq.~\eqref{eq:F(xF)} follows from a wave propagation path integral under the thin lens approximation. In the exponent, $x_F^2$ plays the formal role of $\hbar$, while the Fermat potential plays the role of the action. Geometric images correspond to stationary paths, and geometric optics is analogous to the semiclassical limit. Hence, the $\mathcal{O}(x_F^2)$ bGO correction is formally analogous to an $\mathcal{O}(\hbar)$ semiclassical correction\footnote{This correspondence discussed here is a formal analogy with particle scattering. The considered GW lensing effects arise from classical wave propagation in a gravitational field and do not require a quantum description.}. Expanding to linear order in the lens potential corresponds to the Born approximation. In the point mass limit and for $x_F\ll x_s$, Eq.~\eqref{eq:eqdiff} reproduces the Born-limit amplification correction in Eq.~(28) of Ref.~\cite{CarrilloGonzalez:2025gqm}. In the notation of that reference, the correspondence follows from identifying $r_E^2/b_l^2$ with $x_E^2/x_s^2$ and $\nu b_l^2/(2r_E^2)$ with $x_s^2/(2x_F^2)$.

In elastic electron-nucleus scattering the charge contribution to the Born amplitude is the point charge amplitude multiplied by a normalized form factor, which encodes the Fourier transform of the nuclear charge distribution evaluated at the momentum transfer~\cite{Hofstadter:1956qs}. We identify that central diffractive image has an analogous structure. Here, the projected mass distribution of the lens should be considered in place of the nuclear charge distribution.

To establish correspondence, for a lens with finite total mass $M$  we define the normalized projected mass form factor
\begin{equation}
  \mathcal{S}(\mathbf{k})
  = \frac{\tilde{\kappa}(\mathbf{k})}{\tilde{\kappa}(0)}, 
\end{equation}
where $\tilde{\kappa}(0) = \tilde{\kappa}_p  = \pi x_E^2$, so that $\mathcal{S}(0)=1$. Using Eq.~\eqref{eq:Fdiffgeo}, derived at linear order in the lens potential and with the stationary-phase approximation for $x_F\ll x_s$, we obtain
\begin{equation}
  F_{\rm diff}(f) \simeq \frac{x_E^2}{x_s^2} e^{i x_s^2/(2x_F^2)} \mathcal{S}(\mathbf{k}_*) = F_{\rm diff}^{\rm p} \mathcal{S}(\mathbf{k}_*)~,
  \label{eq:formfactor}
\end{equation}
where $\mathbf{k}_*= \mathbf{x}_s/x_F^2$.
The prefactor is the point mass contribution in Eq.~\eqref{eq:eqdiff}. At fixed total mass and geometry, the profile dependence of the central diffractive contribution is contained in $\mathcal{S}(\mathbf{k}_*)$. The vector $\mathbf{k}_*$ plays the role of the change in the wave vector transverse to the line of sight, as set by the source-lens-observer geometry and the GW frequency.

The projected density adds up the mass along each line of sight, retaining its distribution across the lens plane. By the central Fourier slice theorem, its form factor is
\begin{equation}
  \mathcal{S}(\mathbf{k}) =\frac{\tilde{\rho}(\mathbf{k},k_z=0)}{M},
\end{equation}
where $\tilde{\rho}$ is the three-dimensional Fourier transform of the mass density and $k_z$ is the wave vector component along the line of sight. Thus, the projected form factor describes transverse mass structure without resolving how that mass is distributed in depth. At fixed geometry, $\mathbf{k}_*$ points in a fixed direction and its magnitude grows as $k_*\propto f$. As the GW frequency increases, the signal therefore probes progressively finer transverse structure along that direction.

The above examples of uniform disk in Eq.~\eqref{eq:Fdiff_disk} and for halo-dressed point mass of Eq.~\eqref{eq:Fcentral_dress}  illustrate the form factor factorization we put forth explicitly. For the uniform disk, the factor multiplying the point mass diffractive amplitude is its normalized mass form factor. The dressed point mass expression has a similar structure, with the halo contributing an additional frequency-dependent factor. Since the untruncated power law halo has no finite total mass, this factor is referenced to the central point mass rather than normalized by the total
lens mass.

\section{Isolating Lens Structure with Diffraction}
\label{sec:Distinction}

In this section, we compare extended mass lenses and halo-dressed point mass lenses primarily with point mass reference lenses. For extended mass lenses, we align the low-frequency response where the internal structure is unresolved. For dressed point mass lenses, we instead align the high-frequency geometric-optics response. We then examine the remaining differences in their diffraction patterns that reveal the lens mass distribution.

\begin{figure*}[t]
    \centering
    \includegraphics[width=0.48\textwidth]{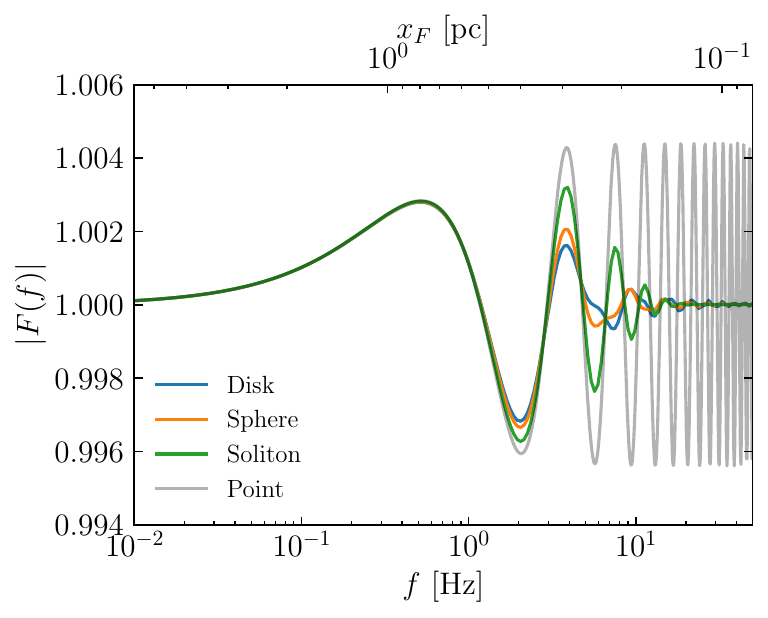}
    \includegraphics[width=0.48\textwidth]{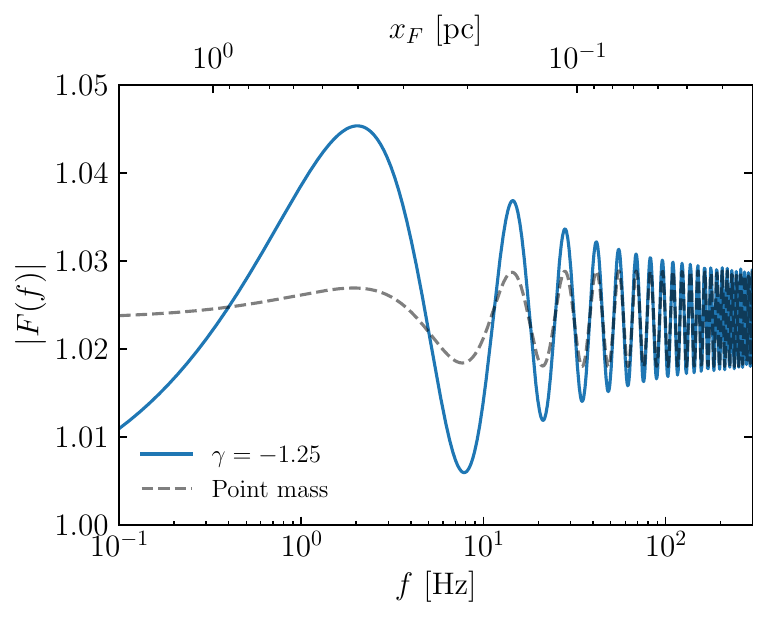}
    \caption{
    Amplification factors showing the profile-dependent diffraction that remains after common lensing features are aligned.
    Left: differences among the diffraction patterns of extended mass lenses after aligning their unresolved low frequency response at the same $M$ and $x_s$, with a common $R$ for the extended mass profiles. The curves separate toward higher frequencies as the extended mass distributions are resolved. We show uniform disk, uniform sphere, and soliton lenses together with a point mass lens, considering $z_l=0.17$, $z_s=0.3$, $M=100~M_\odot$, $R=3x_E$, and $x_s=15x_E$ with $x_E=0.069~{\rm pc}$ as reference parameters.
    Right: halo-induced diffraction after aligning the high frequency geometric optics responses of a power law halo-dressed point mass lens and its effective point mass reference. The curves agree toward higher frequencies and separate toward lower frequencies because of halo diffraction. We use $\gamma=-1.25$, $z_l=0.17$, $z_s=0.3$, $M=20~M_\odot$, and $x_s=0.5~{\rm pc}$.
    The upper horizontal axes show the corresponding Fresnel scale.
}
    \label{fig:profile_comparison}
\end{figure*}

\subsection{Extended mass lenses}
\label{sec:extended}

We consider three representative extended mass lens profiles of a uniform disk, a uniform sphere  and a solitonic profile motivated by ultralight DM~\cite{Schive:2014dra}. Each lens is described by a total mass $M$ and a characteristic radius $R$. We consider $x_E=\sqrt{4d_{\rm eff}M} $  as a common reference length scale. For an extended lens, $x_E$ does not in general coincide with the radius of a critical curve. The density profiles and lens potentials used in the numerical calculations are summarized in Appendix~\ref{app:potentials}.

We first align the common low frequency response. When $x_F\gg R,x_s$, neither the characteristic scale of the extended mass lens nor the source impact parameter is resolved. The leading frequency dependence of $F(f)$ is then determined by the total lens mass. Extended mass lenses with the same $M$ therefore approach the same low frequency response as a point mass. A different value of $M$ would change this response before the lens structure is resolved.

Matching $M$ alone is not sufficient across the transition to geometric optics. The source impact parameter sets the approximate transition scale $x_F\sim x_s$. Different values of $x_s$ shift this transition to different frequencies. Since the diffraction and geometric optics contributions have different frequency dependence, the shift cannot in general be absorbed by an overall amplitude, a constant phase, or a time shift. We therefore use the same $M$ and $x_s$ for the extended and point mass lenses.

At fixed source-lens geometry, the point mass reference lens has only one structural parameter $M$  that is fixed by the low frequency alignment. Optimizing $M$ and $x_s$ of the reference lens in the likelihood could further reduce $\ln\Lambda_{LL'}$. Appendix~\ref{app:source_parameters} examines degeneracies with the intrinsic source parameters, and we leave a detailed joint analysis of lens and source parameters to future work. When comparing two extended mass profiles, we additionally use the same characteristic radius so that their profile shapes are tested at a common mass and size.

After $M$ and $x_s$ are aligned, the models have the same response for $x_F>\max(R,x_s)$. How they separate as $x_F$ decreases depends on the ordering of $R$ and $x_s$. If $R>x_s$, profile dependence first appears for $R>x_F>x_s$ through the aperture-mean convergence in Eq.~\eqref{eq:Fdiff}. Once $R>x_s>x_F$, the geometric optics and bGO contributions provide further profile dependence. If $x_s>R$, the characteristic scale of the extended mass lens enters the central diffractive image when $x_F\lesssim\sqrt{x_sR}$ through $\tilde{\kappa}(k_*)$ in Eq.~\eqref{eq:Fdiffgeo}. The left panel of Fig.~\ref{fig:profile_comparison} shows the latter ordering as the models share their low frequency behavior and separate toward higher frequencies as their extended mass distributions are resolved. The extended mass amplification factors shown in this panel are obtained by numerically evaluating the wave optics integral in Eq.~\eqref{eq:F(f)1d} over the full frequency range. The analytic expressions in Sec.~\ref{sec:waveoptics} are used solely to interpret their frequency dependence.

\subsection{Halo-dressed point mass lenses}
\label{sec:diffdress}

A halo-dressed point mass lens consists of a central point mass and an extended mass distribution. We consider a power law halo with $\kappa_h(x)\propto x^\gamma$, characterized by the mass $M$ of the central point mass component, the halo scale $R_h$, the amplitude parameter $\mathcal{R}$, and the slope $\gamma$. The full density profile and lens potential are given in Appendix~\ref{app:dress}. We use $\gamma=-1.25$ for the fiducial example shown below.

For a halo-dressed point mass lens, the halo extends to much larger scales than the central point mass. 
For a finite halo, the total mass limit of $F(f)$, in which the entire halo is unresolved, would occur only at frequencies far below the range relevant here. 
The low frequency alignment adopted for extended mass lenses is therefore not applicable over the frequency range considered here.

At sufficiently high frequencies, the halo-dressed lens and its effective point mass reference are both described by two geometric images. We therefore align their image magnifications and relative time delay in geometric optics~\cite{GilChoi:2023ahp}. The mapping to the effective point mass parameters is given in Appendix~\ref{app:matching}.

After this alignment, the remaining frequency-dependent difference is produced by halo diffraction. For $x_F<x_s$, its characteristic contribution is the central diffractive image discussed in Sec.~\ref{sec:diffprofile}. For $x_F>x_s$, the halo profile enters through the source-centered diffraction integral in Eq.~\eqref{eq:Fdiff}. The right panel of Fig.~\ref{fig:profile_comparison} shows how the halo-dressed lens approaches the matched geometric optics signal toward high frequencies and departs from it toward lower frequencies. The halo-dressed point mass amplification factor shown in this panel is obtained by numerically evaluating the wave optics integral in Eq.~\eqref{eq:F(f)1d} over the full frequency range. The analytic expressions in Sec.~\ref{sec:waveoptics} are used only to interpret its frequency dependence.

The two comparisons therefore expose profile-dependent diffraction by aligning different limits: the low frequency unresolved response for extended mass lenses and the high frequency geometric optics response for halo-dressed point mass lenses. The reference parameters are held fixed at their a priori aligned values. A broader optimization over lens parameters may introduce additional degeneracies. The same geometric optics matching can be applied to other systems analogous to halo-dressed point mass described here, including the superradiance cloud considered in Sec.~\ref{sec:cloud}.

\section{Statistical Framework for Lensing Forecasts}\label{sec:statistics}
\subsection{Signal detection and distinguishability}

Detection of lensing and distinction between two different lensing models can be quantified by the Bayes factor~\cite{romano2017detection}
\begin{equation}\label{eq:lambda}
    \Lambda_{LL'} = \frac{p(d|\mathcal{H}_L)}{p(d|\mathcal{H}_{L'})}~.
\end{equation}
Here $p(d|\mathcal{H}_L)$ and $p(d|\mathcal{H}_{L'})$ are the marginalized likelihoods for the data $d$ under the lensing hypotheses $\mathcal{H}_L$ and $\mathcal{H}_{L'}$, respectively. The alternative hypothesis $\mathcal{H}_{L'}$ may also correspond to the no-lensing hypothesis $\mathcal{H}_0$. Evaluating the marginalized likelihoods for every lens configuration would require full parameter estimation. We instead use a Laplace approximation, in which each marginalized likelihood is given by its maximum likelihood multiplied by factors determined by the local prior density and posterior volume. 

Assuming that the ratio of the relevant factors does not differ greatly between the two hypotheses, the log Bayes factor is dominated by the difference between their maximum matched-filter likelihoods~\cite{Cornish:2007if,GilChoi:2023ahp,dai2018detecting,romano2017detection}:
\begin{equation}
    \ln \Lambda_{LL'} \simeq \frac{1}{2}\left(\rho_{m,L}^2 - \rho_{m,L'}^2 \right)~ ,
\end{equation}
where
\begin{equation} \label{eq:mf}
    \rho_{m,L}^2 = \max \limits_{\bm{\theta}} \left[(d|d)-(d-h(\bm{\theta};\mathcal{H}_L)|d-h(\bm{\theta};\mathcal{H}_L))\right]
\end{equation}
is the squared matched-filter signal-to-noise ratio (SNR) for the waveform model $h(\bm{\theta};\mathcal{H}_L)$ with model parameters $\bm{\theta}$. The noise-weighted inner product is defined as
\begin{equation}
    (a|b) = 4\mathrm{Re}\int_0^{\infty} df~ \frac{a^*(f)b(f)}{S_n(f)}
\end{equation}
for stationary Gaussian detector noise with power spectral density $S_n(f)$.
Throughout this work, $\ln\Lambda$ denotes this maximum likelihood approximation rather than a full evidence calculation.

The matched-filter SNR and likelihood ratio depend on the particular realization of detector noise response. To simplify the analysis, we adopt the zero noise approximation $d\simeq h_L$. The expectation value of the log likelihood ratio is then
\begin{equation}\label{eq:lambdaexp}
    \langle \ln \Lambda_{LL'} \rangle \simeq \frac{1}{2}\left(\rho_L^2 - \rho_{L'}^2  \right)
     = \frac{1}{2}\min \limits_{\bm{\theta}} \left( h_L-h_{L'}(\bm{\theta})|h_L-h_{L'}(\bm{\theta}) \right)~ ,
\end{equation}
where $\rho_L^2 = \langle\rho_{m,L}^2 \rangle \simeq (h_L|h_L)$ and $\rho_{L'}^2 = \langle\rho_{m,L'}^2 \rangle \simeq \max_{\bm{\theta}}(h_{L'}(\bm{\theta})|2h_L-h_{L'}(\bm{\theta}))$.

In this work, $\bm{\theta}$ includes the overall amplitude, a constant phase shift, and a time shift, which are optimized in the computation of $\ln\Lambda_{LL'}$. For each sampled source and lens configuration, the population forecasts hold the intrinsic BBH and lens parameters fixed in this minimization. The reference lens parameters are set by the alignment prescriptions described in Sec.~\ref{sec:Distinction} and are not varied further in this minimization. This allows us to explore a broad range of lensing configurations without performing full parameter estimation for each case. We note that resulting forecasts quantify discrimination compared to considered reference lens parameters that are fixed and not separately optimized; fitting them would in general reduce $\ln\Lambda_{LL'}$. Thus the results do not strictly establish exclusion of all alternative possible lens parameters or profiles.

We consider lensing detection when 
\begin{equation}\label{eq:detcri}
    \rho_L > 8,\quad\langle \ln \Lambda_{L0} \rangle>3   ~.
\end{equation}
For distinguishability between lenses, we require both
\begin{equation}\label{eq:distcri}
    \rho_L > 8,\quad\langle \ln \Lambda_{L0} \rangle>3, \quad \mathrm{and}\quad \langle \ln \Lambda_{LL'} \rangle>3~.
\end{equation}
We adopt $\rho_L>8$ as a benchmark source-detection threshold, following the single detector sensitivity convention used in Ref.~\cite{abbott2020prospects}. The additional likelihood ratio cuts select detectable lensing and distinguishable lens profiles. We adopt $\langle\ln\Lambda\rangle>3$ as a benchmark threshold. For representative lensing configurations, Ref.~\cite{GilChoi:2023ahp} found false alarm probabilities below $1\%$ at this threshold. Raising the threshold to $5$ or $8$ reduces the lensing probability in our calculations, but the changes are within a factor of few. A similar stability of the forecast scale under a more conservative threshold was found for PBH lensing analysis in Ref.~\cite{Jung:2017flg}. In observational searches, the significance of lensing candidates is assessed against an unlensed background and can depend on search trials and waveform systematics~\cite{LIGOScientific:2023bwz,LIGOScientific:2025cwb}. Calibrating the thresholds for a specific search is beyond the scope of this work.

The thresholds in Eqs.~\ref{eq:detcri} and \ref{eq:distcri} determine the range of source impact parameters that contributes to the lensing cross section used in the probability calculation below. Including additional source parameters in the minimization in Eq.~\eqref{eq:lambdaexp} may reduce $\ln\Lambda$ and shrink this range. Appendix~\ref{app:source_parameters} evaluates this effect for representative benchmark configurations.

\subsection{Unlensed gravitational waveform}

For the purposes of this work, we model the unlensed GW signal using the Newtonian order frequency domain inspiral waveform~\cite{cutler1994gravitational} 
\begin{equation}
\begin{split}
    h_0(f) &= - \sqrt{\frac{5}{24}} ~ \pi^{-2/3} A_p
    \frac{\mathcal{M}_z^{5/6}}{d_L} f^{-7/6} e^{i\Psi(f)}, \\
    \Psi(f) &= \theta_c + 2\pi f t_c - \frac{\pi}{4}
    + \frac{3}{128} (\pi \mathcal{M}_z f)^{-5/3}.
\end{split}
\end{equation}
 Here, $d_L$ is the luminosity distance to the source, and
\begin{equation}
    \mathcal{M}_z = (1+z_s)
    \frac{(m_1m_2)^{3/5}}{(m_1+m_2)^{1/5}}
\end{equation}
is the detector frame chirp mass of a BBH system with companion black hole masses $m_1$ and $m_2$. The factor $A_p$ accounts for the source polarization and detector response. We set $A_p=1$ for simplicity, corresponding to an optimally oriented source for ground-based detectors (see also Ref.~\cite{Finn:1992xs}). For space-based detectors, we retain this waveform normalization and use the effective noise curves with the averaged detector response described below.  The forecasts should be considered as order of magnitude benchmarks. The corresponding lensed waveform is $h_L(f)=F(f)h_0(f)$, where $F(f)$ is the amplification factor introduced in Sec.~\ref{sec:waveoptics}.
We truncate the waveform at the upper frequency cutoff given by the innermost stable circular orbit (ISCO),
\begin{equation}
    f_{\mathrm{ISCO}} = \left[6^{3/2}\pi(m_1+m_2)(1+z_s)\right]^{-1}.
\end{equation}

For ground-based detectors, we use the $40~\mathrm{km}$ low frequency configuration of CE~\cite{srivastava2022science,CE:strain-v8}, the $10~\mathrm{km}$ ET-D sensitivity curve combining the low and high frequency interferometers~\cite{hild2011sensitivity}\footnote{We use the $90^\circ$ opening angle convention of the published ET-D curve without an additional factor for the full triangular observatory.}, and the LIGO $A^\sharp$ sensitivity curve~\cite{ligoAsharp2023}. For the O4-level configuration, we use the arithmetic mean of the Hanford and Livingston noise power spectral densities reported in Ref.~\cite{Capote:2024rmo}, treating it as a hypothetical single detector. We adopt lower frequency cutoffs of $5~\mathrm{Hz}$ for CE, $2~\mathrm{Hz}$ for ET, and $10~\mathrm{Hz}$ for both LIGO configurations.

For space-based detectors such as LISA~\cite{LISA:2017pwj} and Taiji~\cite{Hu:2017mde}, the lower cutoff is instead set by the binary frequency at the beginning of the observation period. Using the inspiral time to coalescence relation for a circular binary~\cite{Cutler:1997ta}, we take
\begin{equation}
    f_{\mathrm{init}} = 0.015 \left(\frac{30~M_\odot}{\mathcal{M}_z}\right)^{5/8} \left(\frac{5~\mathrm{yr}}{t_{\mathrm{obs}}}\right)^{3/8} ~\mathrm{Hz},
\end{equation}
where $t_{\mathrm{obs}}$ is the considered observation time.

For LISA, we use the instrumental noise model of Ref.~\cite{Robson:2018ifk} with an arm length of $2.5\times10^9~\mathrm{m}$. For Taiji, we use an arm length of $3\times10^9~\mathrm{m}$ and nominal displacement and acceleration noise amplitudes of $8~\mathrm{pm}/\sqrt{\mathrm{Hz}}$ and $3\times10^{-15}~\mathrm{m\,s^{-2}}/\sqrt{\mathrm{Hz}}$, respectively~\cite{Liu:2023qap}. We take the Taiji displacement noise to be frequency independent and retain only the factor $1+(0.4~\mathrm{mHz}/f)^2$ in the frequency dependence of its acceleration noise power. Both effective strain noise curves include the sky  and polarization averaged response for two independent low-frequency channels~\cite{Robson:2018ifk}. We omit confusion noise from unresolved Galactic binaries.

\subsection{Calculation of lensing probability}

The lensing probability is computed from the optical depth for isolated lens objects,
\begin{equation}
    \tau(\bm{\theta}_h,z_s;\bm{\theta}_L) = \int_0^{z_s} dz_l~ n_l~ \sigma(z_l;\bm{\theta}_h,z_s,\bm{\theta}_L) (1+z_l)^2 \frac{1}{H(z_l)}~,
\end{equation}
where $n_l$ is the comoving number density of the lens objects and $\sigma(z_l;\bm{\theta}_h,z_s,\bm{\theta}_L)$ is the lensing cross-section. The latter is defined by the area spanned by projected source positions $\bm{x}_s$ that satisfy either the detection criterion in Eq.~\eqref{eq:detcri} or the distinguishability criterion in Eq.~\eqref{eq:distcri}.

We assume a uniform comoving distribution of lenses, $n_l=\mathrm{constant}$, and adopt a flat $\Lambda$CDM cosmology~\cite{ParticleDataGroup:2026mpi}, with the Hubble expansion rate
\begin{equation}
    H(z) = H_0\sqrt{\Omega_m(1+z)^3+\Omega_\Lambda},
\end{equation}
with fiducial parameters $H_0=70~{\rm km~s^{-1}~Mpc^{-1}}$ and $\Omega_m=1-\Omega_\Lambda=0.3$. For lenses of mass $M_{\rm Lens}$ constituting a fraction
$f_{\rm DM}$ of DM, we take the comoving
number density to be
\begin{equation}
    n_l = \frac{f_{\rm DM}\rho_{\rm DM,0}}{M_{\rm Lens}},
    \qquad
    \rho_{\rm DM,0} = \Omega_{\rm DM}\rho_{c,0},
\end{equation}
where $\rho_{\rm DM,0}$ is the present day mean DM density and $\rho_{c,0}=3H_0^2/(8\pi)$ is the present day critical density. We adopt $\Omega_{\rm DM}=0.25$, with $H_0$ denoting the present day Hubble expansion rate.

We evaluate the optical depth numerically using Monte Carlo integration. For each source, we first determine a maximum source impact parameter that encloses the region satisfying the detection or distinguishability criterion. We use the critical source impact parameter evaluated at the lens redshift that maximizes $d_{\rm eff}$ to set the sampling boundary. Lens positions are then drawn uniformly within a cylinder in comoving space whose axis connects the observer and the source and whose transverse radius is given by the corresponding comoving source impact parameter. The required volume is obtained by multiplying the cylinder volume by the fraction of sampled lens positions that satisfy Eq.~\eqref{eq:detcri} or Eq.~\eqref{eq:distcri}. We find that 100 random lens positions provide sufficient precision for the calculations in this work.

We use this Monte Carlo procedure for CE, ET, LISA, and Taiji. For the LIGO configurations, the optical depth is much smaller and the detectable cross section has a more complicated dependence on $z_l$, making direct Monte Carlo sampling inefficient. 
We therefore determine the source impact parameter intervals satisfying the detection or distinction criteria at each lens redshift and integrate their areas to obtain $\sigma(z_l)$. Where a single image condition is imposed, we also exclude the multiple-image region, as specified in Sec.~\ref{sec:prospectextended}.

Although the Monte Carlo procedure samples many lens positions, the optical depth is computed under the isolated-lens approximation: lensing contributions from different lens positions are not combined coherently. The quantity $\tau(\bm{\theta}_h,z_s;\bm{\theta}_L)$ can therefore be interpreted as the expected number of isolated lens objects whose lensing signals satisfy the adopted criterion for a single BBH source. When $\tau\gtrsim1$, the isolated-lens approximation becomes inaccurate and multiple lenses should be treated simultaneously~\cite{Zumalacarregui:2024ocb,Kim:2025njb}. We do not model this regime explicitly in the present work.

For an isolated lens population, the probability that a single BBH source is lensed by at least one relevant object is
\begin{equation}
    p_L(\bm{\theta}_h,z_s;\bm{\theta}_L)
    = 1-e^{-\tau(\bm{\theta}_h,z_s;\bm{\theta}_L)}~.
\end{equation}
The lensing probability averaged over the BBH population is then
\begin{equation}\label{eq:lensprob}
    P_L = \int_0^{z_c} dz_s \int d^m\theta_h~ p_{\rm BBH}(\bm{\theta}_h,z_s) p_L(\bm{\theta}_h,z_s;\bm{\theta}_L),
\end{equation}
where $p_{\rm BBH}(\bm{\theta}_h,z_s)$ is the BBH population distribution. For ground-based detectors, $\bm{\theta}_h$ includes the primary BH mass and the mass ratio $q=m_2/m_1$, and we adopt the Power Law + Peak mass distribution~\cite{talbot2018measuring} with the GWTC-2 parameter values used as fixed forecast inputs~\cite{abbott2021population}. We neglect redshift evolution of the mass distribution and include sources up to $z_c=10$. For space-based detectors, we instead consider equal-mass massive BBHs with a total mass of $10^6~M_\odot$.
For CE, ET, LISA, and Taiji, we evaluate Eq.~\eqref{eq:lensprob} using samples drawn from $p_{\rm BBH}(\bm{\theta}_h,z_s)$,
\begin{equation}
    P_L \simeq \frac{1}{N} \sum_{j=1}^{N} p_L(\bm{\theta}_h^j,z_s^j;\bm{\theta}_L)~.
\end{equation}
For the LIGO configurations, we instead evaluate Eq.~\eqref{eq:lensprob} by direct numerical integration over the BBH population.

The expected number of lensing events is
\begin{equation}
    \langle N_L\rangle = P_L~\mathcal{V}~R_0~T_{\rm obs},
\end{equation}
where
\begin{equation}
    \mathcal{V} = \int_0^{z_c} dz_s~ \frac{dV_c(z_s)}{dz_s} \frac{1}{1+z_s}
\end{equation}
is the effective source volume and $T_{\rm obs}$ is the observation time. For stellar-mass BBHs, we adopt the GWTC-2 value $R_0=28.3~{\rm Gpc}^{-3}~{\rm yr}^{-1}$~\cite{abbott2021population}. For the massive BBHs considered with LISA and Taiji, we instead use a benchmark normalization $R_0=0.01~{\rm Gpc}^{-3}~{\rm yr}^{-1}$, whose corresponding total merger rate is comparable to that of a representative population model~\cite{Klein:2015hvg}, as discussed below. In both cases, we assume a redshift-independent comoving merger rate up to $z_c=10$.

\subsection{Projected sensitivity to lens object abundance}

We estimate the projected sensitivity to the abundance of lens objects with Poisson statistics, expressed as the 90\% projected upper limit expected in the absence of lensing detections.   Assuming a null result, the posterior distribution of DM fraction $f_{\rm DM}$ is
\begin{equation}
    p(f_{\rm DM}~|~N_L=0) = \frac{ \exp\!\left[-\langle N_L(f_{\rm DM})\rangle\right] }{ \int_0^1 df_{\rm DM}~ \exp\!\left[-\langle N_L(f_{\rm DM})\rangle\right] }~,
\end{equation}
where $\langle N_L(f_{\rm DM})\rangle$ is the expected number of lensing events for a given abundance. Adopting a flat prior over $0\leq f_{\rm DM}\leq1$, the $90\%$ projected upper limit is given by
\begin{equation}
    \int_0^{f_{\rm DM}^{90\%}} df_{\rm DM}~ p(f_{\rm DM}~|~N_L=0) = 0.9~.
\end{equation}

The interpretation of the null result depends on the criterion used to define $\langle N_L\rangle$. For the detection criterion in Eq.~\eqref{eq:detcri}, $N_L=0$ corresponds to absence of detected lensing event. For the distinguishability criterion in Eq.~\eqref{eq:distcri}, it instead means that no event is identified as arising from the target lens population. A detectable lensing signal could still be consistent with another lens model, so the distinguishability requirement generally reduces the expected number of identifiable events. The corresponding projected upper limits on $f_{\rm DM}$ are therefore weaker than those considering detection only.

The abundance forecasts below quantify the sensitivity of GW lensing under the adopted lens and population models. We do not combine them with constraints from other observations; the displayed parameter ranges therefore do not represent regions allowed by all existing bounds.

\section{Application to Extended Dark Matter Lenses}\label{sec:prospectextended}

\subsection{Lensing probability for ground-based detectors}

We first estimate the probability that diffraction allows an extended mass lens to be distinguished from a point mass lens. We consider the uniform disk, uniform sphere and solitonic profiles introduced in Sec.~\ref{sec:Distinction}, and compute the lensing probability $P_L$ across the $(M_{\rm Lens},R)$ plane using Eqs.~\eqref{eq:distcri} and \eqref{eq:lensprob}. In each comparison, the extended lens and its point mass reference have the same total mass $M_{\rm Lens}$, source impact parameter $x_s$, and lens redshift $z_l$. For our analysis we impose the single image requirement 
\begin{equation}
 x_c < x_s, 
\end{equation}   
where $x_c$ is the caustic radius defined in Appendix~\ref{app:caustic}. This condition removes multiple image configurations and isolates the regime in which the distinction from a point mass is provided by diffraction.

For each lens model and geometry, we evaluate $F(f)$ by direct numerical integration of Eq.~\eqref{eq:F(f)1d}. We therefore do not rely on the linear potential approximation derived in Sec.~\ref{sec:diffprofile} for the probability calculation. Instead, the diffraction formalism developed there is used to interpret the resulting frequency dependence.

\begin{figure}[t]
    \centering
    \includegraphics[width=0.99\linewidth]{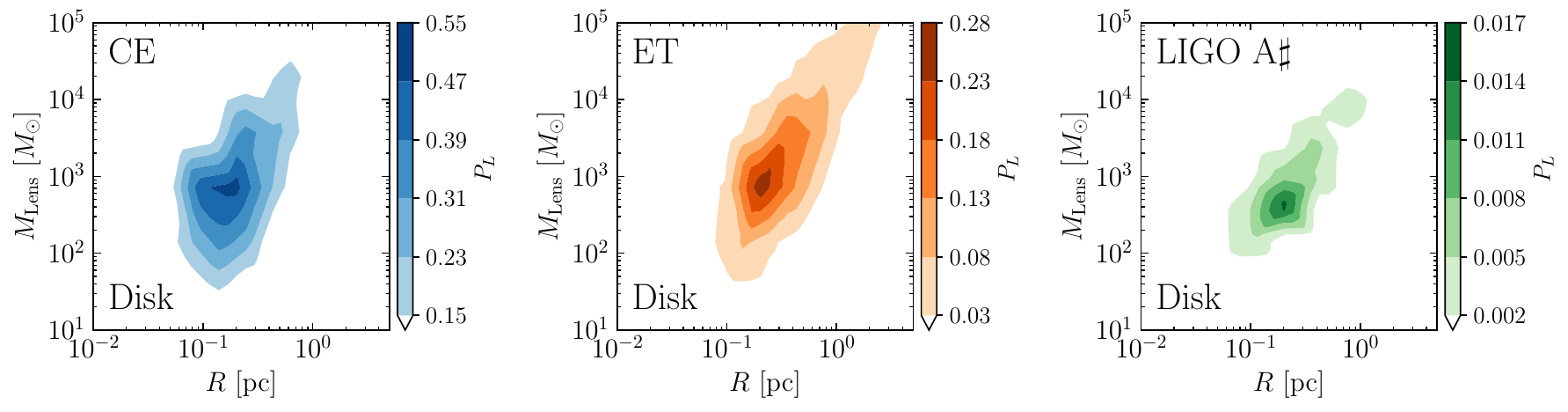}
    \includegraphics[width=0.99\linewidth]{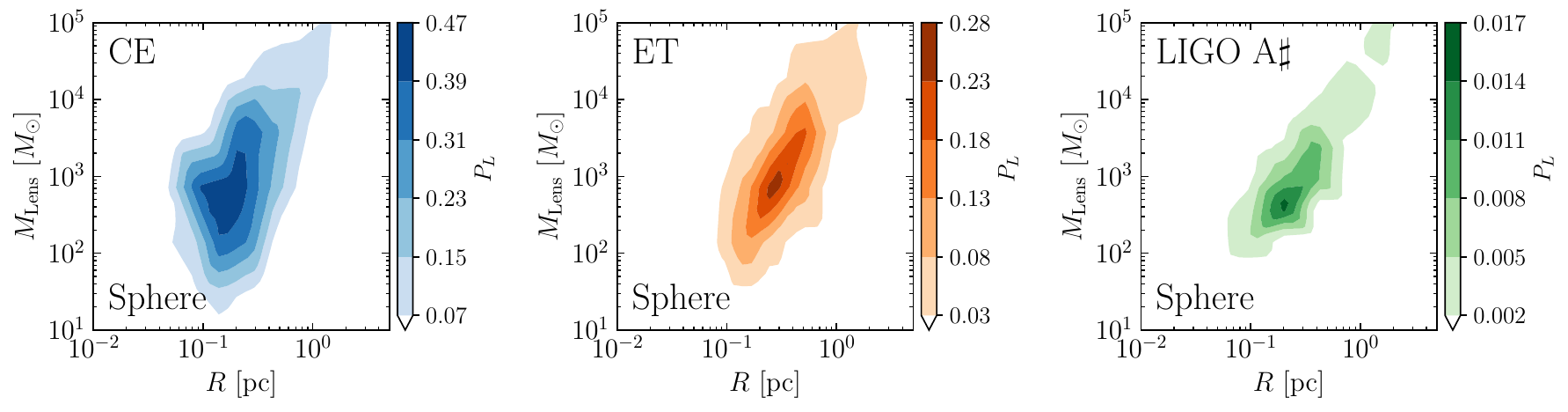}
    \includegraphics[width=0.99\linewidth]{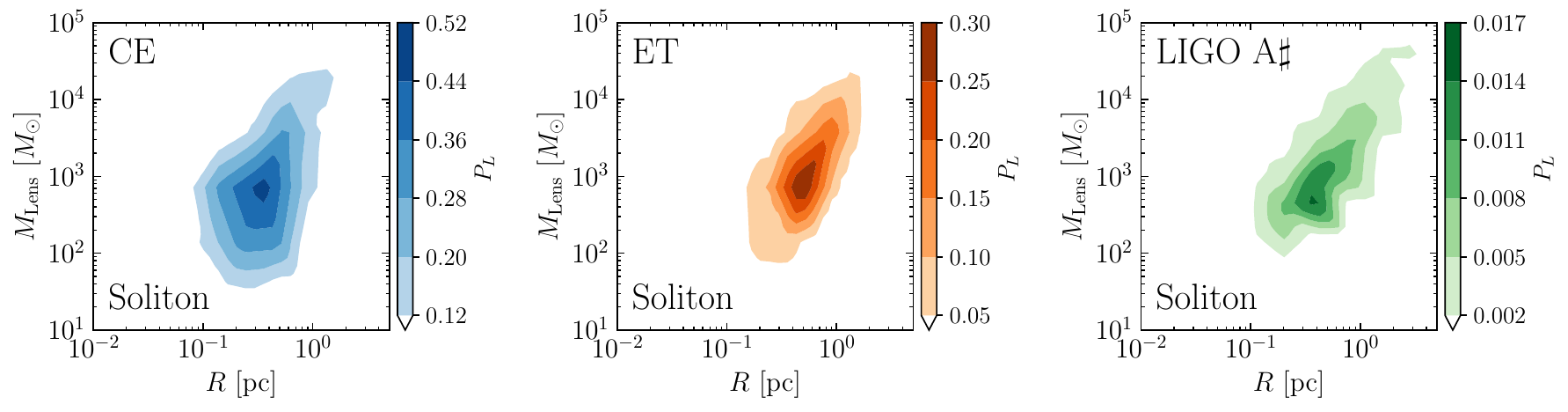}
    \caption{
    Lensing probability in the $(M_{\rm Lens},R)$ plane for the uniform disk, uniform sphere, and soliton profiles, requiring distinguishability from a point mass lens under the single image condition $x_s>x_c$. The columns correspond to CE, ET, and LIGO $A^\sharp$, as indicated by the panel labels. The color scale shows $P_L$ for $f_{\rm DM}=1$, using the source and lens populations described in Sec.~\ref{sec:statistics}.
}
    \label{fig:extlens_prob}
\end{figure}

 Fig.~\ref{fig:extlens_prob} shows the lensing probabilities for CE~\cite{srivastava2022science}, ET~\cite{hild2011sensitivity}, and LIGO $A^\sharp$~\cite{ligoAsharp2023} configurations, assuming $f_{\rm DM}=1$. A common feature of all three considered lens profiles is that the probability is largest around
\begin{equation}
    M_{\rm Lens}\sim O(10^2)~M_\odot, \qquad R\sim O(0.1)~{\rm pc}.
\end{equation}

The preferred scale in $R$ can be understood from the resolution condition discussed in Sec.~\ref{sec:diffprofile}. Diffraction becomes sensitive to structure on the scale $R$ when
\begin{equation}
    k_*R = \frac{x_sR}{x_F^2} \sim 1,
\end{equation}
or equivalently $x_F\sim\sqrt{x_sR}$. For $k_*R\ll1$, the finite lens remains unresolved and its diffraction pattern approaches that of a point mass. In the opposite limit, $k_*R\gg1$, the profile is resolved, but the diffractive contribution is suppressed as $|\tilde{\kappa}(k_*)|$ decreases toward large $k_*$. The distinction probability is therefore largest near the transition between these regimes.

The configurations that contribute most strongly to the probability also typically have $x_s=O(x_F)$. If $x_s=\alpha x_F$ with $\alpha\gg1$, the lens is already deep in the resolved regime at the characteristic Fresnel scale of the detector, where the diffractive contribution is suppressed. Combining $x_s\sim x_F$ with $k_*R\sim1$ therefore gives the approximate condition
\begin{equation}
    R\sim x_F~.
\end{equation}
For stellar-mass BBH sources at cosmological distances, the most relevant frequencies of ground-based detectors correspond to $x_F=O(0.1)~{\rm pc}$. This explains the preferred lens radius seen in Fig.~\ref{fig:extlens_prob}.

The preferred mass scale is set by the reference scale
\begin{equation}
    x_E = \sqrt{4d_{\rm eff}M_{\rm Lens}} \simeq 0.14
    \Big(
        \frac{d_{\rm eff}}{1~{\rm Gpc}}
        \frac{M_{\rm Lens}}{10^2~M_\odot} 
    \Big)^{1/2}   ~{\rm pc}~.
\end{equation}
For $x_E\ll R$, increasing $M_{\rm Lens}$ strengthens the lensing signal and increases the distinction probability. When $x_E\gg R$, the caustic grows and a larger fraction of detectable configurations enters the multiple image regime, which is removed by the single image condition. The probability therefore peaks approximately where $x_E\sim R$. Together with the diffraction condition above, this gives the characteristic scaling
\begin{equation}
    x_F\sim x_E\sim R~.
\end{equation}

The detector frequency range also has an important effect on the lensing probability. ET gives peak probabilities about a factor of two smaller than CE, even though the SNR  of our fiducial stellar-mass BBHs are typically smaller by a factor of $3-4$. ET extends down to $2~{\rm Hz}$, compared with $5~{\rm Hz}$ for CE, providing a broader range of Fresnel scales over which the transition between unresolved and resolved lens structure can be measured. This additional frequency coverage partly compensates for the smaller SNR. LIGO $A^\sharp$, with both a smaller SNR and frequencies down to 10~Hz, allows for substantially smaller probabilities.

The dependence on the detailed lens profile is comparatively mild. The uniform disk, uniform sphere, and soliton produce similar probability contours, with modest differences in the peak probability, its location, and the contour shape. The overall parameter range in which an extended lens can be distinguished from a point mass is therefore determined mainly by the characteristic mass and size of the lens rather than by the detailed inner profile. Below, we examine whether diffraction can also distinguish different extended mass profiles from one another.

\begin{figure}[t]
    \centering
    \includegraphics[width=0.66\linewidth]{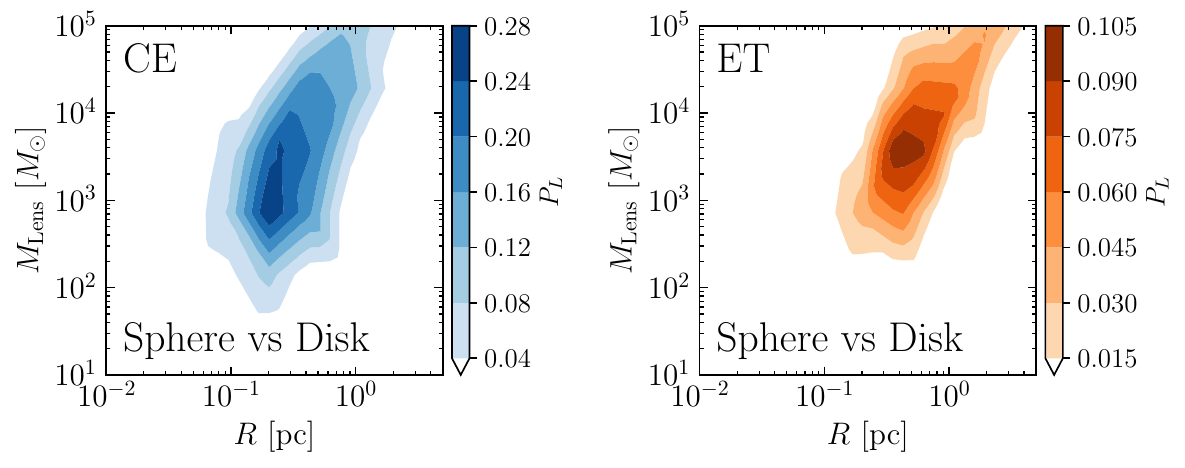}
    \caption{
    Lensing probability in the $(M_{\rm Lens},R)$ plane for distinguishing a uniform sphere lens from a uniform disk model through diffraction. The two profiles are compared at the same $M_{\rm Lens}$, $R$, $x_s$, and $z_l$. Results are shown for CE and ET using the stellar-mass BBH  source and lens populations described in Sec.~\ref{sec:statistics}, with $f_{\rm DM}=1$ and the single image condition $x_s>x_c$.   }
    \label{fig:spheredisk_prob}
\end{figure}

\subsection{Distinguishability and sensitivity to dark matter abundance}

The similarity of the probability contours  in Fig.~\ref{fig:extlens_prob} does not imply that different extended mass profiles are observationally indistinguishable. Distinction from a point mass primarily tests whether the finite size of the lens can be resolved, whereas distinguishing two extended profiles requires sensitivity to differences in their mass distributions within the resolved region.

\begin{figure}[t]
    \centering
        \includegraphics[width=0.8\linewidth]{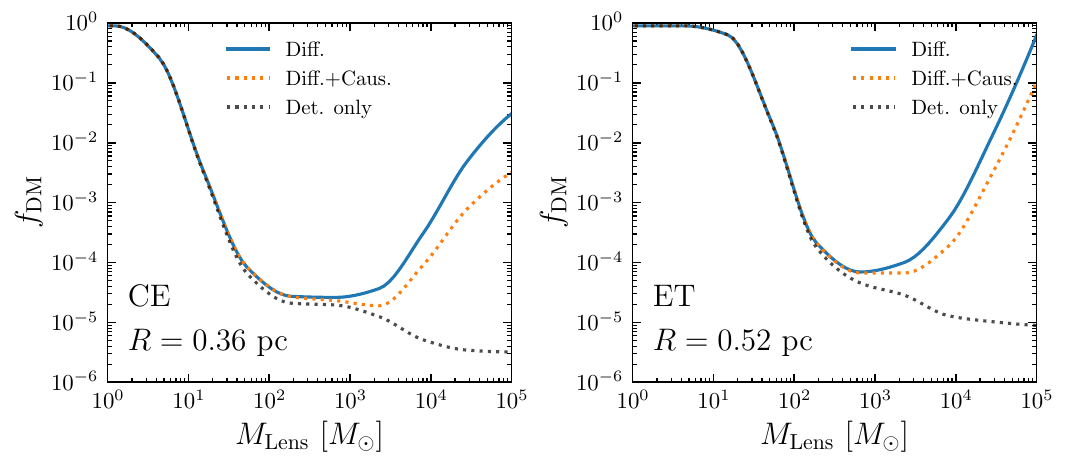}
    \includegraphics[width=0.8\linewidth]{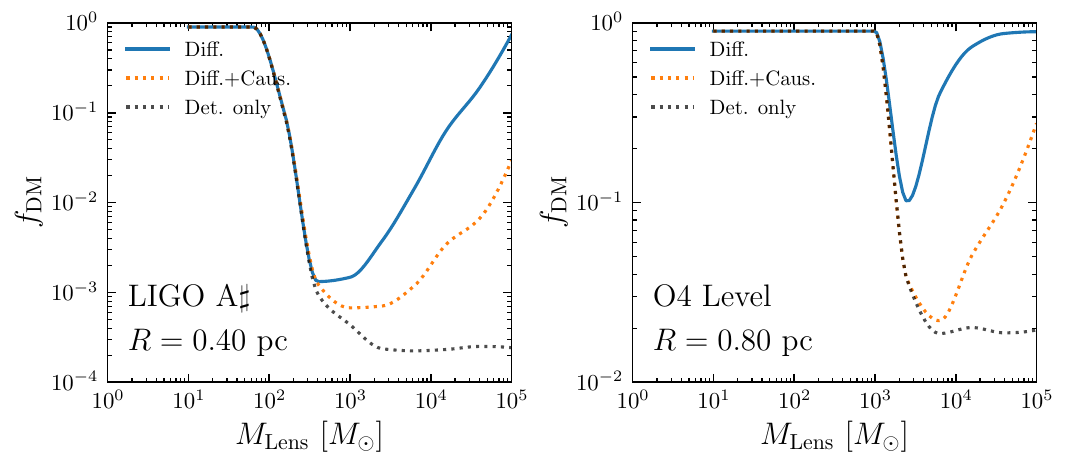}
    \caption{Projected $90\%$ upper limits on the DM fraction $f_{\rm DM}$ for soliton lenses considering $5$ yr observations with CE (top left) and ET (top right). For each detector, $R$ is fixed to the value corresponding to the maximum lensing probability in the $(M_{\rm Lens},R)$ plane, as indicated in the panel. The solid curves require distinction from a point mass lens in the single image regime. The orange dotted curves apply the same distinction criterion while including both single  and multiple image configurations. The black dotted curves show the detection only projection, requiring only distinction from the unlensed signal.
    LIGO $A^\sharp$ (bottom left) and a  LIGO O4-level detector (bottom right) are displayed analogously. For the O4-level case, we use a hypothetical single detector with a noise curve given by the average of the Hanford and Livingston O4 sensitivities and assume  $5$ yr observation time. The result should therefore be interpreted as an illustrative sensitivity projection rather than a constraint based on the actual O4 observing run.    
    }
    \label{fig:fDMCEETLIGO}
\end{figure}

As a representative example, we compare a uniform sphere lens with a uniform disk model. The two models are evaluated at the same $M_{\rm Lens}$, $R$, $x_s$, and $z_l$, so that the comparison isolates the difference in the lens profile.  Fig.~\ref{fig:spheredisk_prob} shows the resulting distinguishability probability for CE and ET. Although the probabilities are somewhat smaller than those for distinguishing the sphere from a point mass in Fig.~\ref{fig:extlens_prob}, they remain of comparable magnitude. Thus, once diffraction resolves the finite lens structure, the signal can retain sensitivity to differences between extended mass profiles.

The preferred scale in $R$ remains similar to that in Fig.~\ref{fig:extlens_prob}, since the profile-dependent differences become measurable when the characteristic lens scale is resolved within the detector band. The probability peak shifts toward somewhat larger $M_{\rm Lens}$, indicating that distinguishing two extended profiles requires a stronger lensing signal than distinguishing an extended lens from a point mass. This is expected because the former comparison depends on differences between two finite size diffraction patterns, whereas the latter is primarily sensitive to the appearance of finite size structure itself.

We next convert the lensing probabilities into projected upper limits on $f_{\rm DM}$ using the procedure described in Sec.~\ref{sec:statistics}. We focus on the soliton profile as a representative physically motivated extended mass model. For each detector, we fix $R$ to the value corresponding to the maximum of the lensing probability in the $(M_{\rm Lens},R)$ plane and vary $M_{\rm Lens}$ along this constant-$R$ slice.

Extended DM objects are constrained by a range of observations, including stellar microlensing searches~\cite{Croon:2020wpr} and interstellar gas heating~\cite{Kim:2025gck}, with further microlensing sensitivity expected from Roman~\cite{DeRocco:2023hij}. These bounds and projections depend on the object size and density profile, and their published curves do not in general correspond to the fixed-$R$ soliton models considered here. We therefore present the GW-lensing sensitivity separately, without overlaying constraints evaluated for different object models.

 Fig.~\ref{fig:fDMCEETLIGO} shows the projected $90\%$ upper limits for CE and ET. The solid curves use the same criterion as in Sec.~\ref{sec:prospectextended}: the soliton lens must be distinguishable from a point mass lens in the single image regime. For this distinction-conditioned projection, the assumed null result means that no event is identified as arising from the target soliton population. Around $M_{\rm Lens}\sim10^2~M_\odot$, the projected limits reach $f_{\rm DM}\sim10^{-5}$ for CE and $f_{\rm DM}\sim10^{-4}$ for ET.

The orange dotted curves show the effect of removing the single image restriction while retaining the distinction requirement. They therefore include both single  and multiple-image configurations. The improvement is largest at higher masses, where $x_E\gtrsim R$ and caustic configurations become increasingly important. Including these events enlarges the region in which the soliton signal can be distinguished from the point mass model.

The black dotted curves show the detection-only projection, for which the lensed signal need only be distinguishable from the unlensed waveform. These curves therefore give the most optimistic reach among the three criteria shown. At low masses, the detection-only and distinction-conditioned projections are similar, indicating that a detectable diffraction signal can also retain information about the finite lens size. Toward higher masses, the detection-only limits become substantially stronger than the distinction-conditioned limits. In this regime, lensing can remain detectable even when the signal is difficult to identify specifically as arising from an extended mass lens.

 Fig.~\ref{fig:fDMCEETLIGO} also displays the corresponding projections for LIGO $A^\sharp$ and the O4-level sensitivity. The same qualitative behavior is found, but the projected limits are weaker than those for CE and ET. LIGO $A^\sharp$ reaches around $f_{\rm DM}\sim10^{-3}$ at its most sensitive masses. For the O4-level configuration, the single image distinction limit reaches only around $f_{\rm DM} \sim 10^{-1}$, while including caustic configurations or adopting the detection only criterion reaches around $f_{\rm DM} \sim 10^{-2}$. The O4-level result is based on the hypothetical $5$ yr single detector configuration observation described in the caption and is not a projection for the actual O4 data set. Across the detector configurations, the strongest sensitivity projections occur near the same region selected by the interplay of the Fresnel scale, the reference scale $x_E$, and the finite lens size discussed in Sec.~\ref{sec:prospectextended}.

\subsection{Lensing probability and sensitivity for space-based detectors}

\begin{figure}[t]
    \centering
\includegraphics[width=0.66\linewidth]{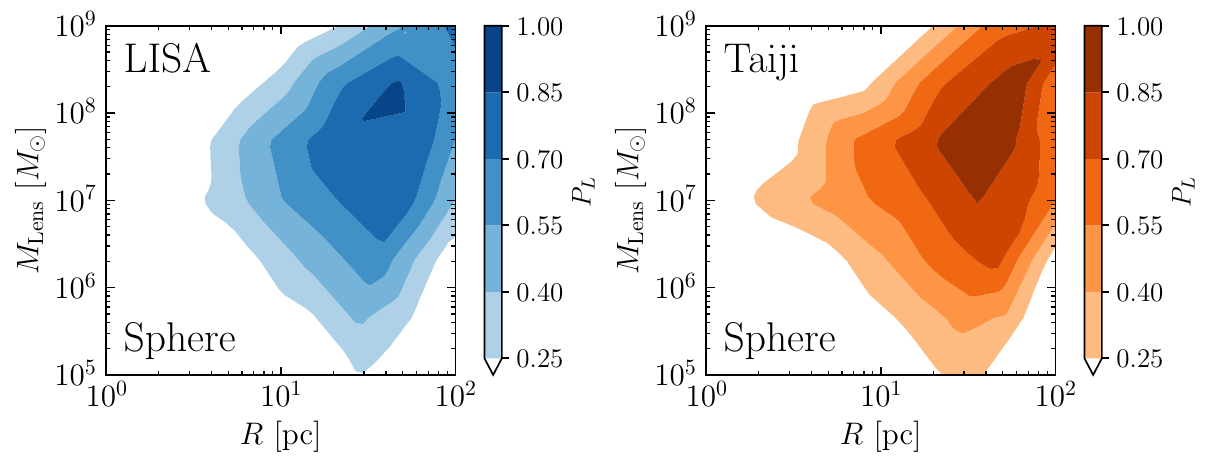}
    \caption{
    Lensing probability in the $(M_{\rm Lens},R)$ plane for distinguishing a uniform sphere lens from a point mass lens with  space-based LISA and Taiji experiments. The source population consists of equal mass BBHs with total mass $10^6~M_\odot$, and $f_{\rm DM}=1$ is assumed.    }
    \label{fig:extlens_prob_space}
\end{figure}

We next analyze space-based detectors LISA~\cite{LISA:2017pwj} and Taiji~\cite{Hu:2017mde}.  We consider space-based detectors separately since typically their lower observing frequencies enable probing different lens masses and spatial scales and require a different source population model compared to ground-based detectors. For these forecasts, we consider source population of equal mass massive BBHs with total mass $10^6~M_\odot$, rather than the stellar mass BBHs used for ground-based detectors. Since the physical interpretation is the same as for the ground-based detectors, we consider only the uniform sphere profile as a representative extended mass lens and its distinction from a point mass lens.

Fig.~\ref{fig:extlens_prob_space} shows the resulting lensing probability in the $(M_{\rm Lens},R)$ plane. As discussed in Sec.~\ref{sec:prospectextended}, the largest probabilities occur approximately where
\begin{equation}
    x_F\sim x_E\sim R~.
\end{equation}
For the characteristic frequencies of LISA and Taiji, $f\sim O(10^{-3})~{\rm Hz}$, and effective distances of order $d_{\rm eff}\sim1~{\rm Gpc}$, this corresponds approximately to
\begin{equation}
    R\sim O(10)~{\rm pc},
    \qquad
    M_{\rm Lens}\sim O(10^7)~M_\odot~,
\end{equation}
consistent with the locations of the probability peaks in Fig.~\ref{fig:extlens_prob_space}.
Near the peak, the isolated lens calculation gives probabilities approaching unity for both LISA and Taiji. As discussed in Sec.~\ref{sec:statistics}, this corresponds to optical depths for which the isolated-lens approximation becomes inaccurate. The probability in this regime should therefore be interpreted as indicating a very high lensing incidence rather than as a precise single lens prediction.

An absolute source rate is required to convert these probabilities into projected abundance limits. Predictions for massive BBH mergers depend on the black hole seeding scenario, including light and heavy seeds, and on the subsequent binary evolution and merger delays~\cite{Klein:2015hvg,Bonetti:2018tpf}. Rather than model the resulting mass and redshift distributions, we use the equal-mass $10^6~M_\odot$ sources specified above and adopt a redshift-independent reference rate $ 
    R_0=0.01~{\rm Gpc}^{-3}~{\rm yr}^{-1}$ .
This corresponds to approximately $\mathcal{V}R_0\sim8$ mergers per observer year integrated up to $z=10$. As one comparison, this is similar to the total merger rate predicted by the heavy-seed model with merger delays in Ref.~\cite{Klein:2015hvg}. We use this value as a benchmark normalization, without adopting a specific population model's mass and redshift distributions.

\begin{figure}[t]
    \centering
    \includegraphics[width=0.8\linewidth]{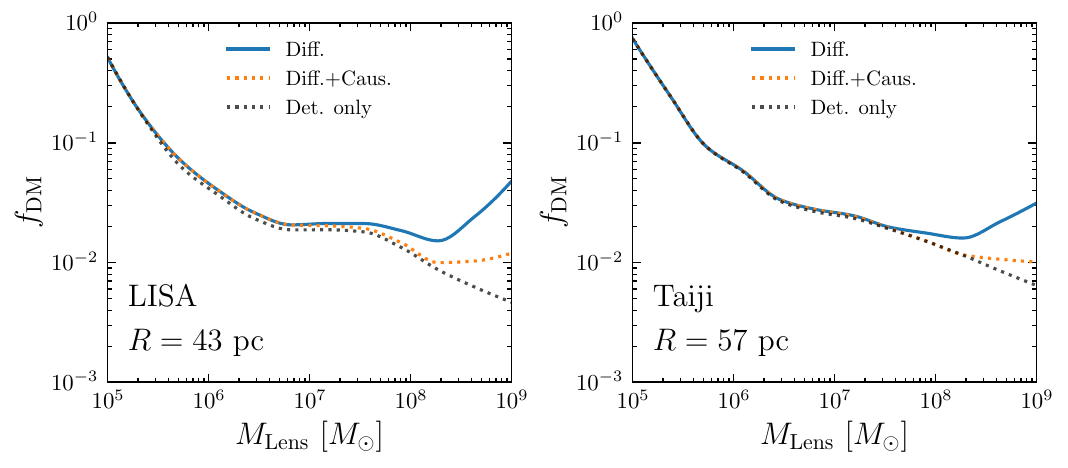}
    \caption{
    Projected $90\%$ upper limits on $f_{\rm DM}$ for uniform sphere lenses from $5$-yr observations with LISA and Taiji. We assume equal-mass massive BBHs with total mass $10^6~M_\odot$ and a fiducial comoving merger rate $R_0=0.01~{\rm Gpc}^{-3}~{\rm yr}^{-1}$. For each detector, $R$ is fixed to the value corresponding to the maximum lensing probability in the $(M_{\rm Lens},R)$ plane. The curve conventions follow Fig.~\ref{fig:fDMCEETLIGO}.    }
    \label{fig:fDMspace}
\end{figure}

Fig.~\ref{fig:fDMspace} shows the resulting projected $90\%$ upper limit sensitivity for a $5$ yr observation period. For each detector, $R$ is fixed to the value corresponding to the maximum lensing probability in Fig.~\ref{fig:extlens_prob_space}. As in the ground-based analysis, Fig.~\ref{fig:fDMspace} shows the sensitivity of GW lensing alone, without incorporating additional constraints such as from the microlensing and gas heating discussed above.

Despite the large per event lensing probability, the projected sensitivity limits are weaker than those obtained for ground-based detectors in their respective optimal parameter regions. The distinguishability  projections reach approximately $f_{\rm DM}\sim O(10^{-2})$. Including multiple image configurations improves the reach, while the detection only projection can extend to $f_{\rm DM}\sim O(10^{-3})$ in the highest mass range shown. For the assumed source population, the main limitation is therefore the small number of available massive BBH merger signal sources rather than the detectability of an individual lensed event.

\section{Application to Halo-Dressed Primordial Black Holes}
\label{sec:prospectdress}

We now specialize the power law halo-dressed point mass model introduced in Sec.~\ref{sec:diffdress} to PBHs surrounded by DM halos. PBHs can act as cosmological seeds, accumulating surrounding
particle DM through gravitational capture to form
halo-dressed PBHs. Dressed PBHs enable distinct gravitational probes of mixed DM scenarios
in which macroscopic PBHs coexist with microscopic particle DM
accumulated in surrounding halos. These systems can be studied with lensing, such as cosmological lensing of fast
radio bursts  considering
the halo-induced enhancement of the lensing
probability~\cite{Oguri:2022fir}. However, diffractive lensing of chirping GWs can identify
the surrounding halo and distinguish such systems from
bare black holes using just a single sufficiently well measured
lensed event~\cite{GilChoi:2023ahp}. Its frequency-dependent distortions can
distinguish a dressed PBH from a bare BH even when their
geometric optics observables are matched, providing an
event-by-event test of the coexistence of PBHs and particle DM. 

The halo profile can depend on its cosmological formation history and DM properties\footnote{For particle DM with efficient annihilation into observable
final states, enhanced annihilation in dense PBH halos can
produce additional signatures and strongly constrain such
coexistence scenarios~\cite{Lacki:2010zf,Adamek:2019gns}.
GW lensing enables efficiently probing scenarios even when such signals are negligible.}. For halos around PBHs, the initial DM phase-space distribution and kinetic decoupling history can lead to different radial density slopes~\cite{Boudaud:2021irr}. More generally, the growth history of a central black hole can affect the surrounding DM distribution~\cite{Ullio:2001fb}, while DM self-interactions can alter the cusp slope through collisional relaxation~\cite{Shapiro:2014oha,Feng:2021qkj}.

Motivated by this dependence, we explore how the halo slope affects the lensing signal using an idealized, spherically symmetric power law,
\begin{equation}
    \rho_h(r)
    =
    \rho_0
    \left(\frac{r}{R_h}\right)^{\gamma-1},
\end{equation}
which gives $\kappa_h(x)\propto x^\gamma$ after projection onto the lens plane. We consider the range
\begin{equation}
    -2<\gamma\leq-1~.
\end{equation}
The lower bound ensures a finite halo mass near the center without an inner cutoff, while the upper bound restricts the model to two geometric images for a nonzero source impact parameter. Shallower profiles can produce four images and require a comparison beyond the two image matching used here. We use the power law without an inner or outer cutoff. Further details of the model and the image multiplicity are given in Appendix~\ref{app:dress}.

The case $\gamma=-1.25$ gives $\rho_h(r)\propto r^{-9/4}$, a scaling obtained in early cosmological infall calculations~\cite{Gott:1975,Bertschinger:1985pd} and supported by later analytic calculations and simulations of cold DM accretion around PBHs~\cite{Adamek:2019gns}. We adopt the benchmark normalization $\rho_0$ and scale $R_h$ used in Ref.~\cite{GilChoi:2023ahp}, motivated by the halo growth estimates of Ref.~\cite{Mack:2006gz}. For the other slopes, we keep $\rho_0$ and $R_h$ fixed at a given $M_{\rm PBH}$ in order to isolate the dependence on $\gamma$. The resulting models therefore provide a controlled extension of the benchmark profile rather than a self-consistent halo growth model for arbitrary $\gamma$.

Following Sec.~\ref{sec:diffdress}, we compare each halo-dressed lens with a matched bare BH lens that reproduces the same geometric optics observables. This allows the additional diffraction produced by the halo to be isolated when distinction between the two models is required. 
Here, $f_{\rm PBH}$ denotes the fraction of the cosmological
DM density in the central PBHs, excluding their surrounding
halos. The comoving lens number density is therefore
$n_l=f_{\rm PBH}\rho_{\rm DM,0}/M_{\rm PBH}$.
The abundance calculation follows the statistical framework of Sec.~\ref{sec:statistics}. For all lens configurations, we evaluate $F(f)$ by direct numerical integration of Eq.~\eqref{eq:F(f)1d}, while using the diffraction analysis of Secs.~\ref{sec:diffprofile} and \ref{sec:diffdress} to interpret the results. Unlike the extended mass analysis of Sec.~\ref{sec:prospectextended}, no single image condition is imposed. The halo dressed-BH configurations considered here lie in the two image regime.

\subsection{Detection sensitivity to dark matter abundance}

We first consider sensitivity using only the detection criterion, for which a lensing event is counted whenever the lensed waveform is distinguishable from the unlensed signal. No distinction from the matched bare BH lens is required. These results therefore give the most optimistic projected abundance limits for the dressed PBH population and extend the $\gamma=-1.25$ calculation of Ref.~\cite{GilChoi:2023ahp} to general power law slopes.
\begin{figure}[t]
    \centering
    \includegraphics[width=0.99\linewidth]{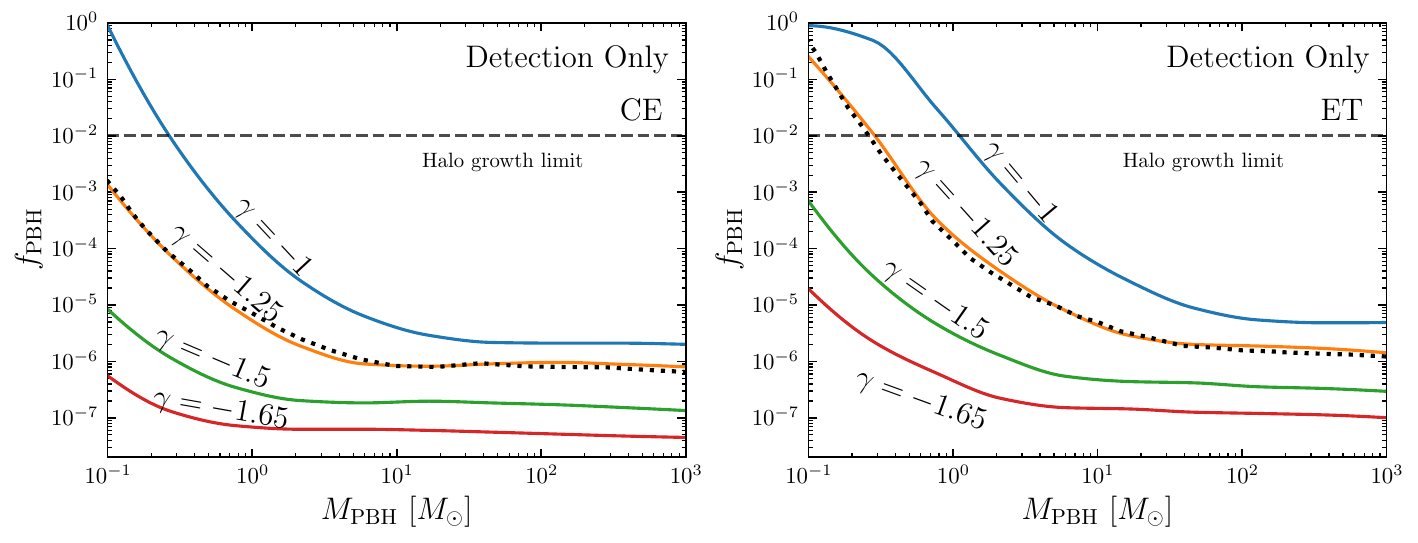}
    \includegraphics[width=0.99\linewidth]{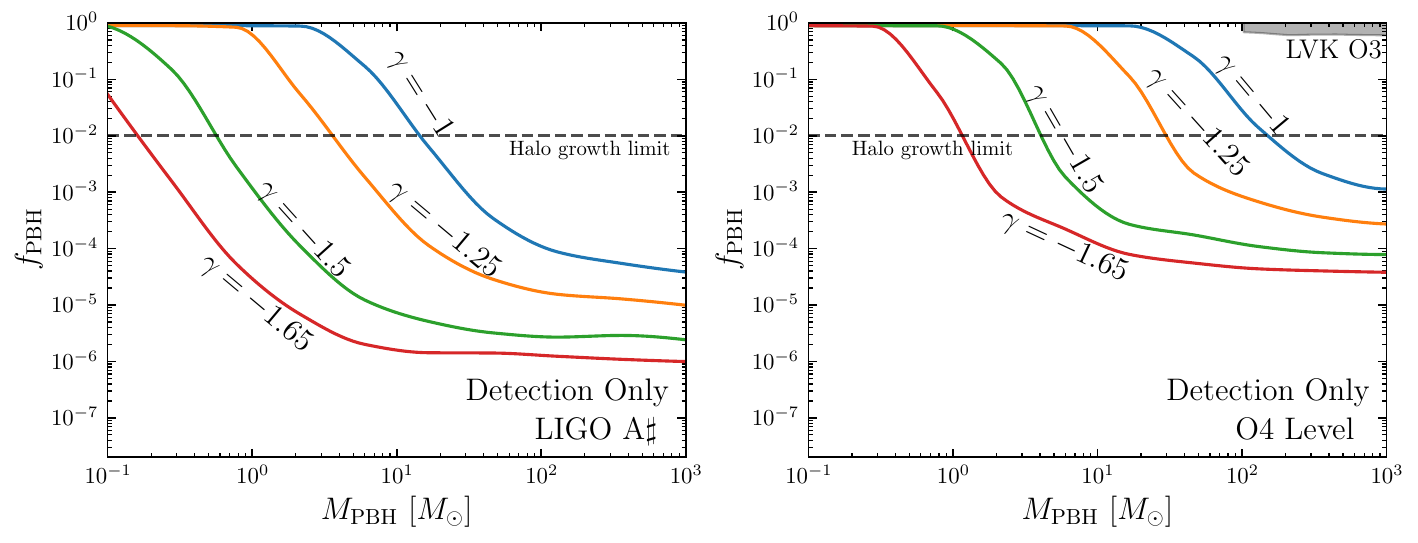}
    \caption{
    Projected $90\%$ upper limits on the halo dressed-PBH abundance $f_{\rm PBH}$ from $5$-yr observations using the detection-only criterion, shown as functions of $M_{\rm PBH}$ for $\gamma=-1$, $-1.25$, $-1.5$, and $-1.65$. The upper panel shows CE and ET, while the lower panel shows LIGO $A^\sharp$ and the O4-level configuration. For $\gamma=-1.25$, the dotted black curve shows the result of Ref.~\cite{GilChoi:2023ahp}. The dashed horizontal line marks the halo growth limit described in the text. The gray region in the O4 panel shows the constraint from bare PBH lensing with LIGO-Virgo O3a data~\cite{LIGOScientific:2021izm}. The O4-level curves are illustrative $5$ yr projections for a hypothetical single detector with the averaged Hanford-Livingston O4 noise curve.
    }
    \label{fig:fDMdressDet}
\end{figure}

Other observations also constrain dressed PBHs, including stellar microlensing~\cite{Cai:2022kbp}, accretion effects on the cosmic microwave background~\cite{Serpico:2020ehh}, and interstellar gas heating~\cite{Kim:2025gck}. Their quantitative comparison with our forecasts requires consistent halo profiles and population assumptions, together with the relevant gas and accretion models for heating and CMB constraints. We present the GW-lensing forecasts separately rather than overlay these bounds without matching their underlying assumptions.

 Fig.~\ref{fig:fDMdressDet} shows the projected $90\%$ upper limits on $f_{\rm PBH}$ for CE, ET, LIGO $A^\sharp$, and the O4-level configuration over the mass range $10^{-1}~M_\odot\lesssim M_{\rm PBH}\lesssim10^3~M_\odot$. The dashed horizontal line marks the halo growth limit as considered in the halo prescription of Ref.~\cite{GilChoi:2023ahp}. This model assumes a halo mass of approximately $M_h\simeq100M_{\rm PBH}$. Extending this relation to the entire PBH population would require more DM in the halos than is available when $f_{\rm PBH}\gtrsim10^{-2}$. Results above this limit therefore lie in a regime where additional considerations regarding halo evolution are required, which is beyond the scope of our work.

The projected limits become stronger as $\gamma$ decreases. At fixed $M_{\rm PBH}$, $\rho_0$, and $R_h$, a more negative $\gamma$ gives a more centrally concentrated halo and enhances its lensing effect. CE and ET provide the strongest projected reach. For $\gamma=-1.65$, the limits reach $f_{\rm PBH}\sim O(10^{-7})$ in the most favorable mass range. For the benchmark slope $\gamma=-1.25$, the result agrees with the previous calculation of Ref.~\cite{GilChoi:2023ahp}, providing a consistency check of the lensing-probability calculation used here.

The projected limits are weaker for LIGO $A^\sharp$ and the O4-level configuration, while retaining the same dependence on $\gamma$. The gray region in the O4 panel shows the existing bare PBH lensing constraint from LIGO-Virgo O3a data~\cite{LIGOScientific:2021izm} for reference. This comparison is not direct, since our O4-level result assumes a hypothetical single detector operating for $5$ yr and additionally includes the lensing enhancement from the surrounding halo.

The detection-only criterion does not determine whether an observed lensing signal contains measurable information about the halo. We therefore next require the halo-dressed lens to be distinguishable from its matched bare BH counterpart.

\subsection{Sensitivity to dark matter abundance with distinguishability}

\begin{figure}[t]
    \centering
    \includegraphics[width=0.99\linewidth]{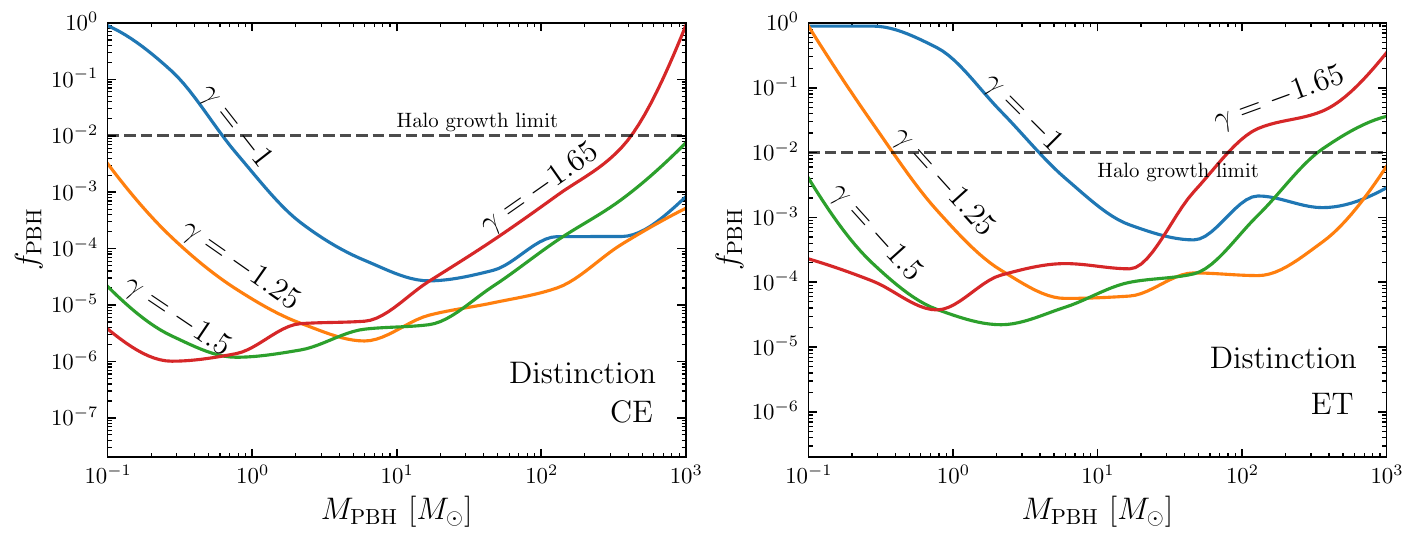}
    \includegraphics[width=0.99\linewidth]{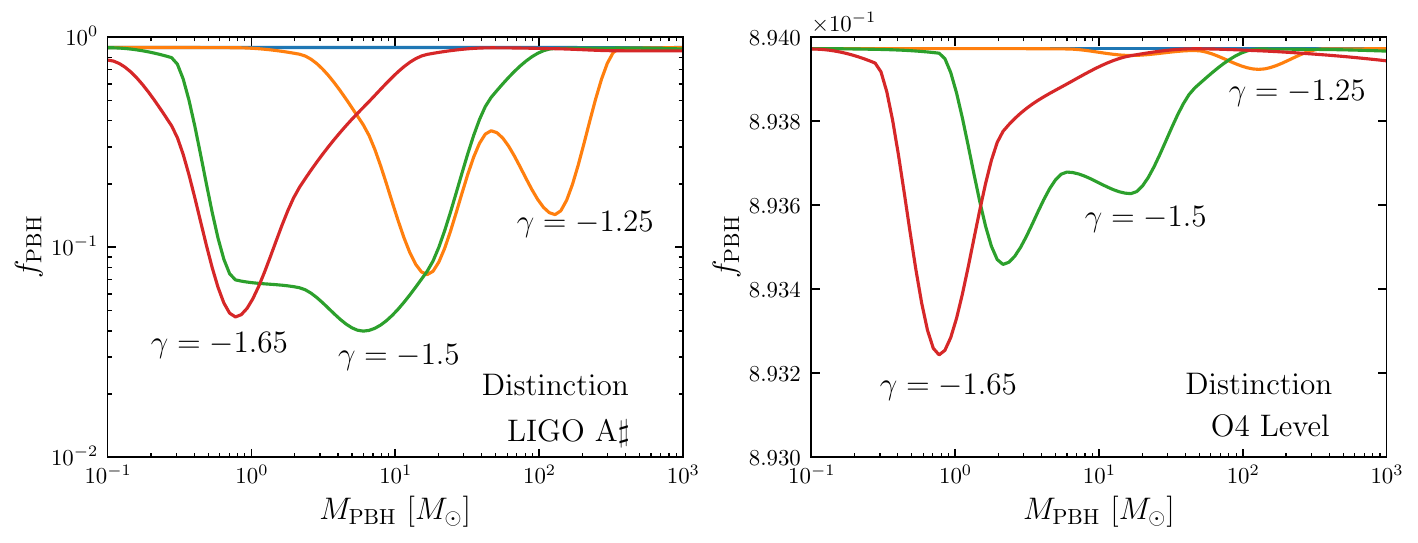}
    \caption{
    Same as Fig.~\ref{fig:fDMdressDet}, but requiring the halo dressed-PBH signal to be distinguishable from the matched bare BH lens in addition to being detectable relative to the unlensed signal. The dashed horizontal line marks the halo growth limit. The benchmark halo growth horizontal boundary is shown in the upper panels and lies below the plotted range of the lower panels.
    }
    \label{fig:fDMdressDist}
\end{figure}

\begin{figure}[t]
    \centering
    \includegraphics[width=0.6\linewidth]{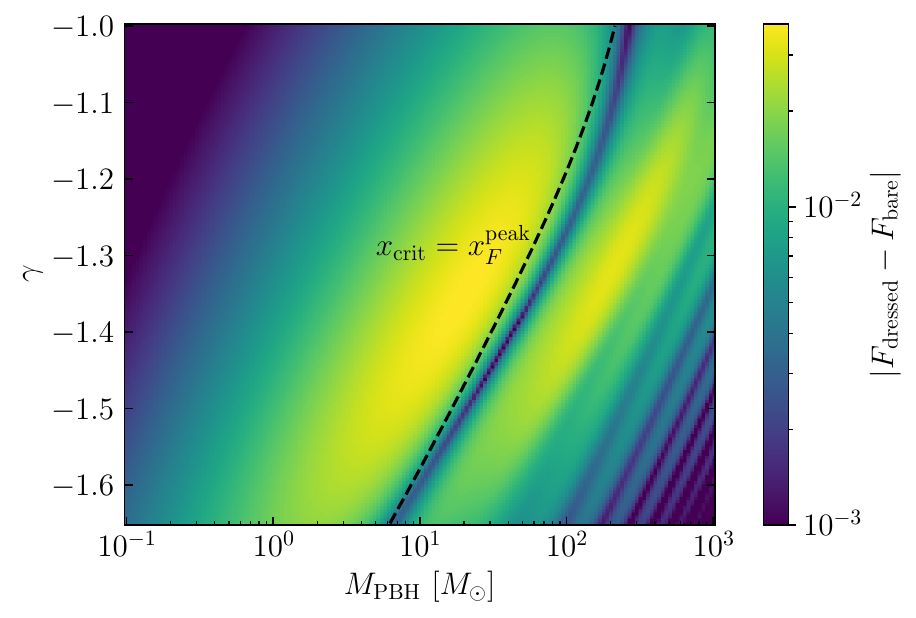}
    \caption{
    Magnitude of the amplification factor difference $|F_{\rm dress}-F_{\rm bare}|$ in the $(M_{\rm PBH},\gamma)$ plane, evaluated at $f^{\rm peak}\simeq16.6~{\rm Hz}$. This frequency maximizes $|h_0(f)|^2/S_n(f)$ for the inspiral waveform of an equal mass BBH with source frame total mass $20~M_\odot$ at $z_s=0.3$, using the CE noise curve. We take $z_l=0.17$ and $x_s=2x_F^{\rm peak}\simeq0.28~{\rm pc}$, where $x_F^{\rm peak}$ is the Fresnel scale at this frequency.
    }
    \label{fig:dressDist}
\end{figure}

We next impose the distinguishability criterion of Sec.~\ref{sec:statistics}. A lensing event is counted only when the halo dressed-BH waveform is distinguishable both from the unlensed signal and from the matched bare BH lens. The corresponding null result therefore means that no event is identified as arising from the target halo dressed-PBH population. Unlike the estimate in Ref.~\cite{GilChoi:2023ahp}, where distinguishability was approximated from the presence of visible diffraction features within the detector band, we evaluate the difference between the halo-dressed and bare BH signals over the full frequency range using the noise-weighted likelihood.

Requiring distinction from a bare BH lens does not remove existing abundance constraints on the same halo dressed-PBH population. The   forecasts with distinguishability instead quantify how sensitively GW lensing can identify that population through the diffraction produced by its surrounding halos.

 Fig.~\ref{fig:fDMdressDist} shows the resulting projected $90\%$ upper limit sensitivity on $f_{\rm PBH}$ for CE, ET, LIGO $A^\sharp$, and the O4-level configuration, using the same halo slopes and PBH-mass range as in Fig.~\ref{fig:fDMdressDet}. The sensitivity with distinguishability is substantially weaker than the results considering just detection. In the most favorable regions, CE reaches $f_{\rm PBH}\sim O(10^{-6})$, ET reaches $O(10^{-5})$, and LIGO $A^\sharp$ reaches approximately $O(10^{-1.5})$. For the O4-level configuration, the projected  upper limits
remain close to the prior  90\% quantile
$f_{\rm PBH}=0.9$ and therefore provide essentially no DM
abundance sensitivity. Thus, a lensing signal can remain detectable even when the halo-induced difference from a bare BH lens is too small to identify the halo-dressed nature of the lens.

The distinction reach also depends more strongly on detector sensitivity than the detection-only reach. For a fixed $\gamma$, the minimum projected limits differ by orders of magnitude among the detectors. This stronger detector dependence arises because distinction is determined by the residual difference between two already lensed waveforms. It therefore depends sensitively on both the SNR and the frequency range over which the halo-induced diffraction can be measured.

The characteristic mass dependence can be understood from the transition between the diffraction and geometric optics regimes. For a given source waveform and detector noise curve, we define $f^{\rm peak}$ as the frequency that maximizes $|h_0(f)|^2/S_n(f)$ and $x_F^{\rm peak}$ as the Fresnel scale at that frequency. Let $x_{\rm crit}$ denote the radius of the critical curve of the halo-dressed lens. The strongest distinction typically occurs near
\begin{equation}
    x_{\rm crit}\sim x_F^{\rm peak}~.
\end{equation}
Fig.~\ref{fig:dressDist} illustrates this behavior for an equal-mass BBH with source-frame total mass $20~M_\odot$ at $z_s=0.3$, observed with CE. For this source and detector configuration, $f^{\rm peak}\simeq16.6~{\rm Hz}$. We take $z_l=0.17$ and $x_s=2x_F^{\rm peak}$.

For $x_{\rm crit}\ll x_F^{\rm peak}$, the lens remains in the diffraction regime at the frequencies carrying most of the signal weight, and increasing $M_{\rm PBH}$ strengthens the difference between the halo-dressed and bare BH signals. When $x_{\rm crit}\gg x_F^{\rm peak}$, the relevant frequencies approach the geometric optics regime, where the two models have been matched and their signals become increasingly similar. The residual difference also develops an oscillatory dependence on $M_{\rm PBH}$ in this regime. The projected sensitivity limits therefore attain their strongest reach near $x_{\rm crit}\sim x_F^{\rm peak}$ and develop the wavy structure visible around their minima.

\section{Application to Ultralight Boson Superradiance Clouds}
\label{sec:cloud}

Black hole superradiance clouds can form when an ultralight
bosonic field around a rotating black hole develops a
superradiant instability
\cite{Detweiler:1980uk,Arvanitaki:2010sy,Arvanitaki:2014wva,Brito:2015oca,Baryakhtar:2017ngi,Baumann:2018vus,Baumann:2019eav,Cardoso:2018tly,Brito:2020lup,Cao:2023fyv}.
The relevant regime is characterized by a boson Compton
wavelength $\lambda_c=1/\mu$ that is comparable to or larger than
the black hole Schwarzschild radius $r_s=2M_{\rm BH}$,
where $\mu$ is the boson mass.
For a mode with real frequency $\omega_R$ and azimuthal
number $m$, superradiant growth occurs when
\begin{equation}
    \frac{\omega_R}{m}<\Omega_H~,
\end{equation}
where $\Omega_H$ is the angular velocity of the black hole horizon. The instability extracts energy and angular momentum from the black hole and can build up a macroscopic bosonic cloud. Such superradiant effects have been analyzed through GW emission, binary dynamics, and black hole spin measurements \cite{Arvanitaki:2010sy,Baryakhtar:2017ngi,Baumann:2018vus,Cardoso:2018tly,Cao:2023fyv,Ding:2025nxe}. Previous investigations have primarily probed the presence
and dynamical effects of these clouds. Here we investigate
whether their spatial mass distribution can be probed
more directly through GW lensing.

Throughout this section, $M_{\rm BH}$ denotes the black hole mass after cloud formation. In the gravitational-atom picture, the cloud has the characteristic length scale
\begin{equation}
    r_b
    = \frac{G M_{\rm BH}}{c^2\alpha^2}
    \simeq
    8.5\times10^{-9}\,{\rm pc}
    \left(\frac{M_{\rm BH}}{10^3\,M_\odot}\right)^{-1}
    \left(\frac{\mu c^2}{10^{-14}\,{\rm eV}}\right)^{-2},
\end{equation}
where
\begin{equation}
    \alpha
    =
    \frac{G M_{\rm BH}\mu}{\hbar c}
    \simeq
    0.075 \left(\frac{M_{\rm BH}}{10^3\,M_\odot}\right)
    \left(\frac{\mu c^2}{10^{-14}\,{\rm eV}}\right)
\end{equation}
is the gravitational fine structure constant. Here we have restored $G$ and $c$ units explicitly. Superradiant growth is typically most efficient for $\alpha=O(0.1)$, for which the cloud size is much larger than the black hole horizon.

Following Ref.~\cite{Cao:2023fyv}, we consider the cloud density as
\begin{equation}
    \rho(\bm r)
    =
    g(r,\theta)\frac{M_c}{r_b^3}
    =
    9.1\times10^{26}\,
    M_\odot\,{\rm pc}^{-3}\, g(r,\theta)
    \left(\frac{\beta}{0.1}\right)
    \left(\frac{\alpha}{0.1}\right)^6
    \left(\frac{M_{\rm BH}}{10^3M_\odot}\right)^{-2},
\end{equation}
where $M_c$ is the cloud mass and $\beta= M_c/M_{\rm BH}$.  We set $\beta=\alpha$ as an optimistic benchmark for the cloud mass\footnote{Neglecting the fractional change in the black hole mass, the saturation estimate is $\beta_{\rm sat}\sim(\alpha/m)\Delta a$, where $\Delta a$ is the decrease in the dimensionless black hole spin~\cite{Arvanitaki:2014wva}. For simplicity, to estimate benchmarks we consider fixed  cloud normalizations without calculating its saturation value.}. The symmetry axis of the cloud is chosen as the $z$ axis. We consider the scalar $|211\rangle$ and $|322\rangle$, vector $|1011\rangle$, and tensor $|1022\rangle$ states.  We use the density profiles given in Ref.~\cite{Cao:2023fyv},

\begin{equation}
\begin{split}
    |211\rangle :\quad
    &g=\frac{1}{64\pi}e^{-u}u^2\sin^2\theta~,\\
    |322\rangle :\quad
    &g=\frac{1}{26244\pi}e^{-2u/3}u^4\sin^4\theta~,\\
    |1011\rangle,\ |1022\rangle :\quad
    &g=\frac{1}{\pi}e^{-2u}~,
\end{split}
\end{equation}
where $u=r/r_b$.

\begin{figure}[t]
    \centering
    \includegraphics[width=0.5\linewidth]{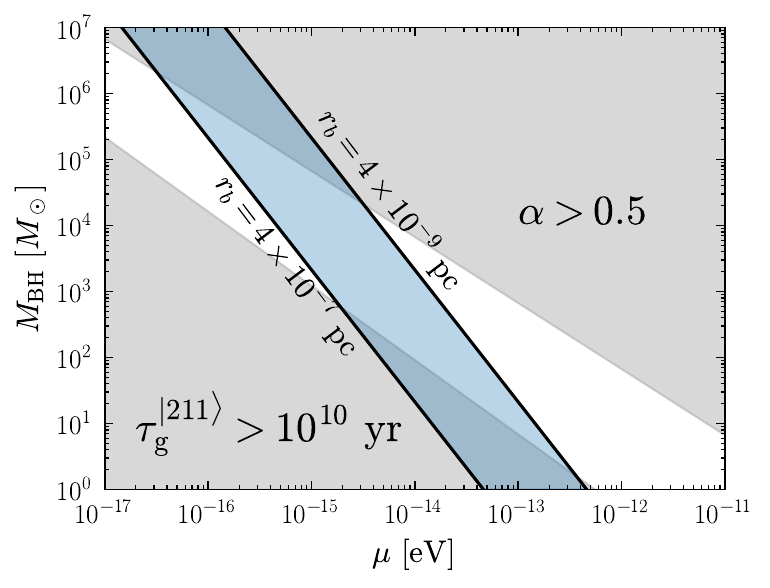}
    \caption{
    Region in the $(\mu,M_{\rm BH})$ plane where the characteristic superradiance  cloud size $r_b$ overlaps the accessible Fresnel scale. We assume $10^{-7}\,{\rm pc}<d_{ls}<10^{-3}\,{\rm pc}$, $f(1+z_l)=10\,{\rm Hz}$, and $z_l\simeq z_s=0.1$. The blue region gives the scale matching estimate for diffraction sensitivity.  The gray regions mark the adopted cuts $\alpha>0.5$ and $\tau_g>10^{10}\,{\rm yr}$ for the scalar $|211\rangle$ state, where $\tau_g$ is the cloud mass e-folding time estimated at $a=1$.
    }
    \label{fig:cloudrb}
\end{figure}

The scalar clouds are not axisymmetric after projection for a general orientation. For simplicity, we assume that the cloud symmetry axis is aligned with the observer's line of sight.  Projecting these density profiles along the line of sight gives the axisymmetric convergence
\begin{equation}
    \kappa_c(\bar x)
    =
    \frac{4d_{\rm eff}M_c}{r_b^2}
    \begin{cases}
        \displaystyle
        \frac{1}{32}\bar x^3K_1(\bar x),
        & |211\rangle~,\\[8pt]
        \displaystyle
        \frac{1}{13122}\bar x^5K_1(2\bar x/3),
        & |322\rangle~,\\[8pt]
        \displaystyle
        2\bar xK_1(2\bar x),
        & |1011\rangle,\ |1022\rangle~,
    \end{cases}
\end{equation}
where $\bar x=x/r_b$ and $K_1$ is the modified Bessel function of the second kind. The total lensing potential is obtained by adding the cloud contribution to that of the central black hole, which is modeled as a point mass lens.

The characteristic cloud size is much smaller than the Fresnel scale for an ordinary lens located at cosmological separation from the source. Cloud diffraction therefore requires the black hole lens to lie close to the GW source. For $d_l\simeq d_s$, one has $d_{\rm eff}\simeq d_{ls}$, and the Fresnel scale becomes
\begin{equation}
    x_F
    \simeq
    \sqrt{\frac{d_{ls}} {2\pi f(1+z_l)}}
    \simeq
    3.9\times10^{-7}\,{\rm pc}
    \sqrt{ \frac{d_{ls}}{10^{-3}\,{\rm pc}} \frac{10\,{\rm Hz}}{f(1+z_l)} }~.
\end{equation}
For $10^{-7}\,{\rm pc}<d_{ls}<10^{-3}\,{\rm pc}$, the corresponding Fresnel scales span approximately $4\times10^{-9}$-$4\times10^{-7}\,{\rm pc}$ at $f(1+z_l)=10\,{\rm Hz}$. The overlap between this range and the characteristic cloud size $r_b$ therefore identifies the region in which the cloud structure can be probed efficiently through diffraction.

Fig.~\ref{fig:cloudrb} shows the resulting scale matching region in the $(\mu,M_{\rm BH})$ plane. This comparison does not by itself impose the detection and distinction thresholds, but it identifies the characteristic band in which the cloud can produce an appreciable diffraction signature. We next test this expectation using the full lensing signal.

The analysis follows the same matching strategy as for the power law halo-dressed BHs in Sec.~\ref{sec:prospectdress}. For each cloud plus BH lens, we construct a matched bare BH lens that reproduces the geometric optics observables. The amplification factor $F(f)$ is obtained by direct numerical integration of Eq.~\eqref{eq:F(f)1d}. We define
\begin{equation}
    \ln\Lambda_{\rm det} = \langle\ln\Lambda_{L0}\rangle,
    \qquad
    \ln\Lambda_{\rm dist} = \langle\ln\Lambda_{LL'}\rangle,
\end{equation}
where the first quantifies detection relative to the unlensed waveform and the second quantifies distinction from the matched bare BH lens. The likelihoods are evaluated using the CE noise curve and the statistical framework of Sec.~\ref{sec:statistics}. For the source configuration considered below, the condition $\rho_L>8$ is satisfied throughout the parameter range of interest.

\begin{figure}[t]
    \centering
    \includegraphics[width=0.5\linewidth]{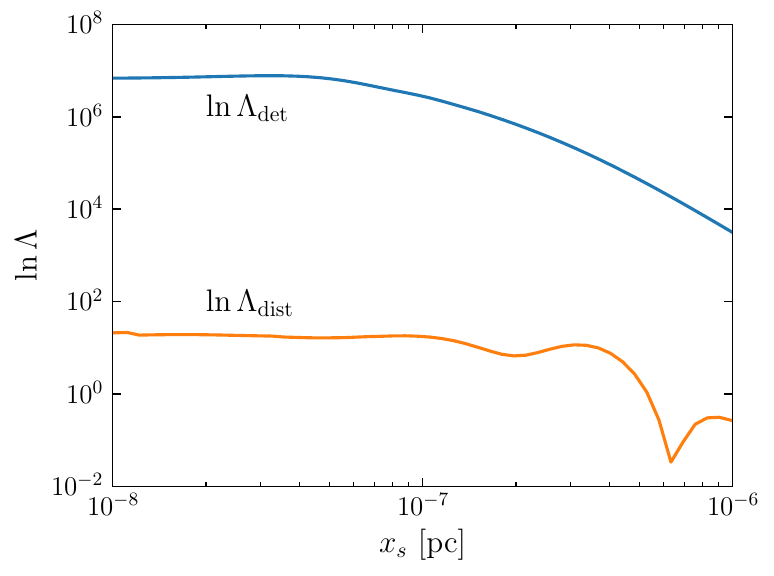}
    \caption{
     Detection and distinction log likelihoods, $\ln\Lambda_{\rm det}$ and $\ln\Lambda_{\rm dist}$, as functions of the source impact parameter $x_s$ for a representative cloud plus BH lens. The cloud occupies the scalar $|211\rangle$ state. We use the CE noise curve, $M_{\rm BH}=10^3\,M_\odot$, $\mu=10^{-14}\,{\rm eV}$, and $d_{ls}=10^{-4}\,{\rm pc}$. The source is an equal mass BBH with total mass $60\,M_\odot$ at $z_s=0.1$. Both likelihoods approach finite values toward small $x_s$.
    }
    \label{fig:lnLxs}
\end{figure}

As illustrated in Fig.~\ref{fig:lnLxs}, both $\ln\Lambda_{\rm det}$ and $\ln\Lambda_{\rm dist}$ saturate toward small $x_s$. We use these saturated values to determine whether a given cloud configuration can in principle satisfy the detection and distinction criteria.  For each $(\mu,M_{\rm BH})$, we evaluate five logarithmically spaced values of $d_{ls}$ in the range
\begin{equation}
    10^{-7}\,{\rm pc}\leq d_{ls}\leq10^{-3}\,{\rm pc}
\end{equation}
and identify the parameter point as accessible when,  for at least one of these separations, the saturated likelihoods satisfy
\begin{equation}
    \ln\Lambda_{\rm det}>3,
    \qquad
    \ln\Lambda_{\rm dist}>3~.
\end{equation}
This procedure tests whether the lensing probability can be non-zero, rather than computing the probability after marginalizing over the source and lens populations.

\begin{figure}[t]
    \centering
    \includegraphics[width=0.495\linewidth]{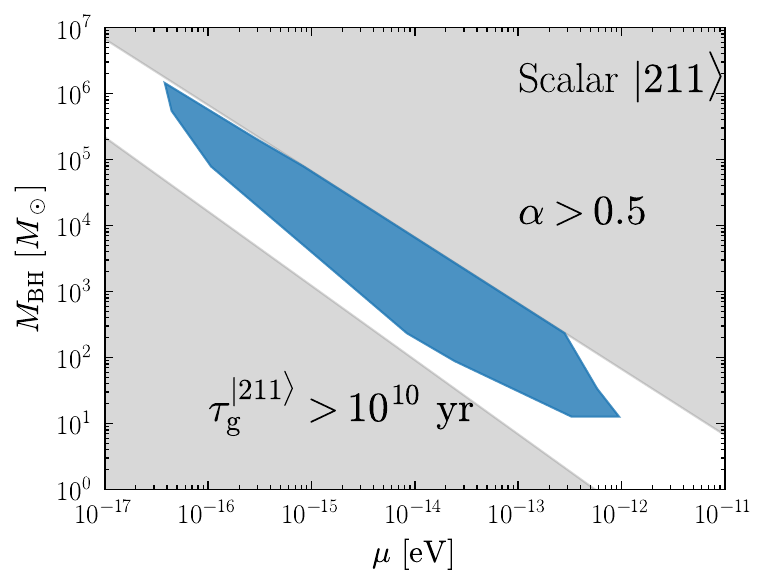}
    \includegraphics[width=0.495\linewidth]{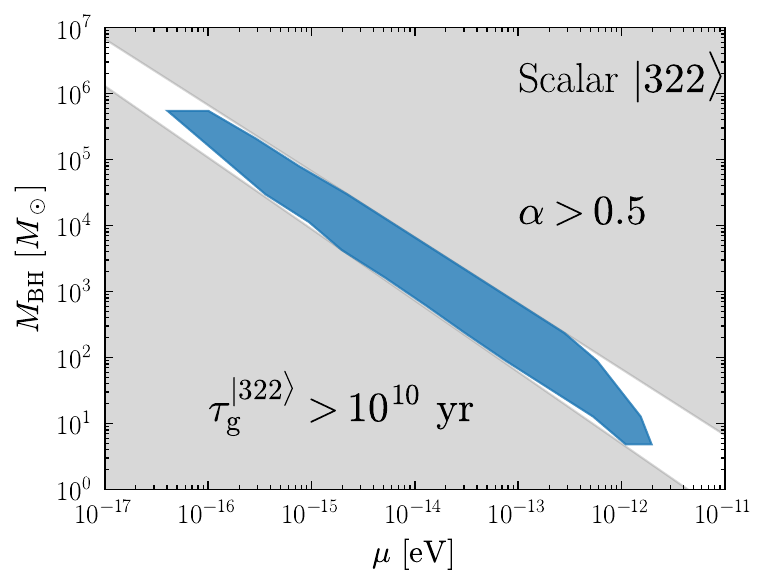}
    \includegraphics[width=0.495\linewidth]{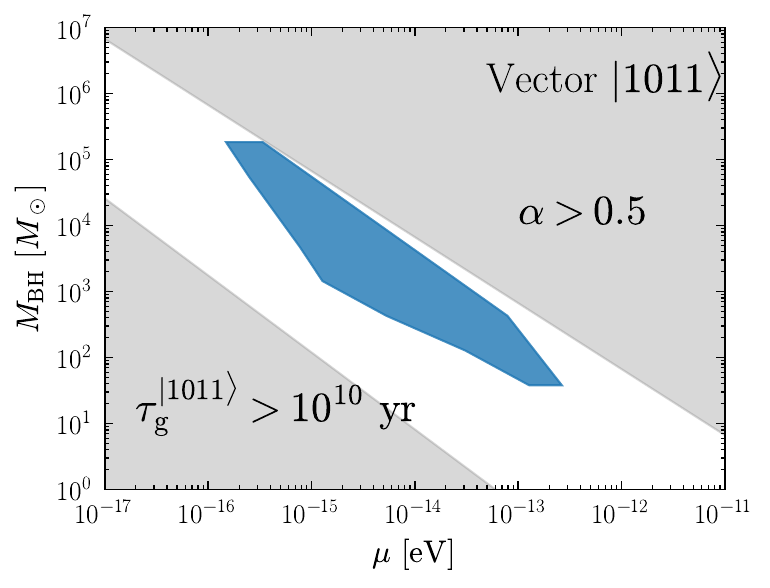}
    \includegraphics[width=0.495\linewidth]{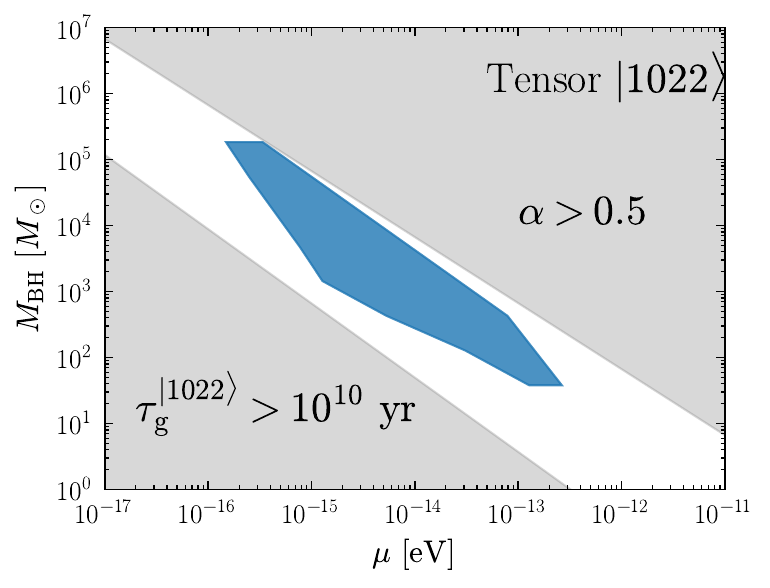}
    \caption{
    Regions in the $(\mu,M_{\rm BH})$ plane where lensing by a black hole surrounded by a superradiance cloud can be both detected and distinguished from a matched bare BH lens with CE. The panels correspond to the scalar $|211\rangle$ and $|322\rangle$, vector $|1011\rangle$, and tensor $|1022\rangle$ states.  We assume an equal-mass BBH source with total mass $60\,M_\odot$ at $z_s=0.1$. We evaluate five logarithmically spaced separations in $10^{-7}\,{\rm pc}\leq d_{ls}\leq10^{-3}\,{\rm pc}$. The cloud mass is normalized by the optimistic benchmark $\beta=\alpha$. Blue regions satisfy $\ln\Lambda_{\rm det}>3$ and $\ln\Lambda_{\rm dist}>3$ for at least one allowed separation using the saturated small-$x_s$ likelihoods.  Gray regions lie outside the adopted range $\alpha<0.5$ and $\tau_g<10^{10}\,{\rm yr}$, where $\tau_g$ is the cloud-mass e-folding time estimated at $a=1$.
    }
    \label{fig:clouddist}
\end{figure}

 Fig.~\ref{fig:clouddist} shows the resulting regions for the four cloud states. We additionally require 
\begin{equation}
    \alpha<0.5,
    \qquad
    \tau_g<10^{10}\,{\rm yr}~.
\end{equation}
The first restriction is motivated by the superradiance bound for $m=1$ at maximal black hole spin and is applied as a common range for all four states. The second requires the cloud growth time to be shorter than a timescale comparable to the age of the Universe and serves as an approximate criterion for selecting the parameters considered here. This cut does not model the full growth to saturation or the subsequent depletion of the cloud.

We define $\tau_g$ as the e-folding time of the cloud mass, $\tau_g=(2\omega_I)^{-1}$, where $\omega_I$ is the imaginary part of the mode frequency. For the scalar $|211\rangle$~\cite{Detweiler:1980uk,Baryakhtar:2017ngi}, scalar $|322\rangle$~\cite{Baumann:2019eav}, vector $|1011\rangle$~\cite{Baryakhtar:2017ngi}, and tensor $|1022\rangle$~\cite{Brito:2020lup} states, the leading small-$\alpha$ expressions are
\begin{equation}
\begin{split}
    \tau_g^{|211\rangle}
    &=
    \frac{24}{a\alpha^8\mu},
    \\
    \tau_g^{|322\rangle}
    &=
    \frac{885735}{64(a+3a^3)\alpha^{12}\mu},
    \\
    \tau_g^{|1011\rangle}
    &=
    \frac{1}{4a\alpha^6\mu},
    \\
    \tau_g^{|1022\rangle}
    &=
    \frac{45}{64(a+3a^3)\alpha^8\mu},
\end{split}
\end{equation}

Here $a$ is the dimensionless black hole spin. We assume $a=1$ for an optimistic estimate of the growth rate. This value is used only to estimate the growth time, not as the black hole spin after cloud formation. The tensor expression retains the general spin dependence at leading order in $\alpha$. These expressions use natural units. Numerical times are obtained by restoring $\hbar/(\mu c^2)$ in place of $1/\mu$.

The accessible regions broadly follow the characteristic band predicted by the scale comparison in Fig.~\ref{fig:cloudrb}. For the scalar $|211\rangle$ and $|322\rangle$ states, the accessible boson masses extend approximately over $\mu\sim10^{-17}$-$10^{-12}\,{\rm eV}$. For the vector $|1011\rangle$ and tensor $|1022\rangle$ states, the corresponding range is approximately $\mu\sim10^{-16}$-$10^{-13}\,{\rm eV}$. The black hole masses fall mainly in the intermediate-mass range. These results show that diffraction can, in principle, distinguish the mass distribution of a superradiance cloud from that of a bare black hole over a restricted region of the $(\mu,M_{\rm BH})$ plane.

The regions of interest found in Fig.~\ref{fig:clouddist} should not be interpreted as strict lensing probabilities. A probability forecast would additionally require analyzing distributions of source impact parameters, source and lens populations, black hole masses and redshifts, and source-lens separations. For the scalar states, a general cloud orientation could also break the axisymmetry assumed here and can modify the lensing signal. Our simplified analysis also assumes a static lens and the scalar wave, thin lens and small angle approximations as above. Application to a physical cloud system requires the cloud extent of several $r_b$ to be small compared with the source-lens separation, and the relevant propagation paths to remain in the weak field regime.  The static, thin lens treatment adopted here is  most reliable at the larger separations. We have also neglected the potential tidal perturbation of the cloud by the source binary, which can induce transitions between cloud states~\cite{Baumann:2018vus,Baumann:2019eav}. We leave detailed analyses of these effects and of other configurations to future work.

Under the optimistic assumptions adopted here, our results indicate that GW diffraction can distinguish a superradiant cloud BH from a matched bare BH, providing sensitivity to the cloud’s spatial mass distribution. This establishes a potential route to probing superradiant cloud structure
through GW lensing, complementary to other searches including 
GW emission, binary dynamics and black hole spins.

\section{Conclusion}
\label{sec:conclusion}

GW diffraction carries information about the internal mass
distribution of a lens beyond its total mass. We have made
this connection explicit through a gravitational lensing
form factor. Within the linear potential and stationary-phase
approximations for $x_s\gg x_F$, the central diffractive image
contribution probes the Fourier transform of the projected
density at $k_*=x_s/x_F^2$. For a finite-mass lens, this
contribution is the point mass Born contribution multiplied
by a normalized mass form factor, paralleling the role of
form factors in elastic scattering from atomic nuclei.
Since $k_*$ grows with frequency, a chirping GW signal
samples the lens structure across spatial scales.
Together with the aperture-mean-convergence description
relevant for $x_F\gtrsim x_s$, this provides a physical
framework for understanding how the projected mass
distribution enters the observed waveform.

Using full numerical wave optics calculations we demonstrate that this structural information can be observable in the benchmark configurations considered. Extended DM lenses are most readily distinguished from point masses of the same mass near $x_F\sim x_E\sim R$, corresponding to characteristic masses near $10^2~M_\odot$ and radii near $0.1~\mathrm{pc}$ for next-generation ground-based detectors, and $10^7~M_\odot$ and $10~\mathrm{pc}$ for LISA and Taiji. Under the adopted population assumptions and reference comparisons, Cosmic Explorer and Einstein Telescope could probe DM fractions of $10^{-5}-10^{-4}$. Different extended profiles can remain distinguishable at the same total mass, characteristic size, and source impact parameter. 

For PBHs engulfed in power law DM halos, matching the geometric optics observables to a bare black hole isolates a halo contribution concentrated toward lower frequencies. Identifying the surrounding halo offers a gravitational
probe of mixed DM scenarios in which macroscopic PBHs
coexist with microscopic particle DM. The most favorable projected five-year limits reach approximately $f_{\rm PBH}\sim10^{-6}$ for Cosmic Explorer and $10^{-5}$ for Einstein Telescope. These limits are weaker than detection-only limits as detecting a compact lens and identifying its surrounding halo are distinct observational tasks. 

Under the optimistic   parameters adopted here,
superradiance clouds around black holes sufficiently close
to the GW source can produce detectable diffraction
distinguishable from matched bare black holes.  The corresponding boson-mass ranges are approximately $10^{-17}$-$10^{-12}~\mathrm{eV}$ for the scalar states considered and $10^{-16}$-$10^{-13}~\mathrm{eV}$ for the vector and tensor states, providing a potential probe of ultralight particle physics. These are conditional detectability regions; event-rate predictions additionally require models of the source and cloud populations and their relative configurations.

Our results motivate a variety of further directions. The form factor interpretation developed here provides a basis for inferring spatial Fourier components of the projected density directly, complementing comparisons between specified lens models. Developing this approach calls for joint inference of source and lens parameters, extensions to non-axisymmetric configurations, and treatments beyond linear order in the lens potential. Potential applications extend beyond DM including to the structure of nuclear star clusters, stellar cusps and massive self-gravitating accretion disks around massive black holes, for which binaries in galactic nuclei provide natural sources. By connecting frequency-dependent waveform distortions to spatial mass structure, GW diffraction offers a gravitational probe of the internal structure of otherwise unresolved astrophysical objects.

\acknowledgments{H.G.C. is supported by the Institute for Basic Science (IBS) under the project code, IBS-R018-D3, and acknowledges financial assistance from the Max Planck Society Supporting Members. S.J. is supported by Grant Korea RS-2024-00342093.
This work was supported by the
World Premier International Research Center Initiative
(WPI), MEXT, Japan. V.T. acknowledges support in part by the MEXT SPReAD Grant No. 26304205.}

\appendix

\section{Lens models and potentials}
\label{app:potentials}

We summarize the lens models and potentials used in the numerical calculations. The lens potential satisfies the two dimensional Poisson equation
\begin{equation}\label{eq:poisson}
    \nabla^2_{\rm 2d}\psi(\bm{x})=2\kappa(\bm{x})~,
\end{equation}
where $\bm{x}$ denotes the physical transverse coordinate on the lens plane. The convergence is related to the projected surface mass density $\Sigma(\bm{x})$ by
\begin{equation}
    \kappa(\bm{x})=4\pi d_{\rm eff}\Sigma(\bm{x})~,
\end{equation}
where
\begin{equation}
    d_{\rm eff}=\frac{d_l d_{ls}}{d_s}
\end{equation}
and $d_l$, $d_s$, and $d_{ls}$ are the angular-diameter distances from the observer to the lens, from the observer to the source, and from the lens to the source, respectively. For an axisymmetric lens, $\psi(\bm{x})=\psi(x)$ with $x=|\bm{x}|$.

\subsection{Point mass lens}
\label{app:ptmass}

For a point mass $M$ located at the origin,
\begin{equation}
    \Sigma(\bm{x})=M\delta^{(2)}(\bm{x})~,
\end{equation}
and hence
\begin{equation}
    \kappa(\bm{x}) =\pi x_E^2\delta^{(2)}(\bm{x})~,
\end{equation}
where
\begin{equation}
    x_E=\sqrt{4d_{\rm eff}M}
\end{equation}
is the Einstein radius of the point mass lens and coincides with the radius of its critical curve. The corresponding lens potential is
\begin{equation}
    \psi(x) =x_E^2\ln\!\left(\frac{x}{x_E}\right)~.
\end{equation}

\subsection{Extended mass lenses}

We characterize each extended lens by its total mass $M$ and characteristic radius $R$. We define
\begin{equation}
    x_E=\sqrt{4d_{\rm eff}M}
\end{equation}
as a reference length scale. For an extended lens, $x_E$ does not in general coincide with the radius of a critical curve.

\subsubsection{Uniform disk}

For a uniform disk,
\begin{equation}
    \Sigma(x)
    =\frac{M}{\pi R^2}\Theta(R-x)~,
\end{equation}
and
\begin{equation}
    \kappa(x)
    =\frac{x_E^2}{R^2}\Theta(R-x)~,
\end{equation}
where $\Theta$ is the Heaviside step function. The lens potential is
\begin{equation}\label{eq:disk}
    \psi(x)=
    \begin{cases}
        \dfrac{1}{2}\dfrac{x_E^2}{R^2}x^2~, & x\leq R~,\\[10pt]
        x_E^2\ln\!\left(\dfrac{x}{R}\right)
        +\dfrac{1}{2}x_E^2~, & x>R~.
    \end{cases}
\end{equation}

\subsubsection{Uniform sphere}

The uniform sphere is defined by the three-dimensional density profile
\begin{equation}
    \rho(r)
    =\frac{3M}{4\pi R^3}\Theta(R-r)~,
\end{equation}
where $r$ is the radial coordinate from the lens center. Its projected surface mass density is
\begin{equation}
    \Sigma(x)
    =\frac{3M}{2\pi R^2} \sqrt{1-u^2}~ \Theta(1-u)~,
\end{equation}
where $u=x/R$. The corresponding convergence is
\begin{equation}
    \kappa(x)
    =\frac{3x_E^2}{2R^2} \sqrt{1-u^2}~ \Theta(1-u)~.
\end{equation}
The lens potential is
\begin{equation}\label{eq:sphere}
\psi(x)=
\begin{cases}
x_E^2\left[
\dfrac{4}{3} -\dfrac{4}{3}\sqrt{1-u^2} +\dfrac{1}{3}u^2\sqrt{1-u^2}
+\ln\!\left(\dfrac{1+\sqrt{1-u^2}}{2}\right) \right], & u\leq1~, \\[8pt]
x_E^2\left[ \ln u+\dfrac{4}{3}-\ln2 \right], & u>1~.
\end{cases}
\end{equation}

\subsubsection{Soliton}

The soliton profile used in this work is motivated by the possible ultralight DM distribution, for which we follow  Ref.~\cite{Schive:2014dra}. We adopt
\begin{equation}
    \rho(r)
    =\frac{1024M}{33\pi^2R^3} \frac{1}{[1+(r/R)^2]^8}~.
\end{equation}
Its projected surface mass density is
\begin{equation}
    \Sigma(x)
    =\frac{13M} {2\pi R^2(1+u^2)^{15/2}}~,
\end{equation}
where $u=x/R$, and the convergence is
\begin{equation}
    \kappa(x) =\frac{13x_E^2} {2R^2(1+u^2)^{15/2}}~.
\end{equation}
The corresponding lens potential is
\begin{equation}
\psi(x)=x_E^2\left[ \ln\!\left(1+\ell\right)
-\frac{1}{\ell} -\frac{1}{3\ell^3} -\frac{1}{5\ell^5} -\frac{1}{7\ell^7} -\frac{1}{9\ell^9} -\frac{1}{11\ell^{11}} \right]~,
\end{equation}
where
\begin{equation}
    \ell=\sqrt{1+u^2}~.
\end{equation}

\subsection{Point mass lens surrounded by a power law halo}
\label{app:dress}

We model a compact object surrounded by a power law halo as the sum of a central point mass and an extended halo. For a central mass $M$,
\begin{equation}
    \kappa_{\rm p}(\bm{x})
    =\pi x_E^2\delta^{(2)}(\bm{x})~,
\end{equation}
and
\begin{equation}
    \psi_{\rm p}(x) =x_E^2\ln\!\left(\frac{x}{x_E}\right)~,
\end{equation}
where
\begin{equation}
    x_E=\sqrt{4d_{\rm eff}M}~.
\end{equation}

We assume a spherically symmetric halo with the density profile, following discussion in the main text
\begin{equation}
    \rho_h(r)
    =\rho_0 \left(\frac{r}{R_h}\right)^{\gamma-1}~.
\end{equation}
We apply this power law at all radii $r>0$, without an inner or outer cutoff. Here $R_h$ is a reference radius at which $\rho_h=\rho_0$, not an imposed outer boundary. The halo mass enclosed within any finite radius is finite for $\gamma>-2$, although the total mass of the untruncated profile diverges at large radii in the range considered here. This idealized profile does not include changes of slope or truncation associated with halo formation and subsequent evolution.

Projecting along the line of sight gives
\begin{equation} 
    \Sigma_h(x)
     = \int_{-\infty}^{\infty} dz~ \rho_h\!\left(\sqrt{x^2+z^2}\right) =
    \frac{M}{R_h^2} \sqrt{\pi}~ \mathcal{R} \left(\frac{x}{R_h}\right)^\gamma
    \frac{\Gamma(-\gamma/2)}{\Gamma[(1-\gamma)/2]}~, 
\end{equation}
where
\begin{equation}
    \mathcal{R} = \frac{\rho_0R_h^3}{M}~.
\end{equation}
The corresponding convergence is
\begin{equation}
    \kappa_h(x) = \frac{x_E^2}{R_h^2} \pi^{3/2}\mathcal{R} \left(\frac{x}{R_h}\right)^\gamma
    \frac{\Gamma(-\gamma/2)}{\Gamma[(1-\gamma)/2]}~.
\end{equation}
Solving Eq.~\eqref{eq:poisson} gives
\begin{equation}
    \psi_h(x) = x_E^2 \frac{2\pi^{3/2}}{(2+\gamma)^2}
    \mathcal{R} \left(\frac{x}{R_h}\right)^{\gamma+2}
    \frac{\Gamma(-\gamma/2)}{\Gamma[(1-\gamma)/2]}~.
\end{equation}
The total lens potential is therefore
\begin{equation}
    \psi(x)
    =
    \psi_{\rm p}(x)+\psi_h(x)~.
\end{equation}

The upper bound $\gamma\leq-1$ ensures a two image configuration for a nonzero source impact parameter. For the total lens potential, the radial eigenvalue of the lensing Jacobian is
\begin{equation*}
    1-\psi''(x) = 1+\frac{x_E^2}{x^2} -\frac{2(\gamma+1)}{\gamma+2}\kappa_h(x)~.
\end{equation*}
For $-2<\gamma\leq-1$, this expression is strictly positive, so there is no radial critical curve. The lens equation is monotonic on each side of the central point mass and gives one geometric image on each side. At exact alignment, the images form an Einstein ring. For $-1<\gamma<0$, the halo term can instead make the radial eigenvalue negative over a finite interval, allowing radial critical curves and four-image configurations for suitable halo normalizations and source positions. We do not consider these configurations here.

For the halo dressed-PBH application in Sec.~\ref{sec:prospectdress}, we adopt
\begin{equation}
    \rho_0 = 0.26~M_\odot~{\rm pc}^{-3}~,
    \qquad
    R_h = 0.61~{\rm pc}
    \left(\frac{97M_{\rm PBH}}{M_\odot}\right)^{1/3}~,
\end{equation}
which gives $\mathcal{R}\simeq5.72$. These values are motivated by a DM halo that could form around a PBH by the time of structure formation, considering redshifts of around $z \simeq 30$~\cite{Mack:2006gz}. This was adopted for the $\gamma=-1.25$ benchmark in Ref.~\cite{GilChoi:2023ahp}. For the other slopes considered in this work, we retain $\rho_0$ and $R_h$ fixed at a given $M_{\rm PBH}$ to isolate the dependence on $\gamma$. This choice is a controlled variation of the halo profile rather than a self consistent halo growth model for arbitrary $\gamma$.

\section{Caustics of extended mass lenses}
\label{app:caustic}

In this appendix, we determine the caustics of the uniform disk, uniform sphere, and soliton lenses used in Sec.~\ref{sec:prospectextended}. For these axisymmetric extended lenses, the radial caustic separates projected source positions that produce one geometric image from those that produce three. We use this boundary to impose the single image condition in the main analysis and thereby exclude interference between multiple geometric images.

The lens equation is
\begin{equation}\label{eq:deflection}
    \bm{x}_s = \bm{x} - \nabla\psi(\bm{x})~.
\end{equation}
Critical curves are defined by the singularity of the lensing Jacobian,
\begin{equation}
    A_{ij}(\bm{x}) = \delta_{ij} - \frac{\partial^2\psi(\bm{x})}{\partial x_i\partial x_j}~,
\end{equation}
with $\det A=0$. Their images under Eq.~\eqref{eq:deflection} define  caustics~\cite{Schneider:1992bmb}. For an axisymmetric lens,
\begin{equation}
    \det A = \left( 1-\frac{1}{x}\frac{d\psi}{dx} \right)
    \left( 1-\frac{d^2\psi}{dx^2} \right)~.
\end{equation}
The tangential critical curve, determined by $\psi'/x=1$, maps to $\bm{x}_s=\bm{0}$. The finite radial caustic relevant to the image multiplicity is obtained from
\begin{equation}
    \psi''(x)=1
\end{equation}
and the corresponding projected source position from Eq.~\eqref{eq:deflection}.

For the uniform disk, Eq.~\eqref{eq:disk} gives
\begin{equation}
    \psi''(x)
    =
    \begin{cases}
        x_E^2/R^2~, & x<R~,\\[6pt]
        -x_E^2/x^2~, & x>R~.
    \end{cases}
\end{equation}
The sharp disk boundary produces a discontinuity in $\psi''$ at $x=R$. We interpret this profile as the limiting case of a disk with a smooth but increasingly sharp boundary. When $x_E>R$, the corresponding radial critical curve approaches $x=R$, giving the caustic radius
\begin{equation}
    x_c = \frac{x_E^2}{R}-R~,
    \qquad
    x_E>R~.
\end{equation}
No finite radial caustic is present for $x_E\leq R$.

For the uniform sphere, defining
\begin{equation}
    u=\frac{x}{R},
    \qquad
    u_E=\frac{x_E}{R}~,
\end{equation}
Eq.~\eqref{eq:sphere} gives
\begin{equation}
    \psi''(x)
    =
    \begin{cases}
    u_E^2
    \left[ -u^{-2} +(2+u^{-2})\sqrt{1-u^2} \right]~, & u\leq1~,\\[8pt] 
    -u_E^2/u^2~, & u>1~.
    \end{cases}
\end{equation}
The maximum occurs at the lens center,
\begin{equation}
    \psi''(0)=\frac{3}{2}u_E^2~,
\end{equation}
so a radial critical curve exists only for $u_E>\sqrt{2/3}$. Its radius $\hat u$ is
\begin{equation}
    \hat{u}
    =
    \frac{ \sqrt{ -1+\sqrt{1-32u_E^6+48u_E^8} }}{2\sqrt{2}~u_E^2}~.
\end{equation}
The corresponding caustic radius is
\begin{equation}
    \frac{x_c}{R}
    =
    u_E^2\hat{u}^{-1}
    \left[ 1-(1-\hat{u}^2)^{3/2} \right] - \hat{u}~,
    \qquad
    u_E>\sqrt{\frac{2}{3}}~.
\end{equation}

\begin{figure}[t]
    \centering
    \includegraphics[width=0.5\linewidth]{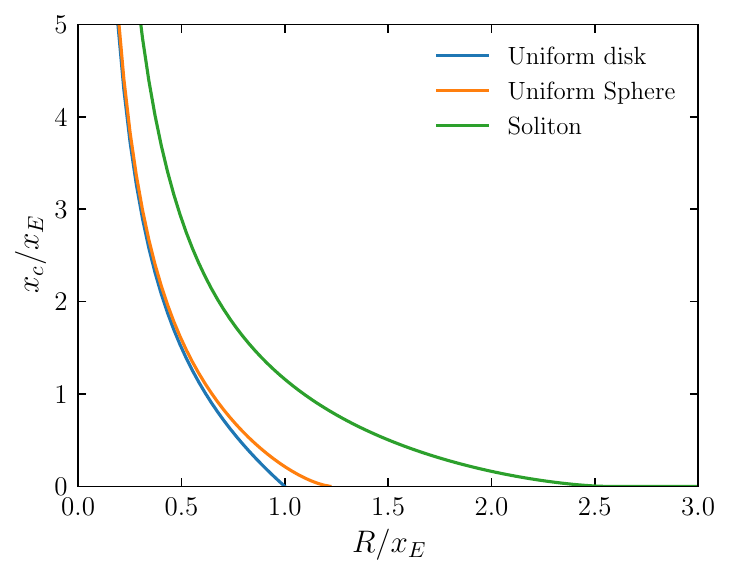}
    \caption{
    Radial caustic radius $x_c/x_E$ as a function of $R/x_E$ for the uniform disk, uniform sphere, and soliton lenses.
    }
    \label{fig:caustics}
\end{figure}

For the soliton profile, the radial critical curve is obtained numerically from $\psi''(x)=1$, and the corresponding $x_c$ follows from Eq.~\eqref{eq:deflection}.  Fig.~\ref{fig:caustics} compares the resulting caustic radii for the three extended mass profiles.

For the same $M$ and $R$, the soliton develops a somewhat larger caustic than the uniform disk and sphere because its mass distribution is more centrally concentrated. The parameter $R$ characterizes the overall scale of each profile but does not correspond to the same enclosed mass radius or density contour across the three models.

\section{Geometric optics matching of two image lenses}
\label{app:matching}

The halo-dressed lenses considered in Sec.~\ref{sec:prospectdress} have two geometric images for a nonzero source impact parameter, as discussed in Appendix~\ref{app:dress}. Following Ref.~\cite{GilChoi:2023ahp}, we match their relative magnification and time delay to those of an effective point mass lens. When the two stationary points are a minimum and a saddle, the amplification factor is
\begin{equation}
    F(f)\simeq |\mu_1|^{1/2}e^{2\pi i f t_1}
    + |\mu_2|^{1/2}e^{2\pi i f t_2-i\frac{\pi}{2}}~,
\end{equation}
where $\mu_{1,2}$ and $t_{1,2}$ are the image magnifications and arrival times, respectively~\cite{nakamura1999wave}. Apart from a common arrival time, the geometric optics signal is characterized by the primary-image magnification $\mu_1$, the magnification ratio
\begin{equation}
    \mu_r=\left|\frac{\mu_2}{\mu_1}\right|~,
\end{equation}
and the relative time delay
\begin{equation}
    \Delta t=t_2-t_1~.
\end{equation}

For a point mass lens, $\mu_r$ and $\Delta t$ are determined by the redshifted lens mass $M_l=M(1+z_l)$ and the normalized source impact parameter $y=x_s/x_E$. This gives a one-to-one mapping between $(\mu_r,\Delta t)$ and $(M_l,y)$~\cite{GilChoi:2023ahp}. The absolute magnification $\mu_1$, however, is not independently adjustable for a pure point mass lens.

To reproduce $\mu_1$ as well, we supplement the point mass with a constant mass sheet of convergence $\kappa_b<1$. The resulting point mass-plus-mass-sheet model provides a physical comparison lens whose geometric optics observables $(\mu_1,\mu_r,\Delta t)$ can be matched to those of the original two image lens.\footnote{
Adding a constant mass sheet corresponds to adding $\kappa_b|\bm{x}|^2/2$ to the lensing potential. In wave optics,
$F(f;\bm{x}_s,\psi,\kappa_b)=(1-\kappa_b)^{-1}
F(f;\tilde{\bm{x}}_s,\tilde{\psi},0)$,
where $\tilde{\bm{x}}_s=\bm{x}_s/\sqrt{1-\kappa_b}$ and
$\tilde{\psi}(\bm{x})=\psi(\bm{x}/\sqrt{1-\kappa_b})$~\cite{Goyal:2025eqo}.
The prefactor changes the overall amplification, while the frequency-dependent phase structure is preserved. This relation should be distinguished from the usual mass-sheet transformation~\cite{Cremonese:2021puh,Chen:2024xal,Meena:2024glj}, which preserves image positions and flux ratios.
}

The mapping between $(\mu_1,\mu_r,\Delta t)$ and $(M_l,y,\kappa_b)$ is
\begin{align}
    \tilde{y} & = \sqrt{\mu_r^{1/2}+\mu_r^{-1/2}-2}~,\\
    M_l &= \frac{\Delta t}{2\left(\sqrt{\mu_r +\mu_r^{-1}-2}-\ln \mu_r\right)}~,\\
    \kappa_b &= 1 - \sqrt{\frac{\mu_+(\tilde{y})}{\mu_1}}~,
\end{align}
where $\tilde{y}=y/\sqrt{1-\kappa_b}$ and $\mu_+$ is the magnification of the primary image of a point mass lens,
\begin{equation}
    \mu_+(y)=\frac{1}{2}+\frac{y^2+2}{2y\sqrt{y^2+4}}~,
\end{equation}
as given in Ref.~\cite{takahashi2003wave}. Equivalently, $y$ and $\kappa_b$ can be written directly in terms of $(\mu_1,\mu_r)$ as
\begin{align}
    y &= \frac{1-\mu_r^{1/2}}{\left[\mu_1\mu_r(1-\mu_r)\right]^{1/4}}~,\\
    \kappa_b &= 1 - \frac{1}{\sqrt{\mu_1(1-\mu_r)}}~.
\end{align}

We use this matched point mass plus mass sheet lens as the comparison model for halo-dressed compact lenses in the main text. Matching the geometric optics observables removes the signal features that can be reproduced by this simpler lens configuration, allowing the remaining frequency-dependent difference to isolate diffraction produced by the surrounding mass distribution.

\section{Impact of intrinsic source parameters on  lensing probability}
\label{app:source_parameters}

In Sec.~\ref{sec:statistics}, the population forecasts perform the minimization in Eq.~\eqref{eq:lambdaexp} only over the overall amplitude, constant phase shift, and time shift. Holding the intrinsic source parameters fixed can underestimate their degeneracy with the lensing-induced waveform distortion and thereby overestimate $\ln\Lambda$ and the inferred lensing probability. Here, we repeat the minimization with selected intrinsic source parameters included for representative benchmark configurations.

\begin{figure*}[t]
    \centering
    \begin{subfigure}[t]{0.48\textwidth}
        \centering
        \includegraphics[width=\linewidth]{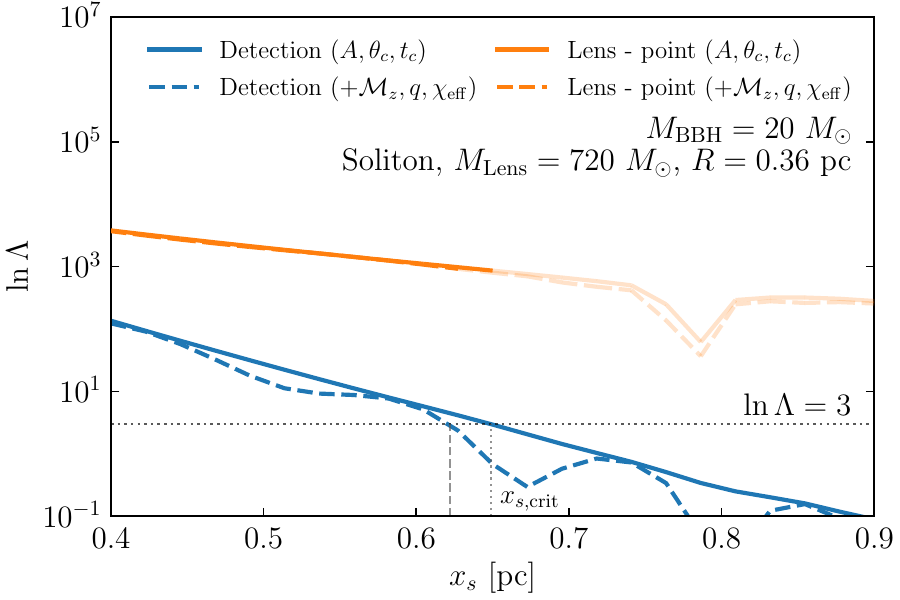}
    \end{subfigure}
    \hfill
    \begin{subfigure}[t]{0.48\textwidth}
        \centering
        \includegraphics[width=\linewidth]{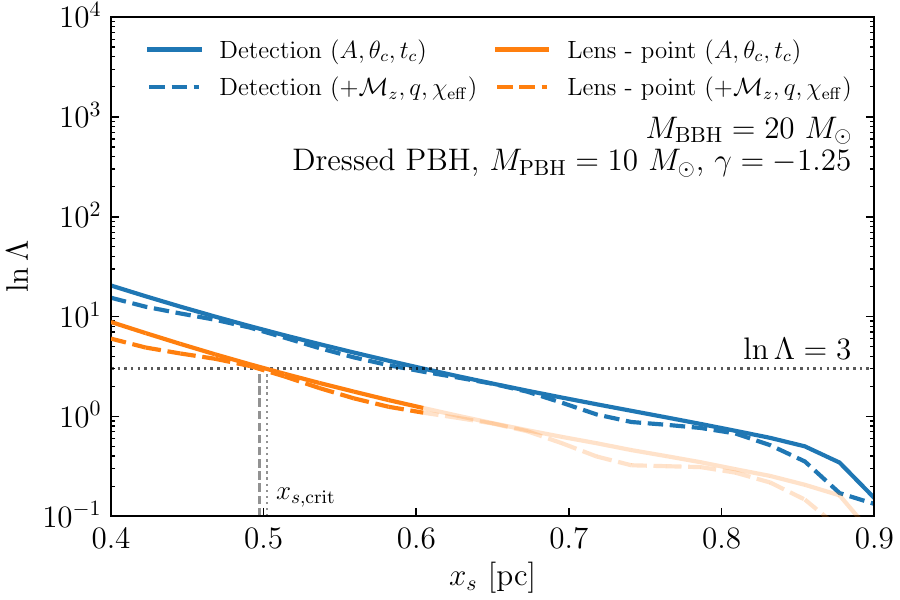}
    \end{subfigure}

    \begin{subfigure}[t]{0.48\textwidth}
        \centering
        \includegraphics[width=\linewidth]{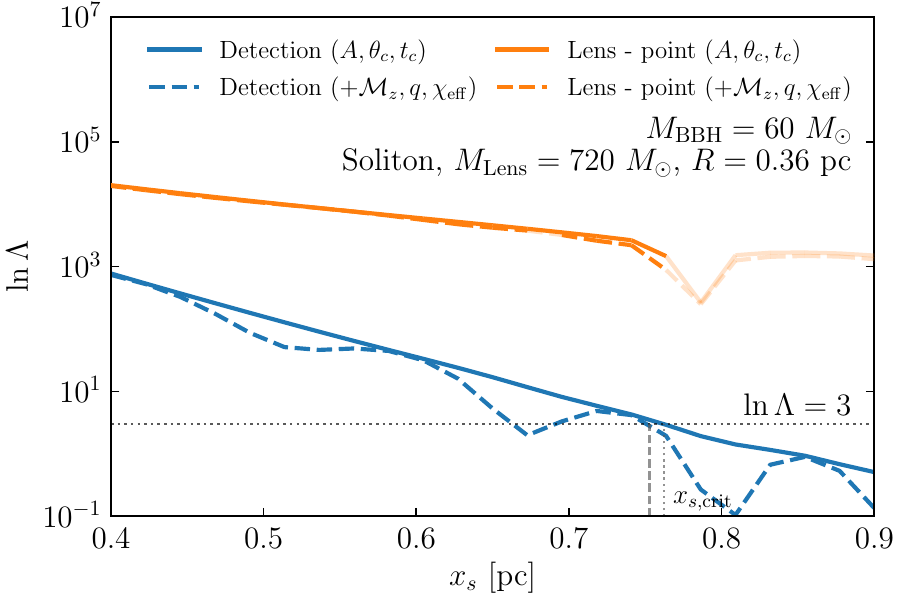}
    \end{subfigure}
    \hfill
    \begin{subfigure}[t]{0.48\textwidth}
        \centering
        \includegraphics[width=\linewidth]{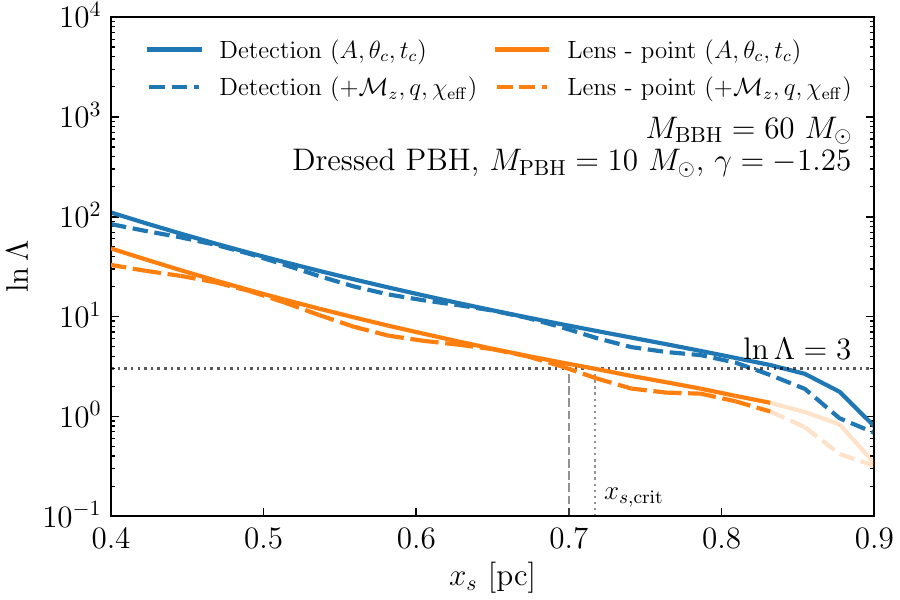}
    \end{subfigure}
    \caption{
    Results of the minimization in Eq.~\eqref{eq:lambdaexp} for the selected soliton and halo dressed-PBH lenses. Blue and orange curves show $\ln\Lambda_{\rm det}$ and $\ln\Lambda_{\rm dist}$, respectively. Solid curves include only the overall amplitude, constant phase shift, and time shift in the minimization. Dashed curves repeat the minimization with $\mathcal{M}_z$, $q$, and $\chi_{\rm eff}$ also included. The horizontal line marks $\ln\Lambda=3$, and the vertical lines mark the outermost critical source impact parameters for which both criteria are satisfied. Portions of the distinction curve are faded where the detection criterion fails.
    }
    \label{fig:source_parameter_optimization}
\end{figure*}

We construct zero noise injections using non-precessing \texttt{IMRPhenomD}  waveforms~\cite{Husa:2015iqa,Khan:2015jqa} and the CE noise curve. The injected BBHs are at $z_s=0.3$ and have source-frame total masses of $20~M_\odot$ and $60~M_\odot$, with $q=1$, $\chi_{\rm eff}=0$, and $\chi_a=0$. Here $q=m_2/m_1$ with $m_1\geq m_2$, $\chi_{\rm eff}=(m_1\chi_{1z}+m_2\chi_{2z})/(m_1+m_2)$, and $\chi_a=(\chi_{1z}-\chi_{2z})/2$, where $\chi_{iz}$ is the dimensionless component spin along the orbital angular momentum.

The lens configurations are selected from favorable regions of the main forecasts. The soliton has $M=719.7~M_\odot$ and $R=0.359~{\rm pc}$, close to the maximum CE lensing probability in Fig.~\ref{fig:extlens_prob}. The halo dressed-PBH lens has $M_{\rm PBH}=10~M_\odot$ and $\gamma=-1.25$, near the minimum projected $f_{\rm PBH}$ for CE in Fig.~\ref{fig:fDMdressDist}. The remaining halo parameters are those of the benchmark in Sec.~\ref{sec:prospectdress}. For distinction, at each source impact parameter the soliton is compared with a point mass of the same mass, whereas the halo-dressed PBH is compared with the effective point mass lens constructed in Appendix~\ref{app:matching}. Both configurations use $z_l=0.1345$, which maximizes $d_{\rm eff}$ for $z_s=0.3$.

 Fig.~\ref{fig:source_parameter_optimization} presents the results of the minimization in Eq.~\eqref{eq:lambdaexp}. The solid curves include only the overall amplitude, constant phase shift, and time shift, as in the main analysis. The dashed curves additionally include the detector frame chirp mass $\mathcal{M}_z$, $q$, and $\chi_{\rm eff}$. The minimization covers $0.7\leq\mathcal{M}_z/\mathcal{M}_{z,{\rm inj}}\leq1.3$, $0.25\leq q\leq1$, and $-0.8\leq\chi_{\rm eff}\leq0.8$. We restrict the non-precessing spins to the equal-spin configuration, $\chi_{1z}=\chi_{2z}=\chi_{\rm eff}$, and therefore keep $\chi_a=0$ while varying $\chi_{\rm eff}$. For each source impact parameter, the dashed curve gives the minimum $\ln\Lambda$ found over this parameter space.

Fig.~\ref{fig:source_parameter_optimization} shows that the lensing cross section decreases by within $\mathcal{O}(10)\%$ across the four cases when the intrinsic source parameters are included in the minimization. Individual values of $\ln\Lambda$ can nevertheless change substantially at fixed $x_s$. Since $\ln\Lambda$ varies steeply with $x_s$ near the threshold, a large change in $\ln\Lambda$ produces a much smaller shift of the critical source impact parameter.

Variations of the intrinsic parameters $\mathcal{M}_z$, $q$,
and $\chi_{\rm eff}$ can partially absorb lensing-induced
phase and amplitude distortions, reducing the residual
waveform differences and hence $\ln\Lambda$. For the configurations tested here the resulting changes in the critical impact parameters  and hence in the cross-sections  remain modest. Thus, our   benchmark results  indicate limited changes in the corresponding cross-sections when the specified intrinsic parameters are included.

The detection and distinguishability statistics have no general ordering, because they use different reference waveforms. In the soliton examples, the high frequency lensing modulation decreases as the extended profile is resolved, while the point mass waveform retains a stronger modulation. The soliton signal can therefore be harder to detect than to distinguish from the point mass signal.

The benchmarks considered here sample lens configurations
among those yielding the strongest projected sensitivities
in Figs.~\ref{fig:extlens_prob} and \ref{fig:fDMdressDist}.
The small changes in their critical source impact parameters
indicate that these sensitivities remain largely unchanged
when the specified intrinsic source parameters are varied.
Establishing the general robustness of the forecasts will
require a joint analysis over broader source and lens
parameter spaces.

\bibliographystyle{JHEP}
\phantomsection
\addcontentsline{toc}{section}{References}
\bibliography{reference}

\end{document}